\documentclass[11pt,a4paper]{article}
\pdfoutput=1

\usepackage{jheppub}
\usepackage{lmodern}
\usepackage{tocvsec2}

\usepackage{amsmath}
\usepackage{mathtools}
\usepackage{slashed}
\usepackage{bm}

\usepackage{float}
\usepackage{pgf}
\usepackage{tikz}
\usepackage{xcolor}

\usepackage{booktabs}

\usepackage[capitalise,noabbrev]{cleveref}
\crefname{equation}{}{}

\hypersetup{bookmarksopen=false}

\makeatletter
\renewcommand*{\maketag@@@}[1]{{\hbox{\m@th\normalfont\normalsize#1}}}%
\makeatother%

\newcommand*{\sqrtsmash}[2][]{\sqrt{\vphantom{#1}\smash[b]{#2}}}

\newcommand{\+}{{\mkern2mu}}  

\newcommand{\mN}{\mathcal N}
\newcommand{\mD}{\mathcal D}

\renewcommand*{\vec}[1]{\boldsymbol{#1}}
\newcommand*{\mode}{\mathcal{Y}}

\newcommand{\scale}{0.6cm}
\pgfmathsetmacro{\range}{4}
\pgfmathsetmacro{\gridgreylevel}{30}
\pgfmathsetmacro{\margin}{0.25}
\newcommand*{\crossdiagramlayout}[1]{%
    \foreach \i in {1,...,\range} {
        \draw[very thin,black!\gridgreylevel] (-\margin,\i) node[left,black] {$\i$} -- ({\range+\margin},\i);
    }
    \foreach \i in {1,...,\range} {
        \draw[ultra thin,black!\gridgreylevel] (\i,-\margin) node[below,black] {$\i$} -- (\i,{\range+\margin});
    }

    \draw[-latex,black] (-\margin,0) node[left] {$0$} -- ([xshift=4pt]{\range+\margin},0) node[below,xshift=2pt] {$p$};
    \draw[-latex,black] (0,-\margin) node[below] {$0$} -- ([yshift=4pt]0,{\range+\margin}) node [left,yshift=2pt] {$q$};

    \node[above] at ({\range/2},{\range+\margin}) {$r=#1$};
}

\newcommand*{\colouredcrossdiagramlayout}[1]{%
    \foreach \i in {0,...,\range} {
        \draw[very thin,black!\gridgreylevel] (-0.5,\i) node[left,black] {$\i$} -- ({\range+0.5},\i);
    }
    \foreach \i in {0,...,\range} {
        \draw[ultra thin,black!\gridgreylevel] (\i,-0.5) node[below,black] {$\i$} -- (\i,{\range+0.5});
    }

    \draw[black] (-0.5,-0.5) -- ({\range+0.5},-0.5);
    \draw[black] ({\range+0.5},-0.5) -- ({\range+0.5},{\range+0.5});
    \draw[black] ({\range+0.5},{\range+0.5}) -- (-0.5,{\range+0.5});
    \draw[black] (-0.5,{\range+0.5}) -- (-0.5,-0.5);

    \node[above] at ({\range/2},{\range+0.5}) {$r=#1$};
    \node[below] at ({\range/2},{-1.4}) {$p$};
    \node[left] at ({-1.2},{\range/2}) {$q$};
}

\title{The complete Kaluza--Klein spectra of $\bm{\mathcal{N}=1}$ and $\bm{\mathcal{N}=0}$ M-theory on $\bm{AdS_4 \times (\text{squashed}\ S^7)}$}

\author[a]{Joel Karlsson}
\author[b]{and Bengt E.W.~Nilsson}
\affiliation[a]{Institute for Theoretical Physics, KU Leuven,\\ Celestijnenlaan 200D, B-3001 Leuven, Belgium}
\affiliation[b]{Department of Physics, Chalmers University of Technology,\\ SE-412 96 G\"oteborg, Sweden}

\emailAdd{joel.karlsson@kuleuven.be}
\emailAdd{tfebn@chalmers.se}

\abstract{%
The squashed seven-sphere operator spectrum is completed by deriving the spectrum of the spin-3/2 operator.
The implications of the results for the $AdS_4$ \mbox{$\mN = 1$}\linebreak supermultiplets obtained from compactification of eleven-dimensional supergravity are analysed.
The weak $G_2$ holonomy plays an important role when solving the eigenvalue equations on the squashed sphere.
Here, a novel and more universal algebraic approach to the whole eigenvalue problem on coset manifolds is provided.
Having obtained full control of all the operator spectra, we can finally determine the irreps $D(E_0, s)$ for all supermultiplets in the left-squashed vacuum.
This includes an analysis of possible boundary conditions.
By performing an orientation flip on the seven-sphere, we also obtain the full spectrum for the non-supersymmetric right-squashed compactification which is of interest in the swampland context and in particular for the $AdS$ swampland conjecture.
Here, a number of boundary condition ambiguities arise making the analysis of dual marginal operators somewhat involved.
This work is a direct continuation of \cite{Nilsson:2018lof} and \cite{Ekhammar:2021gsg}.%
}


\begin{document}
\pagenumbering{roman}
\maketitle
\pagenumbering{roman}
\setcounter{page}{2}

\clearpage

\pagenumbering{arabic}
\section{Introduction}

Recently there has appeared a number of papers discussing various non-supersymmetric $AdS$ flux-solutions in ten- and
eleven-dimensional supergravity aiming at determining their stability properties, perturbatively as well as non-perturbatively.
The interest in this issue is a direct result of the so called swampland program \cite{Vafa:2005ui, Ooguri:2006in}
and in particular the ``$AdS$ swampland conjecture'' \cite{Ooguri:2016pdq,Freivogel:2016qwc} claiming that any $AdS$ flux-compactification without supersymmetry must be unstable, either
perturbatively or, if the solution passes this test, non-perturbatively. In the former case,   the Breitenlohner--Freedman (BF) stability  criterion \cite{Breitenlohner:1982jf,Breitenlohner:1982bm} is easily checked once
the complete linearised spectrum, or at least some crucial parts of it,  is known. In the context of $AdS/CFT$,
this stability corresponds,  in the infinite $N$ limit  of the dual field theory, to unitary conformal dimensions of single trace operators, and the vanishing of the corresponding $\beta$ functions. In non-supersymmetric scenarios, there have been concerns raised in the literature  (see, e.g., \cite{Witten:2001ua,Aharony:2002hx,Dymarsky:2005uh,Dymarsky:2005nc,Pomoni:2009joh,Aharony:2015afa}) regarding the $\beta$ functions of marginal multi-trace operators that might be nonvanishing even if all single-trace $\beta$ functions vanish%
\footnote{We are grateful to the referee for emphasising this issue and to Igor Klebanov for helpful discussions.}.
If one could find an example where this does not happen at leading order in $1/N$, there could still be subleading $1/N$ effects that destroy the conformal fixed point \cite{Berkooz:1998qp} (see also, e.g., \cite{Murugan:2016aty} and references therein). In the bulk, this corresponds to an instability caused by a tadpole of a composite field that destroys the BF-stable non-supersymmetric vacuum. We emphasise that the purpose of this paper is not to add to this discussion. Instead, one of the main results of our work is the possibility to choose boundary conditions  (in compactifications on the squashed seven-sphere) in such a way that all marginal operators on the boundary are eliminated. This possibility was first noted in the singlet sector in \cite{Nilsson:2023ctq} and will be explained in full detail later in this paper.

The above perturbative aspects of the stability question  mean that
one essentially needs to have full control of the whole spectrum. At an even deeper level, one must discuss the possibility of instabilities due to, e.g., \emph{bubbles of nothing} \cite{Witten:1981gj} (see also the generalisation discussed in \cite{Ooguri:2017njy}) and other non-perturbative issues.
Note that even if it is possible to establish stability under all known decay modes in a non-supersymmetric context, it might be very difficult, if not impossible, to produce  a strict proof of stability against
all decay modes, known or not. An interesting attempt in this direction has, however, been made recently in \cite{Giri:2021eob}.

In this line of investigation there have also been efforts to find the complete spectrum of certain compactifications that can be viewed as up-lifted stable extrema of maximal supergravity theories
in  lower dimensions. With the improved knowledge of these spectra it was possible, in some cases, to pinpoint certain massive modes in the Kaluza--Klein spectrum that give rise
to so called brane-jet instabilities \cite{Bena:2020xxb}.

In the same vein, there are also some special non-supersymmetric BF-stable  solutions that have been discussed recently.
These include the interesting case of the non-supersymmetric $G_2$ solution first found in massive type II supergravity, see, e.g., \cite{Lust:2008zd,Borghese:2012qm}.%
\footnote{For some more recent extensive  searches for vacuum solutions, see, e.g., \cite{Bobev:2019dik,Bobev:2020qev}.}
This case seemed for a long time to clash with the $AdS$
swampland  conjecture \cite{Guarino:2020jwv,Guarino:2020flh}  but was
recently \cite{Bomans:2021ara,Dibitetto:2021ltm} shown to be afflicted by the non-perturbative phenomenon of nucleation of a bubble of nothing first described by Witten \cite{Witten:1981gj}.
Another attempt to find a solution of type IIB supergravity in $D=10$ that violates the $AdS$ swampland conjecture is analysed in \cite{Giambrone:2021wsm} based on a  non-geometric S-fold construction. This case is further
discussed in the review \cite{Guarino:2022tlw}, where its chances of surviving all known decay modes is commented upon. It is probably fair to say that its full stability is far from being demonstrated
although it does possess some novel and quite interesting features.

At the point in time when this  is written all known non-supersymmetric $AdS$  flux-compactifications have been shown to be unstable (with a slight caveat for the last case mentioned above) except for the skew-whiffed
squashed seven-sphere solution discussed in \cite{Duff:1986hr}. The reason this solution of eleven-dimensional supergravity has  not been fully analysed until now is that the results in \cite{Nilsson:2018lof,Ekhammar:2021gsg} were still lacking the spin-3/2 eigenvalues on the squashed seven-sphere and that the methods  used in \cite{Malek:2019eaz, Malek:2020mlk} have not been applicable to
vacua that are not lifts from
$\mN=8$ supergravity in four dimensions. However, with the present work and the recent results of \cite{Duboeuf:2022mam}, the shortcomings in both these cases have been overcome.
For further comments on the relation between our work and \cite{Duboeuf:2022mam}, see the end of this introduction.

In this  paper we finish the work started in \cite{Nilsson:2018lof,Ekhammar:2021gsg} by completing the determination of the spectrum in the sense that we have finally
solved also the spin-3/2 equation and thus we have now determined  the spectra of all relevant eigenvalue equations on the squashed seven-sphere. This result has already been reported on in the
MSc thesis \cite{Karlsson:2021oxd} by one of the present authors.
Applying this to the left-squashed case, we then  provide a unique assignment of
masses and energies to the $\mN=1$ supermultiplets, up to some ambiguities related to boundary conditions. As we will see below, this process produces certain degeneracies in the spectrum. We will, towards the end of the
paper,  discuss this further without providing a clear answer as to why this happens. This spectrum degeneracy is apparent both from the supermultiplet analysis and the explicit construction of the two-form mode functions on the squashed seven-sphere.

The stability issue of the non-supersymmetric right-squashed seven-sphere can now be addressed beyond its BF stability
\cite{Duff:1984sv,Duff:1983ajq}. Here, we analyse in detail which  marginal operators can appear that might lead to the kind of perturbative instability mentioned above.
These latter results generalise those for the singlet sector obtained recently in \cite{Nilsson:2023ctq}.

This paper is organised as follows. In \cref{sec:review_and_method}, we give some background to the whole problem of finding the
spectrum of eleven-dimensional supergravity compactified on the squashed seven-sphere, following  \cite{Duff:1986hr}, and
present new universal versions (derived in the MSc thesis \cite{Karlsson:2021oxd}) of some of the key coset space formulas used in \cite{Duff:1986hr} and particularly in the $G_2$ approach of \cite{Ekhammar:2021gsg}. This involves
obtaining a more powerful group theoretic expression for the action of the Weyl tensor on the various isotropy irreps appearing in the spectrum analysis.

In \cref{sec:eigenvalues}, we rederive the operator spectra previously derived in  \cite{Duff:1986hr} and \cite{Ekhammar:2021gsg} and complete this by
obtaining also the so far missing spectrum, namely the spin-3/2 one, following \cite{Karlsson:2021oxd}.
Then, in \cref{sec:eigenmodes}, all two-form modes  are constructed explicitly and their eigenvalues  derived with some intriguing implications.
In \cref{sec:supermultiplet}, we use the now complete information on the operator eigenvalues  on the squashed seven-sphere to give masses and $E_0$ values to all the  supermultiplets.
We find  here that supersymmetry does not prevent a certain amount of ambiguities related to the possibility to choose different boundary conditions in $AdS_4$.
These features are also analysed  for the skew-whiffed theory without supersymmetry which has a rather large number of fields with undetermined boundary conditions. This section ends
with a  discussion on  marginal operators and the implications for the issue of stability.
The eigenvalue degeneracy problem that arose in previous sections is addressed.   Finally, the situation is summarised and commented upon in the concluding \cref{sec:conclusions}.

Some of the results in this paper, in particular the improved methods for computing the
eigenvalues of differential operators on coset spaces reported on in \cref{sec:review_and_method} and the eigenvalues of the  spin-3/2 operator on the Einstein-squashed seven-sphere
derived in \cref{sec:eigenvalues}, have already appeared in the MSc thesis \cite{Karlsson:2021oxd}.
A preliminary study of how the possible values of $E_0$ fit into supermultiplets was also conducted in \cite{Karlsson:2021oxd}.
Here, we complete this analysis by also studying the special cases that arise for non-generic isometry irreps.

During the final stages of this project, there appeared another paper \cite{Duboeuf:2022mam}, obtaining results on the spectrum of $E_0$ for low-lying levels in the bosonic sector of the squashed $S^7$ theory discussed here.
These results, derived in \cite{Duboeuf:2022mam} by entirely different methods from the ones presented here, coincide with ours where they can be compared.
\cite{Duboeuf:2022mam} provides one piece of information not yet obtained by the methods in this paper, namely the connection between $E_0$ values and irreps of the Wess--Zumino multiplets which contain scalar fields related to the Lichnerowicz operator on the squashed $S^7$.
This helps us resolve a final ambiguity in our work related to the assignment of multiplicities for these particular Wess--Zumino multiplets, as explained in \cref{sec:supermultiplet} (see e.g.\ \cref{fig:colour-split}).
This information could have been obtained by repeating the two-form eigenmode construction in \cref{sec:eigenmodes} for the metric modes on the squashed $S^7$, which has not yet been done.
In contrast to \cite{Duboeuf:2022mam}, our methods automatically provide the entire infinite towers in the spectra and we verify the presence of supersymmetry for all supermultiplets with maximum spin $2$, $3/2$ and $1$ by deriving the fermionic spectra independently.
Furthermore, we explicitly construct the complete infinite towers of two-form Laplacian eigenmodes on the squashed $S^7$ for the first time.
We also analyse the role of boundary conditions and find that there are multiple choices that respect supersymmetry in the left-squashed vacuum (see \cref{wesszuminopositivecgmultiplets}).
A limitation of our methods is that they rely on a coset description of the internal manifold which does not seem to be the case for the methods used in \cite{Duboeuf:2022mam}.%

\section{The squashed \texorpdfstring{$S^7$}{S7} eigenvalue problem}%
\label{sec:review_and_method}

To make this paper relatively self-contained we will in the first  subsection reproduce  a few tables from
previous works, the review   \cite{Duff:1986hr} and the more recent \cite{Nilsson:2018lof} and \cite{Ekhammar:2021gsg}.
These tables contain information of general validity and provide useful information for the discussions in this paper,
for instance how many  supermultiplets of each kind that appear in the
$\mN=1$ supersymmetric $AdS_4$ vacuum of eleven-dimensional supergravity on the left-squashed seven-sphere
\cite{Nilsson:2018lof}.
One reason for studying the supersymmetric left-squashed case first is to be able to
establish the correct operator eigenvalue spectra which would be a much more intricate task without the possibility to check the results using supersymmetry.
Having done this, we can then perform a skew-whiffing (i.e., an orientation flip) on the squashed seven-sphere to get the complete field spectrum in the non-supersymmetric
right-squashed case for which the $AdS_4$ stability issue will be analysed in later sections.

In the second subsection we then briefly describe our method, as used in the literature so far, to derive the eigenvalue spectra of the operators on the squashed seven-sphere appearing in the compactification
\cite{Duff:1986hr,Ekhammar:2021gsg}. The main purpose here is to present an improved  formalism by deriving new  more general and powerful versions of some of the coset space  formulas,
the implications of which  are presented in  the next section, largely  following \cite{Karlsson:2021oxd}.

\subsection{Review}
\label{sec:review_and_method:review}

The structure of $\mN=1$ supermultiplets in $AdS_4$ is presented in \cref{tab:AdS4_N=1_supermults}.
\begin{table}[H]
    \centering
    \begin{tabular}{ll}
        \toprule
        \textbf{Type A:} & Wess--Zumino multiplets for $E_0 > 1/2$ \\
            & $D(E_0, 0) \oplus D(E_0 + 1/2, 1/2) \oplus D(E_0 + 1, 0)$ \\[2pt]
        \textbf{Type B:} & Massive higher-spin multiplets for $E_0 > s+1$, $s \ge 1/2$ \\
            & $D(E_0, s) \oplus D(E_0 + 1/2, s+1/2) \oplus D(E_0 + 1/2, s-1/2) \oplus D(E_0 + 1, s)$ \\[2pt]
        \textbf{Type C:} & Massless higher-spin multiplets for $s \ge 1/2$ \\
            & $D(s+1, s) \oplus D(s+3/2, s+1/2)$ \\[2pt]
        \textbf{Type D:} & Dirac singleton \\
            & $D(1/2, 0) \oplus D(1, 1/2)$
        \\ \bottomrule
    \end{tabular}
    \caption{${\mathcal N}=1$ supermultiplets according to Heidenreich \cite{Heidenreich:1982rz}.}
    \label{tab:AdS4_N=1_supermults}
\end{table}

The spectrum of the $AdS_4$ gravity theory based on the  left-squashed seven-sphere contains supermultiplets of all kinds appearing in  \cref{tab:AdS4_N=1_supermults}
except for the singleton one.
However, singletons do appear for the  right-squashed seven-sphere without supersymmetry
due to a possible Higgs/de-Higgs relation of both squashed spectra to the round one as was discussed in \cite{Nilsson:2018lof,Nilsson:2023ctq}.  The former of these also argues that
the compactification on the round $S^7$ must incorporate the ${\mathcal N}=8$  singleton to be consistent. As also established in  \cite{Nilsson:2018lof}
the supermultiplet content  in the left-squashed case  is%
\footnote{\label{footnote:E0}Note that the $Spin(2,3)$-irreps $D(E_0,s)$ are reordered with the highest spin first compared to \cref{tab:AdS4_N=1_supermults}.}
\begin{alignat}{4}
    1 &\times \bigl(D(E_0,2^+) &&\oplus D(E_0-\tfrac{1}{2}, \tfrac{3}{2}) &&\oplus D(E_0+\tfrac{1}{2}, \tfrac{3}{2}) &&\oplus D(E_0, 1^+)\bigr), \\
    6 &\times \bigl(D(E_0,\tfrac{3}{2}) &&\oplus D(E_0-\tfrac{1}{2}, 1^{\pm}) &&\oplus D(E_0+\tfrac{1}{2}, 1^{\mp}) &&\oplus D(E_0, \tfrac{1}{2})\bigr), \\
    6 &\times \bigl(D(E_0,1^-) &&\oplus D(E_0-\tfrac{1}{2}, \tfrac{1}{2}) &&\oplus D(E_0+\tfrac{1}{2}, \tfrac{1}{2}) &&\oplus D(E_0, 0^-)\bigr), \\
    8 &\times \bigl(D(E_0,1^+) &&\oplus D(E_0-\tfrac{1}{2}, \tfrac{1}{2}) &&\oplus D(E_0+\tfrac{1}{2}, \tfrac{1}{2}) &&\oplus D(E_0, 0^+)\bigr), \\
    14 &\times \bigl(D(E_0, \tfrac{1}{2}) &&\oplus D(E_0-\tfrac{1}{2}, 0^{\pm}) &&\oplus D(E_0+\tfrac{1}{2}, 0^{\mp}) \mathrlap{\bigr),}
\end{alignat}
where the multiplicity (number in front) refers to the number of towers  that appear for the given supermultiplet. The reader is advised to consult the appendix in \cite{Nilsson:2018lof}
where a tower is given by a cross diagram where each cross is representing  an isometry group irrep with  the $Sp_2\times Sp_1$ Dynkin labels $(p,q;r)$ (see also \cref{sec:eigenmodes} of this paper).
We emphasise that every isometry irrep in all towers is accounted for by the list of supermultiplets above.

In order to connect the dimensionless energy  values $E_0$ for each spin to the operator eigenvalue spectra (see the next subsection or the next section)   we need two sets of data:
The relation between the $Spin(2,3)$  irrep data  $D(E_0,s)$ and the $AdS_4$ mass parameter $M^2$ (or $M$ for spinor fields) given in \cref{ezerolist} and, secondly,
the relations between the mass $M^2$ or $M$ and the operators on the internal manifold as given in \cref{table:massop}, valid for Freund--Rubin compactifications.%
\footnote{Note the change of notation compared to the review \cite{Duff:1986hr}: $0^{+(1)}\rightarrow 0^+_{\scriptscriptstyle (-)}$ and
$0^{+(3)}\rightarrow 0^+_{\scriptscriptstyle (+)}$ where the new notation used in this paper shows explicitly which branch of $M^2(0^+)$ is used. Note that $M^2(0^+_{\scriptscriptstyle (\pm)})$ is here sometimes written as $M^2_{\scriptscriptstyle (\pm)}(0^+)$.}
Here, $m$ is the inverse of the curvature radius of the internal space (i.e., $R_{(7)} = 42 m^2$), $F_{\mu\nu\rho\sigma} = 3 m \epsilon_{\mu\nu\rho\sigma}$ is the flux of the Freund--Rubin solution and the inverse of the curvature radius of $AdS_4$ is $2m$ (i.e., $R_{(4)} = -12(2m)^2$).
We define the orientation of $AdS_4$ such that $m > 0$ and use conventions in which the bosonic Lagrangian reads \cite{Duff:1986hr}
\begin{equation}
    2 \kappa \mathcal{L}
    = R - \frac{1}{12} F_{MNPQ} F^{MNPQ}
    + \frac{8}{12^4} \epsilon_{M_1 \hdots M_{11}} A^{M_1 M_2 M_3} F^{M_4 \hdots M_7} F^{M_8 \hdots M_{11}}.
\end{equation}

\begin{table}[H]
    \centering
    \renewcommand*{\arraystretch}{1.5} %
    \begin{tabular}{ll}
        \toprule
        $s=2$ & $E_0 = \frac{3}{2} + \frac{1}{2} {\displaystyle\sqrt{(M/m)^2+9}} \geq 3$ \\
        $s=\frac{3}{2}$ & $E_0 = \frac{3}{2} +\frac{1}{2}|{M/m-2}|\geq \frac{5}{2}$ \\
        $s=1$ & $E_0 = \frac{3}{2} + \frac{1}{2} {\displaystyle\sqrt{(M/m)^2 +1 }} \geq 2$ \\
        $s=\frac{1}{2}$ & $E_0 = \frac{3}{2} \pm \frac{1}{2}|{M/m}|\geq 1$ \\
        $s=0$ & $E_0 = \frac{3}{2} \pm \frac{1}{2} {\displaystyle\sqrt{(M/m)^2 +1}} \geq \frac{1}{2}$
        \\ \bottomrule
    \end{tabular}
    \caption{$E_0$ for $AdS_4$ fields of
    given mass $M$ and spin $s$ (in $Spin(2,3)$-irreps $D(E_0,s)$) and the corresponding unitarity bounds.}
    \label{ezerolist}
\end{table}

\begin{table}[ht]
    \centering
    \renewcommand*{\arraystretch}{1.2} %
    \begin{tabular}{ll}
        \toprule
        $\text{spin}^{\text{parity}}$ & Mass operator (giving $M$ or its square)
        \\ \midrule
        $2^{{+}}$ & $\Delta_0$ \\
        $\tfrac{3}{2}_{\scriptscriptstyle (\pm)}$ & $-i\slashed{D}_{1/2} + \tfrac{7m}{2}$ \\
        $1^{-}_{\scriptscriptstyle (\pm)}$ & $\Delta_1 + 12m^2 \pm 6m\sqrt{\Delta_1 + 4m^2}=(\sqrt{\Delta_1 + 4m^2}\pm3m)^2-m^2$ \\
        $1^+$ & $\Delta_2$ \\
        $\tfrac{1}{2}_{\scriptscriptstyle (\pm)}$ & $-i\slashed{D}_{1/2} - \tfrac{9m}{2}$ \\
        $\tfrac{1}{2}_{\scriptscriptstyle (\pm)}$ & $i\slashed{D}_{3/2} + \tfrac{3m}{2}$ \\
        $0^{+}_{\scriptscriptstyle (\pm)}$ & $\Delta_0 + 44m^2 \pm 12m\sqrt{\Delta_0 + 9m^2}=(\sqrt{\Delta_0 + 9m^2}\pm6m)^2-m^2$ \\
        $0^{+}$ & $\Delta_{L} - 4m^2=(\Delta_{L} - 3m^2)-m^2$ \\
        $0^{-}_{\scriptscriptstyle (\pm)}$ & $Q^2 + 6mQ + 8m^2=(Q+3m)^2-m^2$
        \\ \bottomrule
    \end{tabular}
    \caption{Mass operators  in Freund--Rubin compactifications, see for instance \cite{Duff:1986hr}.
    For  spins with two tower assignments, the subscripts $(\pm)$, the plus and minus signs refer to  branches of the $M^2$ formulas or to the positive and negative parts of the spectrum for
    linear operators.  Note the change of notation relative  \cite{Duff:1986hr} where superscripts $(1),(2)$ etc.\ were used instead of the $(\pm)$ notation of this paper.
    }
    \label{table:massop}
\end{table}

With the information in  \cref{ezerolist,table:massop} at hand, we can start discussing  how to obtain the various spectra of the operators on the squashed seven-sphere.
The spectrum of $\Delta_0$ and the Dirac operator were derived for general squashing in \cite{Nilsson:1983ru}.  While that derivation  is a bit different  from the one we present in \cref{sec:eigenvalues}, it essentially coincides with
the one given in \cref{sec:eigenmodes}.
The result for the square of the Dirac operator obtained in \cref{sec:eigenvalues} was derived in \cite{Ekhammar:2021gsg} but then only for the squashed Einstein case, in contrast to \cite{Nilsson:1983ru}.

Turning to the Hodge--de Rham and Lichnerowicz operators, we note that the former ones are positive definite and defined by
\begin{equation}
\Delta_p=\delta d+d\delta,
\end{equation}
whereas the latter, which is not manifestly positive definite, acts on transverse and traceless metric modes and is defined by
\begin{equation}
\Delta_L h_{ab}=-\Box h_{ab}-2R_a{}^c{}_b{}^dh_{cd}+2R_{(a}{}^ch_{b)c}=-\Box h_{ab}-2W_a{}^c{}_b{}^dh_{cd}+14m^2h_{ab}.
\end{equation}
In the last expression, the presence of the Weyl tensor $W_{abcd}$ gives rise to complications when solving
the eigenvalue equations. This problem was addressed in \cite{Ekhammar:2021gsg} case by case and will in this paper be discussed again in the next subsection
where we derive a universal group theoretic formula for the Weyl tensor eigenvalues reducing the work needed and providing a much better understanding of
the whole issue.

\subsection{Improved methodology }

We now turn to the improved formalism that will facilitate the eigenvalue analysis.
This subsection follows \cite{Karlsson:2021oxd}, where some additional details can be found.

Since we are interested in eigenvalue equations involving the Laplacian we start from its standard expression  $\Delta=d\delta + \delta d$ and rewrite it as
\begin{equation}
\label{universallaplacian}
\Delta\equiv -\Box-[D_{a},D_b]\Sigma^{ab}=-\Box - R_{abcd}\Sigma^{ab}\Sigma^{cd},
\end{equation}
which we will refer to as ``the universal Laplacian'' since it cannot only act on $p$-forms but any tensor field. Here and below, $\Sigma^{ab}$ are the $Spin(7)$ generators. To see how the right-hand side arises just consider the following $p$-form identities
\begin{align}
    \nonumber
    &\Delta_p\alpha_{a_1...a_p}
    = (d\delta + \delta d)\alpha_{a_1...a_p}
    =-\Box \alpha_{a_1...a_p}-p[D_{[a_1}, D_{|b|}]\alpha^b{}_{a_2...a_p]}
    \\
    & =-\Box \alpha_{a_1...a_p}-[D_{c_1}, D_{c_2}]p\delta^{c_1c_2}_{[a_1 |b|}\alpha^b{}_{a_2...a_p]}
    =-\Box \alpha_{a_1...a_p}-[D_{c_1}, D_{c_2}]\Sigma^{c_1c_2}\alpha_{a_1...a_p}.
\end{align}
This operator is the relevant one for transverse $p$-forms and becomes the Lichnerowicz operator when acting on traceless and transverse
second rank symmetric tensors. It also appears when squaring the Dirac operator acting on   spinors as well as on  transverse and gamma-traceless vector-spinors. In fact,
\begin{equation}
(i\slashed D)^2\psi=\Delta\psi+\frac{21}{4}m^2\psi,\qquad\quad
(i\slashed D)^2\psi_a=\Delta\psi_a-\frac{3}{4}m^2\psi_a,
\end{equation}
for Einstein spaces with $R_{ab} = 6m^2 \delta_{ab}$.

The next ingredient  we need is the coset structure $G/H$ of the internal manifold on which we perform the compactification. As usual
we will use a reductive coset which means that the structure constants satisfy (see, e.g., \cite{Duff:1986hr})
\begin{equation}
[H,H]= H,\qquad [H,T]=T,\qquad [T,T]=H\oplus T.
\end{equation}
We use the index split $A=(a,i)$ where $A$ and $i$ are adjoint indices of $G$ and $H$, respectively,  while $a$ is
a vector index in the tangent space of the coset ($a=1,2,...,dim(G)-dim(H)$). Hence, the independent non-zero structure constants are
\begin{equation}
f_{AB}{}^C=(f_{ij}{}^k, f_{ia}{}^b,  f_{ab}{}^i, f_{ab}{}^c).
\end{equation}
It is then straightforward to show that the Riemann tensor of the coset manifold $G/H$ can be expressed in terms of the structure constants in the following
way \cite{Bais:1983wc}:
\begin{equation}
\label{eq:coset_riemann}
R_{cda}{}^b=f_{cd}{}^if_{ia}{}^b+\frac{1}{2} f_{cd}{}^ef_{ea}{}^b+\frac{1}{2} f_{[c|a|}{}^ef_{d]e}{}^b.
\end{equation}
Usually, one lowers and raises indices on the structure constants using the Killing form and its inverse.
In this paper, we do \emph{not} do this.
Instead, we assume (already in \cref{eq:coset_riemann}) that the coset is normal homogeneous, i.e., that there exists a $G$-invariant $g_{AB}$ which is block-diagonal over $H \oplus T$ with blocks $\delta_{ab}$ and $g_{ij}$ and use these and their inverses to raise and lower indices.%
\footnote{This assumption is not strictly necessary here, see \cite{Karlsson:2021oxd}, but will be crucial later. The squashed seven-sphere is normal homogeneous.}

The key equation needed to compute spectra of operators on the coset is the so called \emph{coset master equation} which expresses the $H$-covariant derivative algebraically in terms of the $G$ generators in the tangent space directions $T_a$ \cite{Salam:1981xd, Duff:1986hr}:
\begin{equation}
D_aY+\frac{1}{2} f_{abc}\Sigma^{bc}Y=-T_aY.
\end{equation}
This is a schematic equation, see \cref{eq:coset_master_rho} for a more precise version.
The reason why this is not precise is that the left-hand side is   well-defined only for a $Spin(7)$ tensor field $Y$ while the right-hand side requires a field $Y$ transforming in a representation of $G$.
The equation is, however, still very useful since both $Spin(7)$ and $G$ representations can be decomposed into $H$-irreps.
Hence, it can be used literally with tensor fields if it is first manipulated in a way such that the operators become block-diagonal over $H$-irreps, as we will see below.

To get a clear picture of  the above relation between a differential operator and algebraic quantities like group generators we consider a $G$ group element representing a point on the coset manifold with coordinates $y^a$
\begin{equation}
L_y=e^{y\cdot T},
\end{equation}
where $y\cdot T$ is just a sum over products of coordinates $y^a$ and tangent space generators $T_a$. Then clearly the one-form $L_y^{-1}dL_y$ can be expanded as follows:
\begin{equation}
\label{dL}
L_y^{-1}dL_y=e^aT_a + \Omega^iT_i,
\end{equation}
where $d=dy^m\partial_m$, $e^a$  are one-form vielbeins on the coset space and $\Omega^i$ are $H$ one-form gauge fields.
By harmonic analysis on the coset, fields transforming in an irrep of $H$ can be expanded using $\rho(L^{-1}_y)$, where $\rho$ runs over all irreps of $G$ containing the $H$-irrep \cite{Salam:1981xd}.
The equation \cref{dL} can then be expressed as
\begin{equation}
\label{eq:coset_master_rho}
\check D_a\rho(L^{-1}_y)_P{}^{(Q)}\equiv(\partial_a+\Omega_a^iT_i)\rho(L^{-1}_y)_P{}^{(Q)}=-(T_a)_P{}^R\rho(L^{-1}_y)_R{}^{(Q)},
\end{equation}
where $\check{D}_a$ is a $H$-covariant derivative.
Note that we have here been careful and indicated  with parentheses that the upper index $(Q)$ is not affected either by the generators $T_a$ or the $H$-covariant derivative $\check D_a$.
We emphasise also here that the upper $G$-irrep index in  parentheses is the isometry index while the lower $G$-irrep index without parentheses should be split into $H$-irreps
so that the $H$-covariant derivative can act according to each of the $H$-irreps in the split.

However, in the present application to the squashed seven-sphere, constructed as the coset $Sp_2\times Sp_1^{C}/(Sp_1^A\times Sp_1^{B+C})$,  the tangent space group $Spin(7)$ is not the same as the isotropy group $H=Sp_1^A\times Sp_1^{B+C}$.
This means that in order to apply the fundamental coset equation in this case, with all eigenvalue equations expressed in terms of $Spin(7)$-covariant derivatives on the tangent space,
one has to split each $Spin(7)$-irrep of the fields on the coset into $H$-irreps, apply the fundamental coset equation and reconstruct the tangent space irreps of the fields.
This is done as follows (see \cite{Duff:1986hr}). Acting with an exterior derivative on \cref{dL} we get a Maurer--Cartan type equation which can be divided into two equations along the different
generator directions. We find
\begin{equation}
\label{eq:coset_maurer-cartan}
de^a=-\frac{1}{2} e^b\wedge e^c f_{bc}{}^a-e^b\wedge \Omega^i \, f_{bi}{}^a,\quad d\Omega^i=-\frac{1}{2} e^a\wedge e^b f_{ab}{}^i-\frac{1}{2} \Omega^j\wedge \Omega _k f_{jk}{}^i.
\end{equation}
Comparing the first of these equations to the definition of $\omega^a{}_b$, the torsion-free spin connection, $0=de^a+\omega^a{}_b\wedge e^b$, we find
\begin{equation}
\omega_{[bc]}{}^a=-\frac{1}{2} f_{bc}{}^a-\Omega^i_{[b}f_{|i|c]}{}^a,
\end{equation}
which implies
\begin{equation}
\omega_{abc}=-\frac{1}{2} f_{abc}-\Omega_a^if_{ibc}.
\end{equation}

Inserting this result into the definition of the $H$-covariant derivative, $\check{D}_a = \partial_a+\Omega_a^iT_i$,  we see that it can be written in terms of the ordinary tangent space covariant derivative as follows
\begin{equation}
\label{eq:checkD_def}
\check{D}_a = D_a+\frac{1}{2} f_{abc}\Sigma^{bc},
\end{equation}
when acting on an $so(7)$ representation.
Note that, using the $H$-covariant derivative, the coset master equation reads simply
\begin{equation}
\check{D}_aY=-T_aY.
\end{equation}

We now address the point raised above, namely that we must first get an equation where the operators split over $H$-irreps before we can apply it to tensor fields.
The left-hand side has already been addressed by writing the equation using $\check{D}$ instead of $D$ but $T_a$ cannot be restricted to an irrep of $H$.
Hence, we must return to the $\rho$ irreps of $G$ since any action of
a $T_a$ will mix all $H$-irreps in the $G$-irrep. We then see that
\begin{equation}
T_aT_b\rho(L_y^{-1})=-T_a(\partial_b+\Omega^i_bT_i)\rho(L_y^{-1})=-(\partial_b+\Omega^i_bT_i)T_a\rho(L_y^{-1})-\Omega^i[T_a, T_i]\rho(L_y^{-1}).
\end{equation}
Using $[T_a, T_i]=-f_{ia}{}^bT_b$ and \cref{eq:coset_master_rho} we find that the operations are being reversed, i.e.,
\begin{equation}
T_aT_b\rho(L_y^{-1})=\check{D}_b\check{D}_a\rho(L_y^{-1}).
\end{equation}
Note that the commutator term above is needed to get the derivative $\check{D}_b$ to also act on the index $a$ on the second derivative.
Here, we see that if we contract with $\delta^{ab}$, the operator on the left-hand side becomes the difference between two Casimirs and hence split over $H$-irreps.
Thus, since every tensor field $Y$ can be split into $H$-irreducible parts and the above equation applies to each $H$-irrep separately, we get
\begin{equation}
\label{boxcasimirs}
-\check{\Box}Y=-T_a T^a Y=(C_g - C_h)Y,
\end{equation}
where $C_g=-g^{AB}T_A T_B$ and $C_h = -g^{ij} T_i T_j$ are Casimir operators of the two groups in the coset $G/H$.

The operator equations we aim to solve below contain the universal Laplacian $\Delta$, which in turn can be written using $\Box=D^aD_a$.
Hence, we need the explicit relation between $\Box$ and $\check{\Box}$.
It can be obtained as follows: Consider first
\begin{equation}
\check{\Box}=\check{D}^a\big(D_a+\frac{1}{2} f_{abc}\Sigma^{bc}\big)=\check{D}^aD_a+\frac{1}{2} f^{ade}\Sigma_{de}\check{D}_a,
\end{equation}
where we used the fact that $\delta^{ab}$, $f_{abc}$ and $\Sigma^{ab}$ (in any irrep) are all $H$-invariant objects and hence commute with the $H$-covariant derivative
$\check{D}^a$. Expanding also the remaining $\check{D}^a$ operator this expression can be written
\begin{equation}
\check{\Box}={\Box}+f_{abc}\Sigma^{bc}\check{D}^a-\frac{1}{4}f_{abc}f^{ade}\Sigma^{bc}\Sigma_{de}.
\end{equation}
Note that in the last term on the right-hand side the two $so(7)$ generators act successively on any object to the right of them. However, the order of them is not important since the action of the
first one on the second one vanishes as is easy to check (in any irrep).

Finally, in order to obtain a useful  expression for  the universal Laplacian $\Delta$, whose eigenvalues we are interested in, we combine
the last equation above with \cref{universallaplacian} and \cref{boxcasimirs}. The result is the following key formula
\begin{equation}
\label{keyeigenvalueformula}
\Delta = C_g + f_{abc}\Sigma^{ab}\check D^c - \frac{1}{4}(3f_{abc}f^a{}_{de}-2f_{abd}f^a{}_{ce})\Sigma^{bc}\Sigma^{de}.
\end{equation}

The next step is to introduce some explicit expressions obtained from the  analysis of the squashed seven-sphere as the coset $G/H$ with $G=Sp_2\times Sp_1^{C}$ and $H=Sp_1^A\times Sp_1^{(B+C)}$
where the group $Sp_2$ has been split into $Sp_1^A\times Sp_1^B$ making it possible to form the diagonal subgroup $Sp_1^{B+C}$ appearing in $H$, see, e.g., \cite{Bais:1983wc}.
There is a one-parameter family of squashed seven-spheres with this isometry group but, except for the round sphere, only one is Einstein.
This is the squashed sphere we are concerned with here and it has a special role in the one-parameter family;
it is not only normal homogeneous but standard homogeneous ($g_{AB} = -\kappa_{AB}/6$) \cite{Karlsson:2021oxd}.

The metric on the squashed seven-sphere can be written as \cite{Duff:1986hr}
\begin{equation}
    ds^2 = d\mu^2 + \frac{1}{4} \sin^2{\!\mu}\, (\sigma - \tilde{\sigma})^2 + \frac{\lambda^2}{4} \bigl((\sigma + \tilde{\sigma}) + \cos{\mu}\,(\sigma - \tilde{\sigma})\bigr)^2,
\end{equation}
where $0 \leq \mu \leq \pi$, $\lambda$ is the squashing parameter and $\sigma^i$ and $\tilde{\sigma}^i$ are left-invariant one-forms on two copies of $SU(2) \simeq S^3$, satisfying $d \sigma^i = -\frac{1}{2} \epsilon^{ijk} \sigma_j \sigma_k$.
The round seven-sphere has $\lambda = 1$ while the Einstein-squashed one, which we focus on in what follows, has $\lambda = 1/\sqrt{5}$.
Note that the above metric is dimensionless.
To make $R_{(7)} = 42 m^2$ with the Einstein-squashed metric, one has to use units $m = 3/(2\sqrt{5})$.
This is implicit in what follows.

From the metric, we see that an orthonormal frame is given by
\begin{equation}
    e^{\hat{i}} = \frac{\lambda}{2} \bigl((\sigma^i + \tilde{\sigma}^i) + \cos{\mu}\,(\sigma^i - \tilde{\sigma}^i)\bigr),
    \qquad
    e^0 = d\mu,
    \qquad
    e^i = \frac{1}{2} \sin{\mu}\, (\sigma^i - \tilde{\sigma}^i).
\end{equation}
The orientation defined by this frame and the index split $a = (\hat{i}, 0, i)$ is that of the left-squashed seven-sphere.
The right-squashed seven-sphere is obtained by flipping the orientation.
For equations that depend on the orientation, we focus on the left-squashed sphere if nothing else is specified.

One key feature of the Einstein-squashed seven-sphere is that the holonomy of the derivative entering in the Killing spinor equation is $G_2$ \cite{Awada:1982pk,Duff:1986hr}
and that the structure constants become related to the octonions \cite{Bais:1983wc} via
\begin{equation}
\label{eq:squashed_S7_structure_constants}
f_{ab}{}^c=-\frac{1}{\sqrt{5}}a_{ab}{}^c,
\end{equation}
where $a_{abc}$ are given by the octonionic multiplication table (see \cref{app:octonions}). Now we can also see one reason why the covariant derivative $\check{D}_a$ is of interest to us:%
\footnote{The squashed seven-sphere has an $H$-structure, i.e., a reduction of the structure group $Spin(7)$ to $H$, due to it being a coset $G/H$. Since $H \subset G_2$, it follows that the $H$-covariant derivative $\check{D}_a$ of the $G_2$-invariant $a_{abc}$ vanishes. The $H$-structure has torsion, see \cref{eq:checkD_def}, \cref{eq:squashed_S7_structure_constants}, which is related to the fact that the squashed seven-sphere is only a weak $G_2$ manifold \cite{Joyce:2000book}.}
\begin{equation}
\check{D}_a a_{bcd}=0.
\end{equation}
Below we will take full advantage of this fact when expressing the eigenvalue equations we want to solve
entirely in terms of $G_2$ quantities making the equations and the analysis of them fully $G_2$-covariant. The first steps in this direction were developed in \cite{Ekhammar:2021gsg}. The usefulness of this formulation is
related to the following fact: When splitting the indices of tangent space irreps on the tensor fields into $H$-irreps it goes via $G_2$ (as realised already in \cite{Duff:1986hr})
\begin{equation}
Spin(7)\quad\longrightarrow \quad G_2 \quad\longrightarrow \quad H=Sp_1^A\times Sp_1^{B+C}.
\end{equation}

As mentioned above the appearance of $G_2$ has many implications  and can be incorporated into the formalism
in a significant way as we will now show. First, consider the Weyl tensor which will appear in the eigenvalue analysis later. It can be expressed as follows:
\begin{equation}
W_{ab}{}^{cd}=R_{ab}{}^{cd}-\frac{9}{10}\delta_{ab}^{cd}=(T^i)_{ab}(T_i)^{cd}+\frac{1}{10}a_{ab}{}^ea_e{}^{cd}+\frac{1}{10}a_{[a}{}^{ce}a_{b]e}{}^{d}-\frac{9}{10}\delta_{ab}^{cd}.
\end{equation}
Using the identity
\begin{equation}
a_{[a}{}^{ce}a_{b]e}{}^{d} = - \delta_{ab}^{cd} + c_{ab}{}^{cd},
\end{equation}
we find that this can be written in terms of the $G_2$ $\bf{21}\rightarrow \bf{14}$ projector $(P_{14})_{ab}{}^{cd}$ as
\begin{equation}
W_{ab}{}^{cd}=(T^i)_{ab}(T_i)^{cd}-\frac{6}{5}(P_{14})_{ab}{}^{cd}.
\end{equation}
Since the Weyl tensor is the source of the $G_2$ holonomy it vanishes when contracted by $a_{abc}$ over the first or last pair of indices.
Naively, one might have expected it to be proportional to the projector $P_{14}$ onto the adjoint of $G_2$ but, as we see here, there is an extra term.
Since $H$ is a subgroup of $G_2$, this extra term is not in contradiction with the $G_2$ holonomy.
Furthermore, since the projector  $(P_{14})_{ab}{}^{cd}$ actually defines $G_2$ inside $SO(7)$ we find that
\begin{equation}
\label{masterweyl}
W_{abcd}\Sigma^{ab}\Sigma^{cd}=\frac{6}{5}C_{g_2}-C_h,
\end{equation}
where
\begin{equation}
C_{g_2}=-(P_{14})_{abcd}\Sigma^{ab}\Sigma^{cd},\quad
C_h = -T_i T^i = 2 C_{sp_1^A} + \frac{6}{5} C_{sp_1^{B+C}}.
\end{equation}
This  expression for the Weyl tensor is one of the key results of \cite{Karlsson:2021oxd} since it eliminates in one stroke the rather cumbersome analysis to obtain the Weyl tensor eigenvalues for the various
tensors on the squashed seven-sphere performed in \cite{Ekhammar:2021gsg}.

In the application of the master coset formula to the computation of operator eigenvalues the first step is, as done above, to square it to produce $\Delta$.
However, as we saw in \cref{keyeigenvalueformula}
to actually find the eigenvalues the appearance of the operator $a_{abc}\Sigma^{ab}\check{D}^c$ seems to  require a second squaring procedure.
Inserting the octonionic information from the coset construction above into \cref{keyeigenvalueformula} we get
\begin{equation}
\label{masterdelta}
-\frac{1}{\sqrt 5}a_{abc}\Sigma^{ab}\check{D}^c=\Delta - C_g-\frac{6}{5}C_{so(7)}+\frac{3}{2}C_{g_2},
\end{equation}
where the $g = sp_2 \oplus sp_1^C$ Casimir is (see \cref{app:gtwoprojectors:Casimirs} for Casimir conventions)
\begin{equation}
\label{eq:method:C_g}
    C_g(p,q;r) = - T_A T^A = 2 C_{sp_2}(p,q) + 3 C_{\smash[t]{sp_1^C}}(r).
\end{equation}
This group theoretic expression for $a_{abc}\Sigma^{ab}\check{D}^c$ is in the spirit of the Weyl tensor result above and constitutes the second key formula appearing in the improved formalism applied to the squashed seven-sphere.

In addition to these two key equations there is another very useful one, namely the Ricci identity for  $G_2$ derivatives
\begin{equation}
\label{checkricci}
    [\check D_a, \check D_b]=(T^i)_{ab}T_i-\frac{1}{\sqrt 5}a_{ab}{}^c\check D_c
    = \Bigl(W_{ab}{}^{cd}+\frac{6}{5}(P_{14})_{ab}{}^{cd} \Bigr) \Sigma_{cd}-\frac{1}{\sqrt 5}a_{ab}{}^c\check D_c.
\end{equation}
An especially useful form of this Ricci identity is
\begin{equation}
a_a{}^{bc}\check D_b \check D_c=-\frac{3}{\sqrt 5}\check D_a.
\end{equation}

In the next section we will derive the eigenvalue spectra of all the operators relevant for the squashed seven-sphere compactification of eleven-dimensional
supergravity. The cases of $\Delta_p$ for $p=0, 1, 2, 3$, $\Delta_L$ and $\slashed D_{1/2}$ have been obtained before in \cite{Ekhammar:2021gsg}
(for $\Delta_p$ with $p=1$ already  in \cite{Yamagishi:1983ri},  $\slashed D_{1/2}$ in \cite{Nilsson:1983ru} and  for $\Delta_L$ in \cite{Duff:1986hr}) but the derivations will
here be streamlined quite a bit using the novel group theoretic formulas, \cref{masterweyl} and \cref{masterdelta}.
In addition, with these new forms of the key equations, we can rather easily add the so far missing one (presented also in \cite{Karlsson:2021oxd}), namely  $i\slashed D_{3/2}$, to the list,
thereby making it complete. It is then possible to conduct a search for the full set of supersymmetry irreps as indicated in the previous papers on this subject
\cite{Nilsson:2018lof} and \cite{Ekhammar:2021gsg}. This will be the goal of \cref{sec:supermultiplet}.

It might be of interest to the reader to know already here
that the supersymmetry analysis will come short of providing a full understanding of the supersymmetry multiplets: There will remain both ambiguities and
a degeneracy  of eigenvalues which  will require further studies. An initial result in this context will be presented in \cref{sec:eigenmodes}
where we give a novel construction of all modes of the $Y_{ab}$ as well as their eigenvalues.

\section{Derivation of all operator eigenvalues}%
\label{sec:eigenvalues}

In this section we start, in \cref{sec:eigenvalues:everything_but_3/2}, by briefly showing how to derive previously known results using the more powerful methods
presented in \cref{sec:review_and_method}.  Then, in \cref{sec:eigenvalues:spin_3/2} we apply these methods to the novel case of spin 3/2. This case was left undone in \cite{Ekhammar:2021gsg} but
the analysis  of the spin-3/2 equation on the squashed seven-sphere, including finding its eigenvalues, have already been reported on in \cite{Karlsson:2021oxd}.
If the reader find the following account too brief more details can be found in  \cite{Karlsson:2021oxd}.
The \emph{master formula} used repeatedly below is \cref{masterdelta}.
\subsection{All operator equations  except spin 3/2}%
\label{sec:eigenvalues:everything_but_3/2}
\subsubsection{Zero-forms}
The equation to be solved in this case is
\begin{equation}
\Delta_0Y=\kappa^2_0 Y.
\end{equation}
The master formula \cref{masterdelta} for $\Delta$ immediately gives
(using $C_{g_2}=C_{so(7)}=0$)
\begin{equation}
\label{eq:eigenvalue:0-form}
\Delta_0^{(1)}=C_g,
\end{equation}
when applied to scalar modes.%
\footnote{We will in the following denote the eigenvalues by a superscript on the operator, as in \cref{eq:eigenvalue:0-form}.}
Note that $a_{abc}\Sigma^{ab}\check{D}^c Y=0$ in \cref{masterdelta} since $Y$ is a scalar and the action of $\Sigma^{ab}$ on the $c$-index on $\check{D}^c$ also vanishes
due to the contraction with $a_{abc}$.

\subsubsection{One-forms}
The equation for transverse one-forms ($D^aY_a=0$ or, equivalently, $\check{D}^a Y_a = 0$) reads
\begin{equation}
\Delta_1Y_a=\kappa^2_1 Y_a.
\end{equation}
Again using \cref{masterdelta} we get, now with $C_{so(7)}(\vec{7})=3$ and $C_{g_2}(\vec{7})=2$,
\begin{equation}
\label{oneformdone}
\check{\mD}_1Y_a\equiv a_a{}^{bc}\check{D}_c Y_b= \sqrt{5} \biggl(C_g-\kappa^2_1+\frac{3}{5}\biggr)Y_a.
\end{equation}
The standard procedure at this point is to square the operator, i.e., act with $\check{\mD}_1$ again, which gives
\begin{equation}
\label{oneformdonesquare}
    (\check{\mD}_1)^2Y_a=-\check\Box Y_a+\check{D}^b\check{D}_aY_b+c_a{}^{bcd}\check{D}_b\check{D}_cY_d
    = \Bigl(C_g - C_h+\frac{24}{5}-\frac{1}{\sqrt 5}\check{\mD}_1\Bigr)Y_a,
\end{equation}
where we have used that $Y_a$ is transverse.
We have also used the Ricci identity on one-forms
\begin{equation}
[\check{D}_a, \check{D}_b]Y_c=\Bigl(W_{abc}{}^d+\frac{6}{5}(P_{14})_{abc}{}^d\Bigr)Y_d-\frac{1}{\sqrt 5}a_{abd}\check{D}^dY_c,
\end{equation}
and the following two direct implications of it:
\begin{equation}
[\check{D}^b, \check{D}_a]Y_b=\frac{12}{5}Y_a+\frac{1}{\sqrt 5}\check{\mD}_1Y_a,
\end{equation}
and
\begin{equation}
c_a{}^{bcd}\check{D}_b\check{D}_cY_d=\frac{12}{5}Y_a-\frac{2}{\sqrt 5}\check{\mD}_1 Y_a.
\end{equation}
It is interesting to note that $C_h$ reappeared in the above calculation of $(\check{\mD}_1)^2Y_a$ but that the formula for the Weyl tensor \cref{masterweyl} applied to one-forms gives rise to a contracted
Weyl tensor which vanishes and hence gives (also obtained in \cite{Duff:1986hr})
\begin{equation}
\biggl(\frac{12}{5}-C_h\biggr)Y_a=0.
\end{equation}
Hence, the eigenvalue equation \cref{oneformdonesquare} becomes
\begin{equation}
C_g+\frac{12}{5}-\biggl(C_g-\kappa^2_1+\frac{3}{5}\biggr)=5\biggl(C_g-\kappa^2_1+\frac{3}{5}\biggr)^2,
\end{equation}
which is solved by $\kappa^2_1 = \Delta^{(1)_{\pm}}_1$, where
\begin{equation}
\Delta^{(1)_{\pm}}_1=C_g+\frac{7}{10}\pm\frac{1}{\sqrt{5}}\sqrt{C_g+\frac{49}{20}}.
\end{equation}

\subsubsection{Two-forms}
We now turn to two-forms and the eigenvalue equation
\begin{equation}
\Delta_2 Y_{ab}=\kappa_2^2 Y_{ab}.
\end{equation}
In the previous literature the additional technical issues that start to appear here were handled on a case-by-case basis, leading to poor understanding and long calculations.
In \cite{Nilsson:2018lof}, for instance, equations like \cref{masterdelta} were derived for the different types of fields separately.
Moreover, they still contained $C_h$ and the Weyl tensor, presenting further technical difficulties, since the latter appears when acting with $\Delta_p$ on a $p$-form, cf.\ \cref{universallaplacian}, and its expression in terms of Casimirs, \cref{masterweyl}, was not known.
Using the novel version of the key equations \cref{masterweyl} and \cref{masterdelta} this computation will be much easier as we now show.

In the improved formalism of this paper
 \cref{masterdelta} takes the following form when  applied to transverse
two-forms\footnote{The reason for the square bracket on the index 2, here and below, is that the free indices are antisymmetrised.
The same differential operator but without antisymmetrisation or with symmetrised indices will be used below.}
\begin{equation} \label{eq:eigs:2-forms:masterdelta}
\check{\mD}_{[2]}Y_{ab}\equiv a_{[a}{}^{cd}\check{D}_{|d}Y_{c|b]}=\frac{\sqrt{5}}{2}(C_g-\kappa^2_2+3P_7)Y_{ab}.
\end{equation}
Here we have used $C_{so(7)}(\mathbf{21})=5$, $C_{g_2}(\mathbf{7})=2$ and $C_{g_2}(\mathbf{14})=4$. Note that the reason $P_7$ appears in the last expression is that
the two-form splits under $Spin(7)\rightarrow G_2$ into $\bf{14}$ and $\bf{7}$ which have different $C_{g_2}$-eigenvalues. Note also that the transversality constraint reads,
with $G_2$-covariant derivatives,
\begin{equation}
\check{D}^bY_{ba}=\frac{1}{2\sqrt 5}a_{abc} Y^{bc}.
\end{equation}

In order to proceed it is very convenient to define a second  linear operator
\begin{equation}
\tilde\mD_{[2]}Y_{ab}\equiv a_{[a}{}^{cd}\check{D}_{b]}Y_{cd}\equiv -\check{D}_{[a}Y_{b]},
\end{equation}
where we in the last equality have defined the vector $Y_a$ associated to the two-form by
\begin{equation}
Y_a\equiv a_a{}^{bc}Y_{bc}.
\end{equation}
Note that this vector field is not transverse a priori (as it is in the one-form case above), a fact that will become important below.

One can quite easily derive a relation between the two linear derivative operators defined above. It reads
\begin{equation}
c_{a b}{}^{cd}\check{\mD}_{[2]}Y_{cd}=\frac{3}{\sqrt 5}P_7Y_{ab}+2\check{\mD}_{[2]}Y_{ab}-2 \tilde{\mD}_{[2]}Y_{ab}.
\end{equation}
By writing $c_{ab}{}^{cd}$ in terms of $\delta_{ab}^{cd}$ and $(P_7)_{ab}{}^{cd}$, this can be rewritten as
\begin{equation}
\label{eq:eigs:2-forms:tildemD_relation}
\tilde\mD_{[2]}Y_{ab}=(2-3P_7)\check{\mD}_{[2]}Y_{ab}+\frac{3}{2\sqrt 5}P_7Y_{ab}.
\end{equation}
This equation implies that, if $Y_{ab}$ has no ${\bf{7}}$-part,  it will satisfy $\check{\mD}_{[2]}Y_{ab}=0$ since, in that case, $\tilde\mD_{[2]} Y_{ab} = 0 = P_7 Y_{ab}$ and
$2-3 P_7$ is invertible when acting on two-forms.

Combining \cref{eq:eigs:2-forms:masterdelta,eq:eigs:2-forms:tildemD_relation} leads to the following much simpler equation for the eigenvalues
\begin{equation}
\check{D}_{[a}Y_{b]}=-\frac{\sqrt {5}}{2}\left(2\bigl(C_g-\kappa^2_2\bigr)-3\biggl(C_g-\kappa^2_2+\frac{4}{5}\biggr)P_7\right)Y_{ab}.
\end{equation}
The reason this is simpler than the original eigenvalue equation is that it becomes  an eigenvalue equation for one-forms when the $G_2$ part $\mathbf{7}$ of the $Spin(7)$-irrep $\mathbf{21}$
is extracted by contracting with $a_{abc}$. Doing this we find (for $\check{\mD}_1$ see the one-form analysis above)
\begin{equation}
  \label{eq:eigs:2-form:1-form-eq}
  \check{\mD}_1 Y_a = - \frac{\sqrt{5}}{2} \left(C_g - \kappa_2^2 + \frac{12}{5}\right) Y_a.
\end{equation}

To deal with the lack of transversality of the one-form $Y_a$ we contract the last equation above by $\check{D}^a$ and find   that either $\check{D}^a Y_a = 0$
or (again denoting the eigenvalues by indexed operators)
\begin{equation}
\Delta_2^{(1)}=C_g+\frac{18}{5}.
\end{equation}
In the former case, $\check{D}^a Y_a = 0$, we repeat the one-form analysis starting from \cref{eq:eigs:2-form:1-form-eq} and find that either $Y_a = 0$ or
\begin{equation}
\Delta_2^{(2)_{\pm}}=C_g+\frac{11}{5}\pm\frac{2}{\sqrt 5}\sqrt{C_g+\frac{49}{20}}.
\end{equation}
When $Y_a = 0$ then also $\check{\mD}_{[2]}Y_{ab}=0$, as remarked above, and we see from \cref{eq:eigs:2-forms:masterdelta} that the last set of eigenvalues are  given by
\begin{equation}
\Delta_2^{(3)}=C_g.
\end{equation}

\subsubsection{Lichnerowicz}
We now turn to the case of   symmetric transverse and traceless rank-2 tensors which we will denote as $X_{ab}$.
Thus $X_{ab}$ satisfies
\begin{equation}
\Delta X_{ab}=\kappa^2_LX_{ab},\qquad X^a{}_a=0,\qquad D^aX_{ab}=0 \ \Longleftrightarrow\ \check D^aX_{ab}=0.
\end{equation}
Using $C_{so(7)}({\bf 27})=7$ and $C_{g_2}({\bf 27})=\frac{14}{3}$, the master formula becomes
\begin{equation}
\label{masterlichn}
\check\mD_{(2)}X_{ab} \equiv a_{(a}{}^{cd}\check D_{|d}X_{c|b)} = \frac{\sqrt 5}{2}\biggl(C_g-\kappa^2_L+\frac{7}{5}\biggr)X_{ab},
\end{equation}
where we have introduced the differential operator $\check\mD_{(2)}$,
which clearly is the object we need to study and if possible relate its action on $X_{ab}$ to $\kappa_L^2$ and $X_{ab}$ itself.

It turns out, however, that it is much more convenient to start by considering the operator
\begin{equation}
  (\check{\mD}_2 X)_{ab} \equiv a_{a}{}^{cd} \check{D}_d X_{c b},
\end{equation}
without any kind of symmetrisation. The reason for this is that
one can rather easily derive a nice equation for  its square $(\check\mD_2{}^2X)_{ab}$.

To proceed we note that $(\check\mD_2X)_{ab}$ is traceless and has vanishing ${\bf{7}}$-part and hence
contains only the irreducible parts ${\bf{14}}$ and ${\bf{27}}$. The latter, $(\check\mD_{(2)}X)_{ab}$, was  defined above while the  former will from now on be denoted
\begin{equation}
  Y_{ab}\equiv (\check\mD_{[2]}X)_{ab}=(\check\mD_2X)_{ab}-kX_{ab},
\end{equation}
where $k$ is the coefficient on the right-hand side of the master equation, \cref{masterlichn}.
It is easy to show that $\check{D}^a(\check\mD_2X)_{ab}=0$, which by the above equation implies that $Y_{ab}$ is transverse ($D^a Y_{ab} = 0$) since $X_{ab}$ is transverse and $Y_{ab}$ has no $\mathbf{7}$-part.

It is now fairly easy to compute $(\check\mD_2{}^2X)_{ab}$.
In fact, from \cref{masterweyl} we find
\begin{equation}
    2W_a{}^c{}_b{}^d X_{cd}= \biggl(\frac{28}{5}-C_h \biggr) X_{ab}
\end{equation}
(found earlier in \cite{Ekhammar:2021gsg}) which together with  the Ricci identity for $\check D_a$ makes it possible to show that $C_h$ cancels and that
\begin{equation}
\label{eq:eigenvalues:Lich:squared_op}
(\check\mD_2{}^2X)_{ab}=-\check\Box X_{ab}+\check D^c\check D_aX_{cb}+c_a{}^{cde}\check D_c\check D_dX_{eb}=C_gX_{ab}-\frac{1}{\sqrt 5}(\check\mD_2X)_{ab}.
\end{equation}
Finally, by combining this equation with the one defining $Y_{ab}$ above we find
\begin{equation}
\label{keylichneq}
    \biggl(k^2 + \frac{k}{\sqrt{5}} - C_g\biggr) X_{ab}
    = - \biggl(\check\mD_2 + k + \frac{1}{\sqrt{5}}\biggr) Y_{ab}.
\end{equation}
To solve this equation we consider first its antisymmetric part which reads
\begin{equation}
\label{eq:eigenvalues:Lich:Y!=0_case}
\check\mD_{[2]}Y_{ab} = - \biggl(k + \frac{1}{\sqrt{5}}\biggr) Y_{ab}
=-\frac{\sqrt 5}{2}\biggl(C_g-\kappa^2_L+\frac{9}{5}\biggr)Y_{ab}.
\end{equation}
However, $\check\mD_{[2]}Y_{ab}=0$ since $Y_{ab}$ has only a $\mathbf{14}$-part, as already noted above when solving for the two-form eigenvalues. Thus either  $Y_{ab}=0$ or $\kappa_L^2=\Delta_L^{(1)}=C_g+\frac{9}{5}$.

When $Y_{ab}=0$, we have, by \cref{keylichneq},
\begin{equation}
\label{eq:eigenvalues:Lich:Y=0_case}
\biggl(k^2+\frac{k}{\sqrt 5}-C_g\biggr)X_{ab}=0.
\end{equation}
Inserting the expression for $k$ above and solving this equation give the remaining two sets of eigenvalues, $\Delta^{(2)_{\pm}}_L$.

To summarise, we have found that the  eigenvalues of the Lichnerowicz operator are just the following three sets (reported without derivation already in \cite{Duff:1986hr})
\begin{align}
  &\Delta^{(1)}_L=C_g+\frac{9}{5},\\
  &\Delta^{(2)_{\pm}}_L=C_g+\frac{8}{5}\pm\frac{2}{\sqrt 5}\sqrt{C_g+\frac{1}{20}}.
\end{align}

\subsubsection{Three-forms}
Following the by now standard procedure, we start by presenting the master formula applied to three-forms, in the irrep ${\bf{35}}$ of  $so(7)$. Thus,
since its $G_2$-content is given by ${\bf 35}\rightarrow {\bf 1}\oplus{\bf 7}\oplus{\bf 27}$ we need $C_{so(7)}({\bf 27})=7$, $C_{g_2}({\bf 7})=2$ and $C_{g_2}({\bf 27})=\frac{14}{3}$. Using these facts
and the eigenvalue equation for transverse three-forms
\begin{equation}
    \Delta Y_{abc} = \kappa^2_3 Y_{abc},\qquad\quad
    D^aY_{abc}=0,
\end{equation}
the master formula becomes
\begin{equation}
\label{masterthreeformeq}
3a_{[a}{}^{de}\check{D}_{|e}Y_{d|bc]}=\sqrt{5}\left(C_g-\kappa^2_3+\frac{36}{5}-3P_7-7P_{27}\right)Y_{abc}.
\end{equation}
With some care when imposing transversality we can make use of previous results, in particular the ones related to Lichnerowicz discussed just above.

To split the three-form into its $G_2$ parts according to ${\bf 35}\rightarrow {\bf 1}\oplus{\bf 7}\oplus{\bf 27}$ we use
 the projection operators in  \cref{app:gtwoprojectors}.
This  leads to
\begin{equation}
Y_{abc}=\frac{1}{42}a_{abc}Y+\frac{1}{24}c_{abc}{}^d Y_d + \frac{3}{4}a_{[ab}{}^d X_{c]d},
\end{equation}
where
\begin{equation}
Y\equiv a^{abc}Y_{abc},\quad Y_a\equiv -c_a{}^{bcd}Y_{bcd},\,\,\,X_{ab}\equiv a_{(a}{}^{cd}Y_{b)cd}-\frac{1}{7}\delta_{ab}Y.
\end{equation}

The transversality condition $D^aY_{abc}=0$  in terms of $\check{D}_a$ translates into
\begin{equation}
\check{D}^c Y_{c ab} = \frac{1}{\sqrt 5} a_{[a}{}^{cd}Y_{b]cd}.
\end{equation}

The key equation \cref{masterthreeformeq} above splits as the three-form itself into three pieces. Considering first the parts ${\bf 1}$ and ${\bf 7}$ we get
\begin{gather}
\label{masterthreeformonepart}
    3\check{D}^a Y_a={\sqrt 5} \biggl(C_g-\kappa^2_3+\frac{36}{5}\biggr)Y,
    \\
\label{masterthreeformsevenpart}
    \frac{4}{7}\check{D}_aY-\frac{1}{6}a_a{}^{bc}\check{D}_bY_c-3\check{D}^bX_{ba}=-\biggl(C_g-\kappa^2_3+\frac{21}{5}\biggr) Y_a.
\end{gather}
To extract the eigenvalues connected to these two equations we note that the term $\check{D}^bX_{ba}$ in the latter one is also occurring in the ${\bf 7}$ part of the transversality condition
and can thus be replaced by terms involving only $Y$ and $Y_a$. Doing this, \cref{masterthreeformsevenpart} gives rise to an equation which we contract by $\check{D}^a$ and use
\cref{masterthreeformonepart} to eliminate $\check{D}^a Y_a$. The final result is
\begin{equation}
\frac{9}{5}C_gY=\biggl(C_g-\kappa^2_3+\frac{9}{5}\biggr)\biggl(C_g-\kappa^2_3+\frac{36}{5}\biggr)Y,
\end{equation}
which, if $Y\neq 0$ gives the eigenvalues $\kappa^2_3 = \Delta_3^{(3)_{\pm}}$,
\begin{equation}
\Delta_3^{(3)_{\pm}}=C_g+\frac{9}{2}\pm\frac{3}{\sqrt 5}\sqrt{C_g+\frac{81}{20}}.
\end{equation}

If instead $Y=0$ we see from the above equations that $\check{D}^a Y_a=0$, which implies ${D}^aY_a=0$, and that $Y_a$ satisfies an
equation which, apart from having different constants, is identical to the one-form one above. Repeating the steps
taken there we find the following result for the eigenvalues associated with $Y_a$ in the present case:
\begin{equation}
\Delta_3^{(2)_{\pm}}=C_g+\frac{5}{2}\pm\frac{1}{\sqrt 5}\sqrt{C_g+\frac{49}{20}}.
\end{equation}

Finally, if instead of getting the previous set of eigenvalues also $Y_a=0$ then only $X_{ab}$ is left in $Y_{abc}$. Now the transversality condition gives
\begin{equation}
\check{D}^bX_{ba}=0,\qquad\quad P_{14}(\check{\mD}_{[2]}X)_{ab}=0,
\end{equation}
and the master equation \cref{masterthreeformeq} reduces to
\begin{equation}
    (\check{\mD}_2 X)_{ab} = -{\sqrt{5}} \biggl(C_g-\kappa^2_3+\frac{1}{5}\biggr)X_{ab}.
\end{equation}
The situation has now become similar to the one for the Lichnerowicz operator but with the additional information that what we there called $Y_{ab} = (\check{\mD}_{[2]}X)_{ab}$ vanishes since $P_{14}(\check{\mD}_{[2]}X)_{ab}=0$ and the $\vec{7}$-part of $Y_{ab}$ always vanish for transverse $X_{ab}$, as noted above.
Hence, we only get the case analogous to \cref{eq:eigenvalues:Lich:Y=0_case}, not \cref{eq:eigenvalues:Lich:Y!=0_case}.
Changing the constants of the Lichnerowicz analysis appropriately, this gives the two eigenvalues in \cref{eq:eigenvalues:Lich:a}.

Thus we have found the following  three branches of eigenvalues for the three-forms:
\begin{align}
\label{eq:eigenvalues:Lich:a}
    &\Delta_3^{(1)_{\pm}}=C_g+\frac{1}{10}\pm\frac{1}{\sqrt 5}\sqrt{C_g+\frac{1}{20}},\\
    &\Delta_3^{(2)_{\pm}}=C_g+\frac{5}{2}\pm\frac{1}{\sqrt 5}\sqrt{C_g+\frac{49}{20}},\\
    &\Delta_3^{(3)_{\pm}}=C_g+\frac{9}{2}\pm\frac{3}{\sqrt 5}\sqrt{C_g+\frac{81}{20}}.
\end{align}
When we summarise these results at the end of this section we will instead give the eigenvalues of the linear operator $Q$ related to $\Delta_3$ by
$\Delta_3=Q^2$. This will force us to choose one of two possible signs which, however, can be done using supersymmetry as will be clear from
the analysis in \cref{sec:supermultiplet}.
\subsubsection{Spin 1/2}
When we now turn to operator equations on the squashed seven-sphere for modes in spinorial tangent space representations
we follow the strategy in \cite{Ekhammar:2021gsg} of squaring the corresponding Dirac equation. For the spin-1/2 case we thus consider
\begin{equation}
 \Delta\psi=\kappa_{1/2}^2 \psi,
\end{equation}
where the universal Laplacian  in \cref{universallaplacian} is related to the square of the Dirac operator by
\begin{equation}
 \Delta\psi=(i\slashed D)^2\psi-\frac{189}{80} \psi.
 \label{diracunivlaplace}
\end{equation}
As in \cite{Ekhammar:2021gsg} we will make heavy use of the $G_2$-invariant Killing spinor $\eta$ which satisfies
\begin{equation}
\check{D}_a\eta=0.
\end{equation}
Using $\eta$, we can define the projection operators needed to split the $Spin(7)$ spinor irrep ${\bf 8}$ into $G_2$ parts according to
${\bf 8}\rightarrow {\bf 1}\oplus {\bf 7}$:
\begin{equation}
P_1\equiv \eta\bar\eta,\qquad\quad P_7\equiv \Gamma^a\eta\bar\eta\Gamma_a.
\end{equation}
These  projection operators satisfy $P_1+P_7=1$ (which is just a Fierz identity) which leads to the following splitting of the spinor:
\begin{equation}
  \psi = Y \eta + i Y_a \Gamma^a \eta.
\end{equation}
where we have defined two real tensor fields
\begin{equation}
  Y \equiv \bar\eta \psi,\qquad\quad Y_a \equiv -i \bar\eta \Gamma_a \psi.
\end{equation}

Using the fact that $C_{so(7)}({\bf 8})=21/8$ we find that the master formula reads
\begin{equation}
a_{abc}\Gamma^{ab}\check{D}^c\psi=4\sqrt 5\left(C_g - \kappa_{1/2}^2 +\frac{63}{20}-3P_7\right)\psi.
\end{equation}
This equation can be projected onto its scalar and vector parts leading to the following two equations
\begin{gather}
\label{eq:eigenvalues:spin_1/2:scalar_eq}
  3 \check{D}^a Y_a = 2\sqrt{5} \left(C_g - \kappa_{1/2}^2 + \frac{63}{20}\right) Y,\\
\label{eq:eigenvalues:spin_1/2:vector_eq}
  3 \check{D}_a Y + a_a{}^{bc} \check{D}_c Y_b = -2\sqrt{5}  \left(C_g - \kappa_{1/2}^2 + \frac{3}{20}\right)Y_a.
\end{gather}
Here, we see immediately that if $Y = 0$ then $Y_a$ is a transverse one-form satisfying an equation of the form $\check{\mD}_1 Y_a \propto Y_a$.
Thus, we first want to use the two equations above to derive the eigenvalues in the $Y \neq 0$ case to then be able to reuse the  analysis of the one-forms.

This is done as follows. We contract the latter of the two equations above by $\check{D}^a$ and use the first equation  to eliminate the $\check{D}^aY_a$ term.
This gives
\begin{equation}
\frac{9}{20}C_gY=\left(C_g-\kappa_{1/2}^2+\frac{9}{20}\right)\left(C_g+\kappa_{1/2}^2+\frac{63}{20}\right)Y.
\end{equation}
Thus either $Y=0$ or the eigenvalue $\kappa_{1/2}^2$ is given by the solution to this equation, which reads
\begin{equation}
    \Delta_{1/2}^{(1)} = C_g+\frac{9}{2}\pm\frac{3}{2\sqrt 5}\sqrt{C_g+\frac{81}{20}}.
\end{equation}

On the other hand, when $Y=0$ the above equations \cref{eq:eigenvalues:spin_1/2:scalar_eq,eq:eigenvalues:spin_1/2:vector_eq} reduce to the equations for a transverse one-form analysed previously.
Reusing that computation we find
\begin{equation}
    \Delta_{1/2}^{(2)} = C_g+\frac{1}{10}\pm\frac{1}{2\sqrt 5}\sqrt{C_g+\frac{49}{20}}.
\end{equation}

The final step is to solve for the eigenvalues of the Dirac operator $i\slashed{ D}_{1/2}$ from  \cref{diracunivlaplace}.
The result is given as $i\slashed{D}_{1/2}^{(1)_{\pm}}$ and $i\slashed{D}_{1/2}^{(2)_{\pm}}$ in \cref{sec:eigenvalues:summary}.

\subsection{The novel case of spin 3/2}%
\label{sec:eigenvalues:spin_3/2}

The last operator equation to  discuss is the spin-3/2 one.
This is also the one that has not been solved previously in the literature (see however \cite{Karlsson:2021oxd})
but which can now be dealt with using the more efficient methods presented in the previous section.
As we will see below the analysis of the spin-3/2 eigenvalue equation will contain most of the intricate features encountered,
and resolved, in connection with the cases discussed above.

The starting point here is the same as for the other cases above:
Write out the master formula, that is \cref{masterdelta}, using $C_{so(7)}({\bf{48}})=\frac{49}{8}$, $C_{g_2}({\bf{7}})=2$, $C_{g_2}({\bf{14}})=4$ and $C_{g_2}({\bf{27}})=\frac{14}{3}$,
\begin{equation}
\label{spinthreehalfeq}
a_a{}^{bc}\check{D}_c\psi_b + \frac{1}{4}a_{de}{}^c\Gamma^{de}\check{D}_c\psi_a = \sqrt{5}\left(C_g- \kappa_{3/2}^2 +\frac{147}{20}-3P_7-6P_{14}-7P_{27}  \right)\psi_a,
\end{equation}
which is satisfied for transverse and gamma-traceless vector-spinors $\psi_a$, i.e.,  $D^a\psi_a=0$ and $\Gamma^a\psi_a=0$. Here $\kappa_{3/2}^2$ is the eigenvalue of the universal Laplacian
which is related  the square of the spin-3/2 operator $i\slashed{D}_{3/2}$, i.e., we have,
\begin{equation}
\Delta\psi_a=\kappa_{3/2}^2\psi_a,\qquad\quad \Delta \psi_a = ({i\slashed{D}})^2\psi_a + \frac{27}{80} \psi_a.
\label{spinthreehalfeqs}
\end{equation}
When giving the result of this subsection in the summary at the end of this section it is the eigenvalues of ${i\slashed{D}}_{3/2}$ that are quoted.
This step requires taking the square root of the result for $({i\slashed{D}}_{3/2})^2$ giving a sign ambiguity that is resolved by requiring that
the eigenvalues must be consistent with supersymmetry in the case of the left-squashed $S^7$ compactification. This is analysed in detail in \cref{sec:supermultiplet}.

\subsubsection{Irreducible \texorpdfstring{$G_2$}{G2}-components and transversality}
The strategy we will adopt to analyse the above spin-3/2 equation will rely on the fact that the $G_2$-content in ${\bf 48}$ is $ {\bf 7}\oplus {\bf 14} \oplus {\bf 27}$ which
corresponds  precisely to a traceless two-tensor without symmetrisations: The antisymmetric part, denoted $Y_{ab}$, corresponds to ${\bf 7}\oplus {\bf 14}$ while the traceless symmetric one,
$X_{ab}$,  corresponds to ${\bf 27}$. This makes it possible to  translate
the present spin-3/2 problem into problems involving two-tensors and hence to use  knowledge obtained from the previous subsection.
As in the spin-1/2 case above, we can use the Killing spinor $\eta$ (and the Fierz identity $\Gamma_a\eta\bar\eta\Gamma^a=1-\eta\bar\eta$) to define the
split \cite{Bais:1983wc, Duff:1984sv, Duff:1986hr}
\begin{equation}
\label{eq:eigs:spin-3/2:split}
\psi_a\equiv Y_a\eta + i\Gamma^bZ_{ba}\eta,
\end{equation}
where the two (real) tensors are given by
\begin{equation}
Y_a \equiv \bar\eta \psi_a,\qquad\quad Z_{ab} \equiv -i \bar\eta \Gamma_a \psi_b.
\end{equation}
Note that the two-tensor appearing here has no specified symmetry on the indices. However, from the gamma-tracelessness of the spin-3/2 field, $\Gamma^a\psi_a=0$, we see that $Z_a{}^a=0$.
Thus, the only remaining issue is whether or not  there is an independent ${\bf 7}$ piece in  $Z_{ab}$. To clarify this point we compute
\begin{equation}
a_a{}^{bc}Z_{bc} = \bar\eta\Gamma^{abc}\eta \bar\eta\Gamma_b\psi_c = -\bar\eta\Gamma^{ac}\psi_c = Y^a,
\end{equation}
where we have used the above Fierz identity for $\eta$ together with $\Gamma^a\psi_a=0$ and the relation $a_{abc}=i\bar\eta\Gamma_{abc}\eta$ (see \cref{app:octonions}).
Thus it is convenient to define the $G_2$ irreducible parts of $Z_{ab}$ as follows (for the projection operators, see \cref{app:gtwoprojectors})
\begin{equation}
\label{eq:eigs:spin-3/2:irr_comps}
X_{ab}\equiv Z_{(ab)},\qquad Y_{ab}\equiv Z_{[ab]},\qquad Y^7_{ab}=P_7Y_{ab},\qquad Y^{14}_{ab}=P_{14}Y_{ab},
\end{equation}
where $Y^7_{ab}=\frac{1}{6}a_{ab}{}^cY_c$.
Thus we see that the content of independent tensors in the spin-3/2 field $\psi_a$ is exactly one vector ($\vec{7}$), one $\vec{14}$-part of an antisymmetric two-tensor and one traceless symmetric two-tensor ($\vec{27}$).

Before attempting to find the spin-3/2 eigenvalues $\kappa_{3/2}^2$  defined in \cref{spinthreehalfeq}, we also need to analyse the transversality condition of $\psi_a$.
In terms of $\check{D}_a$, the transversality condition reads
\begin{equation}
    \check{D}^a \psi_a
    = - \frac{1}{8\sqrt{5}} a^{abc} \Gamma_{ab} \psi_c.
\end{equation}
By contracting with $\bar\eta$ and $\bar\eta \Gamma_a$, we can extract its $\vec{1}$ and $\vec{7}$ parts, respectively,
\begin{equation}
\label{eq:eigenmodes:spin_3/2:transversality}
    \check{D}^a Y_a = 0,
    \qquad\quad
    \check{D}^b Z_{ab}
    = \check{D}^b X_{ba} - \check{D}^b Y_{ba}
    = \frac{1}{\sqrt{5}} Y_a,
\end{equation}
where we have used
\begin{equation}
\label{eq:eigenmodes:spin_3/2:useful_contraction}
    i \bar\eta \Gamma_a \Gamma^{cd} \psi_b
    = a_a{}^{cd} Y_b
    + c_a{}^{cde} Z_{eb}
    - \delta_a^c Z^d{}_b
    + \delta_a^d Z^c{}_b,
\end{equation}
which follows from \cref{eq:eigs:spin-3/2:split}. These equations will be used below.
\subsubsection{Deriving the eigenvalues}
To find the eigenvalues associated to the various tensors obtained in the above decomposition of $\psi_a$,
we start by converting \cref{spinthreehalfeq} into an equation for $Z_{ab}$.%
\footnote{The eigenmodes and eigenvalues are not strictly associated to the $G_2$-irreducible pieces since $\Delta$ is block-triangular rather than block-diagonal over these, as we will see below.}
Contracting it with $\bar\eta \Gamma_a$ and
using \cref{eq:eigenmodes:spin_3/2:useful_contraction}, we find
\begin{equation}
\label{eq:eigenmodes:spin_3/2:Z_quad_fund_eq}
    -\frac{3}{2\sqrt{5}} \check{D}_a Y_b
    + \frac{1}{\sqrt{5}} a_b{}^{c d} \check{D}_{d} Z_{a c}
    - \frac{1}{2\sqrt{5}} a_a{}^{cd} \check{D}_{d} Z_{c b}
    = \biggl(C_g - \kappa_{3/2}^2 + \frac{7}{20} + P_{14} + 4 P_{7} \biggr) Z_{ab}.
\end{equation}
Here we note that the first term contains $Y_a$.

In order to get further equations involving $Y_a$ we need to discuss the various divergences (see, e.g., \cref{eq:eigenmodes:spin_3/2:transversality}). Note first that
\begin{equation}
\label{eq:eigenmodes:spin_3/2:DY7_DopY_relation}
    \check{D}^b Y^7_{ba}
    = \frac{1}{6} a_a{}^{cb} \check{D}_b Y_c
    = \frac{1}{6} \check{\mD}_1 Y_a,
\end{equation}
and that \cref{eq:eigenmodes:spin_3/2:transversality} relates the divergences of $X_{ab}$ and $Y_{ab}$ to $Y_a$.
To get other equations relating the divergences of the two-index tensors, we can contract \cref{eq:eigenmodes:spin_3/2:Z_quad_fund_eq} with $\check{D}^a$ and $\check{D}^b$.
This will lead to a situation resembling the analysis of one-forms carried out in the previous subsection. Note that  $Y_a$ satisfies the transversality
condition \cref{eq:eigenmodes:spin_3/2:transversality} which implies $D^aY_a=0$.

By contracting \cref{eq:eigenmodes:spin_3/2:Z_quad_fund_eq} with $\check{D}^b$ we get a second relation between the divergences of $Y^7_{ab}$, $Y^{14}_{ab}$ and $X_{ab}$.
Combining this with the transversality condition for $\psi_a$, \cref{eq:eigenmodes:spin_3/2:transversality}, we can solve for two of them as
\begin{alignat}{2}
\label{eq:eigenmodes:spin_3/2:divergence:a}
    &\check{D}^b Y^{14}_{ba}
    &&= -3 \check{D}^b Y^7_{ba}
    + \frac{\sqrt{5}}{3} \biggl(C_g - \kappa_{3/2}^2 + \frac{81}{20} \biggr) Y_a,
    \\
\label{eq:eigenmodes:spin_3/2:divergence:b}
    &\check{D}^b X_{ba}
    &&= - 2 \check{D}^b Y^7_{ba}
    + \frac{\sqrt{5}}{3} \biggl(C_g - \kappa_{3/2}^2 + \frac{93}{20} \biggr) Y_a.
\end{alignat}
Next, we contract \cref{eq:eigenmodes:spin_3/2:Z_quad_fund_eq} with $\check{D}^a$.
The calculation is a bit involved, more details can be found in \cite{Karlsson:2021oxd}, but again gives an equation relating the divergences of the two-index tensors.
Using \cref{eq:eigenmodes:spin_3/2:divergence:a,eq:eigenmodes:spin_3/2:divergence:b} to eliminate $\check{D}^b Y^{14}_{ba}$ and $\check{D}^b X^{14}_{ba}$ and \cref{eq:eigenmodes:spin_3/2:DY7_DopY_relation} to convert $\check{D}^b Y^7_{ba}$ to $\check{\mD}_1 Y_a$, the result can be written as
\begin{equation}
    \biggl(C_g - \kappa_{3/2}^2 + \frac{109}{40}\biggr) \check{\mD}_1 Y_a
    = \frac{\sqrt{5}}{2} \biggl[
        \biggl(C_g - \kappa_{3/2}^2 + \frac{57}{20}\biggr)^2
        - \frac{1}{4} \biggl(C_g + \frac{12}{5} \biggr)
    \biggr] Y_a.
\end{equation}
To find the eigenvalues from this equation we recall the analysis of the one-form above, in particular \cref{oneformdone,oneformdonesquare}.
It is then easy to see that in the present case it leads to a fourth-order equation in $\kappa_{3/2}^2$ with solutions
\begin{equation}
    \Delta_{3/2}^{(3)} = C_g + \frac{14}{5} \pm \frac{1}{2\sqrt{5}} \sqrt{C_g + \frac{49}{20}},
    \qquad
    \Delta_{3/2}^{(4)} = C_g + \frac{31}{10} \pm \frac{5}{2\sqrt{5}} \sqrt{C_g + \frac{49}{20}},
\end{equation}
if $Y_a \neq 0$.

If, on the other hand, $Y_a = 0$, then both $X_{ab}$ and $Y^{14}_{ab} = Y_{ab}$ are transverse by \cref{eq:eigenmodes:spin_3/2:divergence:a,eq:eigenmodes:spin_3/2:divergence:b}.
Now, the situation is a bit similar to part of the three-form analysis, where we also had $\vec{14}$ and $\vec{27}$ pieces.
To proceed, we pick out the symmetric and antisymmetric pieces of \cref{eq:eigenmodes:spin_3/2:Z_quad_fund_eq}, giving
\begin{gather}
\label{eq:eigenmodes:spin_3/2:separated:sym}
    \check{\mD}_{(2)} X_{ab} - 3 (\check{\mD}_{(2)} Y)_{ab}
    = 2\sqrt{5} \biggl(C_g - \kappa_{3/2}^2 + \frac{7}{20}\biggr) X_{ab},
    \\
\label{eq:eigenmodes:spin_3/2:separated:asym}
    -3 (\check{\mD}_{[2]} X)_{a b} + \check{\mD}_{[2]} Y_{a b}
    = 2\sqrt{5} \biggl(C_g - \kappa_{3/2}^2 + \frac{27}{20}\biggr) Y_{ab}.
\end{gather}
Recall, again, from the two-form analysis the very useful fact that a transverse two-form without $\vec{7}$-part satisfies $\check{\mD}_{[2]} Y_{ab} = 0$.
Thus, we can form a linear combination of \cref{eq:eigenmodes:spin_3/2:separated:sym,eq:eigenmodes:spin_3/2:separated:asym} only containing the operator $\mD_2$ as
\begin{equation}
\label{eq:eigenmodes:spin_3/2:combined_X_Y}
    (\check{\mD}_2 X)_{ab} - 3 (\check{\mD}_2 Y)_{ab}
    = 2\sqrt{5} \biggl(C_g - \kappa_{3/2}^2 + \frac{7}{20}\biggr) X_{ab}
    - \frac{2\sqrt{5}}{3} \biggl(C_g - \kappa_{3/2}^2 + \frac{27}{20}\biggr) Y_{ab}.
\end{equation}
To see why this is useful, we recall another piece of analysis from the previous subsection, namely for the Lichnerowicz operator, and in particular the equation \cref{eq:eigenvalues:Lich:squared_op}
\begin{equation}
\label{eq:eigenvalue:D2^2}
    (\check{\mD}_2{}^2 X)_{ab} = C_g X_{ab} - \frac{1}{\sqrt{5}} (\check{\mD}_2 X)_{ab},
\end{equation}
for transverse traceless symmetric rank-2 tensors $X_{ab}$.
A short calculation shows that the same relation also holds for transverse two-forms.
If we act with $\check{\mD}_2$ on \cref{eq:eigenmodes:spin_3/2:combined_X_Y}, use this relation and then pick out the antisymmetric part, we get an equation relating $Y_{ab}$ to $(\check{\mD}_{[2]} X)_{a b}$, namely
\begin{equation}
    C_g Y_{ab}
    = -\frac{2\sqrt{5}}{3} \biggl(C_g - \kappa_{3/2}^2 + \frac{9}{20}\biggr) (\check{\mD}_{[2]} X)_{a b}.
\end{equation}
Importantly, since we used \cref{eq:eigenvalue:D2^2}, this equation is linearly independent from \cref{eq:eigenmodes:spin_3/2:separated:asym} and they can be combined into
\begin{equation}
    C_g Y_{ab}
    = \frac{20}{9} \biggl(C_g - \kappa_{3/2}^2 + \frac{9}{20}\biggr) \biggl(C_g - \kappa_{3/2}^2 + \frac{27}{20}\biggr) Y_{ab}.
\end{equation}
Thus, either $Y_{ab} = 0$ or
\begin{equation}
    \Delta_{3/2}^{(2)} = C_g + \frac{9}{10} \pm \frac{3}{2\sqrt{5}} \sqrt{C_g + \frac{9}{20}}.
\end{equation}
In the last case, $Y_{ab} = 0$, when only the $\vec{27}$-part is non-vanishing \cref{eq:eigenmodes:spin_3/2:separated:sym,eq:eigenmodes:spin_3/2:separated:asym} gives us the Lichnerowicz situation reviewed above  but with the additional  information that $(\check{\mD}_{[2]} X)_{ab} = 0$.
As it turns out, this is exactly the situation from the three-form analysis with only a non-vanishing $\vec{27}$-part, also reviewed in the previous subsection.
The implication from that analysis is then, again, that we only get two eigenvalues, given in \cref{eq:eigenmodes:spin_3/2:kappa^2:d} below.

In summary, we have found that all eigenvalues of $\Delta_{3/2}$ fall into one of the following sets:
\begin{align}
    &\Delta_{3/2}^{(3)} = C_g + \frac{14}{5} \pm \frac{1}{2\sqrt{5}} \sqrt{C_g + \frac{49}{20}},
    \\
    &\Delta_{3/2}^{(4)} = C_g + \frac{31}{10} \pm \frac{5}{2\sqrt{5}} \sqrt{C_g + \frac{49}{20}},
    \\
    &\Delta_{3/2}^{(2)} = C_g + \frac{9}{10} \pm \frac{3}{2\sqrt{5}} \sqrt{C_g + \frac{9}{20}},
    \\
\label{eq:eigenmodes:spin_3/2:kappa^2:d}
    &\Delta_{3/2}^{(1)} = C_g + \frac{2}{5} \pm \frac{1}{2\sqrt{5}} \sqrt{C_g + \frac{1}{20}}.
\end{align}
Using \cref{spinthreehalfeqs}, and supersymmetry to fix the sign ambiguity, it is possible to determine the eigenvalues of ${i\slashed D}_{3/2}$, which are denoted ${i\slashed D}_{3/2}^{(i)_{\pm}}$, $i=3,4,2,1$, in the summary below.

\subsection{Summary: The entire squashed \texorpdfstring{$S^7$}{S7} operator eigenvalue spectrum}
\label{sec:eigenvalues:summary}

In \cref{sec:supermultiplet} we will  use the results of this section, and the next, to specify the content of irreps
$D(E_0,s)$ of all supermultiplets in the left-squashed vacuum.
To facilitate that discussion we here summarise the results on the left-squashed seven-sphere operator spectra we have obtained in this section:
\begin{alignat}{2}
\label{eq:eigs:summary:Delta_0:1}
    & \Delta_0^{(1)} &&= \frac{m^2}{9}\,20C_g,
    \\[4pt]
\label{eq:eigs:summary:Delta_1:1}
    & \Delta_1^{(1)_\pm} &&= \frac{m^2}{9}\,\Bigl(20C_g+14\pm 2\sqrtsmash[C_G]{20C_g+49}\Bigr) =\frac{m^2}{9}\,\Bigl(\sqrtsmash[C_G]{20C_g+49}\pm1\Bigr)^2-4m^2,
    \\[4pt]
\label{eq:eigs:summary:Delta_2:1}
    & \Delta_2^{(1)} &&= \frac{m^2}{9}\,(20C_g+72),\\
\label{eq:eigs:summary:Delta_2:2}
    & \Delta_2^{(2)_\pm} &&= \frac{m^2}{9}\,\Bigl(20C_g+44\pm 4\sqrtsmash[C_G]{20C_g+49}\Bigr) =\frac{m^2}{9}\,\bigl(\sqrtsmash[C_G]{20C_g+49}\pm2\bigr)^2-m^2,\\
\label{eq:eigs:summary:Delta_2:3}
    & \Delta_2^{(3)} &&= \frac{m^2}{9}\,20C_g,
    \\[4pt]
\label{eq:eigs:summary:Delta_L:1}
    & \Delta_L^{(1)} &&= \frac{m^2}{9}\,(20C_g+36),\\
\label{eq:eigs:summary:Delta_L:2}
    & \Delta_L^{(2)_\pm} &&= \frac{m^2}{9}\,\Bigl(20C_g+32\pm 4\sqrtsmash[C_G]{20C_g+1}\Bigr) =\frac{m^2}{9}\,\Bigl(\sqrtsmash[C_G]{20C_g+1}\pm2\Bigr)^2+3m^2,
    \\[4pt]
\label{eq:eigs:summary:Q:1}
    & Q^{(1)_\pm} &&= \frac{m}{3}\Bigl(-1\pm\sqrtsmash[C_G]{20C_g+1}\Bigr),\\
\label{eq:eigs:summary:Q:2}
    & Q^{(2)_\pm} &&= \frac{m}{3}\Bigl(1\pm \sqrtsmash[C_G]{20C_g+49}\Bigr),\\
\label{eq:eigs:summary:Q:3}
    & Q^{(3)_\pm} &&=  \frac{m}{3}\Bigl(3\pm \sqrtsmash[C_G]{20C_g+81}\Bigr),
    \\[4pt]
\label{eq:eigs:summary:D_1/2:1}
    & i{\slashed D}_{1/2}^{(1)_\pm} &&= \frac{m}{3}\biggl(\frac{3}{2}\pm \sqrtsmash[C_G]{20C_g+81}\biggr),\\
\label{eq:eigs:summary:D_1/2:2}
    & i{\slashed D}_{1/2}^{(2)_\pm} &&= \frac{m}{3}\biggl(-\frac{1}{2} \pm \sqrtsmash[C_G]{20C_g+49}\biggr),
    \\[4pt]
\label{eq:eigs:summary:D_3/2:1}
    & i{\slashed D}_{3/2}^{(1)_\pm} &&= \frac{m}{3}\biggl(\frac{1}{2}\pm \sqrtsmash[C_G]{20C_g+1}\biggr),\\
\label{eq:eigs:summary:D_3/2:2}
    & i{\slashed D}_{3/2}^{(2)_\pm} &&= \frac{m}{3}\biggl(-\frac{3}{2}\pm \sqrtsmash[C_G]{20C_g+9}\biggr),\\
\label{eq:eigs:summary:D_3/2:3}
    & i{\slashed D}_{3/2}^{(3)_\pm} &&= \frac{m}{3}\biggl(\frac{1}{2}\pm \sqrtsmash[C_G]{20C_g+49}\biggr),\\
\label{eq:eigs:summary:D_3/2:4}
    & i{\slashed D}_{3/2}^{(4)_\pm} &&= \frac{m}{3}\biggl(\frac{5}{2}\pm \sqrtsmash[C_G]{20C_g+49}\biggr),
\end{alignat}
Note that all linear operators have a sign ambiguity from the calculations above since those analyses used quadratic operators and gave the squares of the eigenvalues of the linear operators. However, as we will see in \cref{sec:supermultiplet} using supersymmetry, and for spinors already in \cref{sec:eigenmodes}, these
signs can be determined uniquely and these results has been used when quoting the eigenvalues above.%
\footnote{There is one exception were this sign ambiguity is not resolved uniquely, namely the $(p,q;r) = (0,1;0)$ isometry irrep in which $Q^{(1)_-} = -10 m/3$ and $-Q^{(1)_+} = -8 m/3$ give rise to the same $M^2$ using the formula in \cref{table:massop}. We conjecture that the former is the correct eigenvalue based on the methods of the next section but either way this has no bearing on the $AdS_4$ analysis of \cref{sec:supermultiplet}.}

In  \cref{sec:supermultiplet} we will  fit all these eigenvalues into the supermultiplets that were established to be present in the left-squashed seven-sphere spectrum in
\cite{Nilsson:2018lof}. These  supermultiplets are one spin $2^+$, six spin $3/2$, six spin $1^+$, eight spin $1^-$ and 14 Wess--Zumino supermultiplets.
This procedure shows that there are degenerate $E_0$ values of supermultiplets as well as operator eigenvalues that are associated with more than one set of modes.
This curious fact requires a deeper analysis of these modes. One such
analysis is carried out for two-forms in the next section proving explicitly that the degeneracy deduced from the supersymmetry analysis is realised in this case.
\section{Associating isometry irreps with eigenvalues through eigenmodes}%
\label{sec:eigenmodes}

The purpose of this section is to provide a better understanding of the connection between isometry irreps and eigenvalues, i.e., which eigenvalue formulas from the previous section apply to which isometry irreps, through individual eigenmodes in the various operator spectra.
The eigenvalue results  in the previous section  and in previous works like
\cite{Ekhammar:2021gsg,Karlsson:2021oxd} were obtained using methods that did not rely on writing down explicit modes
and hence could not give  a  precise relation  to the spectra of isometry irreps, which was derived in full detail  in \cite{Nilsson:2018lof}.
Exceptions to this where explicit modes have been constructed are spinors \cite{Nilsson:1983ru}, one-forms \cite{Yamagishi:1983ri}
and the entire singlet sector \cite{Nilsson:2023ctq}.

Lacking this connection between eigenvalues and isometry irreps has led to a potential problem, namely that
there seems to be fewer distinguished eigenvalues than isometry irreps. For instance, for two-forms, which is the first instant where this phenomenon arises,
we found in the previous section four different eigenvalues that should  be related to five different modes all having the same isometry irrep, namely the five
$r=p$ cross diagrams for two-forms in the appendix of \cite{Nilsson:2018lof}. Similar properties can be identified for the Lichnerowicz operator (three eigenvalues and six modes), three-forms (six eigenvalues and eight modes)
and vector-spinors (eight eigenvalues and twelve modes).
However, the eigenvalues found in the previous section are consistent with supersymmetry for all modes in the left-squashed vacuum, as will be clear from the next section, indicating that there might not be a problem after all.

To get a better understanding of this issue  we will construct all eight sets of two-form modes and compute their individual eigenvalues in this section.
The results obtained provide strong support for the existence of these degeneracies which are made explicit in the next section.
We start the discussion in this section by reviewing the one-form and spinor cases which will also define our notation and explain our approach,
which is in the spirit of the spinor analysis in \cite{Nilsson:1983ru}.

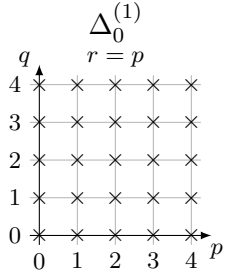
\begin{figure}[H]
    \centering
    \begingroup
    \renewcommand{\scale}{0.50cm}  %
    \footnotesize
    \newcommand{\fig}{%
        \begin{tikzpicture}[x=\scale, y=\scale]
            \crossdiagramlayout{p}
            \foreach \p in {0,...,\range} {
                \foreach \q in {0,...,\range} {
                    \node at (\p,\q) {$\times$};
                }
            }
        \end{tikzpicture}
    }
    \newcommand*{\eig}[1]{{\normalsize$\Delta_0^{#1}$}}%
    \begin{tabular}{c}
        \eig{(1)}   \\
        \fig
    \end{tabular}
    \endgroup
    \caption{Scalar cross diagram. Each cross corresponds to an isometry irrep $(p,q;r)$, for $p\ge 0$, $q\ge 0$,  of eigenmodes $\phi$ of $\Delta_0$ with eigenvalues
    $\Delta_0^{(1)} =\frac{m^2}{9} \,20C_g$, as given in \cref{eq:eigs:summary:Delta_0:1}.}
    \label{fig:modes:0-form:cross_diagrams}
\end{figure}

The modes will, in all cases, be constructed by letting differential operators $\mode$ (of orders zero%
\footnote{The zeroth order differential operators are just constants, like $\mode^{(2)i}_a = s_a{}^i$ in \cref{eq:modes:1-form:basis}.}%
, one and two) act on a scalar mode functions $\phi$.
From \cite{Nilsson:2018lof}, we know that the isometry irreps in the scalar Fourier expansion are $(p, q; p)$ with $p \geq 0$ and $q \geq 0$ (all of multiplicity one) which we illustrate diagrammatically in a
cross diagram in \cref{fig:modes:0-form:cross_diagrams}. As is well-known, the zero-form eigenvalues are $\Delta_0^{(1)}=\frac{m^2}{9}\,20C_g$ (see \cref{sec:eigenvalues}).
In what follows, it is understood that these cross diagrams should be infinitely extended in the two positive directions and that
they are always made large enough to account for all special cases that arise for small $p$ or $q$.
\subsection{Warm-up: One-form and spinor modes}
\subsubsection{One-form modes}
We want to construct the transverse, $D^a Y_a = 0$, eigenmodes $Y_a$ of $\Delta_1$.
We will again make use of \cref{masterdelta} but this time write it as
\begin{equation}
\label{eq:modes:1-form:masterdelta}
  \Delta Y_a = C_g Y_a - \frac{1}{\sqrt{5}} a_a{}^{bc} \check{D}_c Y_b + \frac{3}{5} Y_a,
\end{equation}
which tells us that we only need  to act with one derivative on $Y_a$ to obtain  the eigenvalues of $\Delta$.

The first step in this analysis is to construct a basis for all one-form modes spanning the entire set of  isometry irreps.  Equivalently, we can write down an
expression for the most general one-form mode in the isometry irrep $(p,q;r)$ which can be done
using a scalar $\phi$ and a one-form differential operator $\mode_a$.
Since scalars only exists for $r=p$, see, e.g., \cite{Nilsson:2018lof}, the scalar will be in the isometry irrep $(p,q;p)$ which implies that we
have to consider $\mode_a$ in non-trivial representations of $sp_1^C$ to get one-forms with $r \neq p$.

Not yet enforcing transversality, the one-form isometry irreps from \cite{Nilsson:2018lof} consists of one $\vec{1}_{sp_1^C}$ and two $\vec{3}_{sp_1^C}$.
With this, we mean that the isometry irreps are obtained by decomposing $(p,q;p) \otimes (\vec{1}_{sp_1^C} \oplus (2 \times \vec{3}_{sp_1^C}))$ into irreps, i.e.,
there are three cross diagrams with $r = p$ and two each with $r = p \pm 2$ by standard addition of angular momentum.%
\footnote{There are sometimes special cases for small $p$ or $q$ where there are fewer isometry irreps than indicated here. These details will provide a consistency check presented below.}
Hence, we need one singlet $\mode_a$ and two triplets $\mode^i_a$, for which we use
\begin{equation}
\label{eq:modes:1-form:basis}
  \mode^{(1)}_a = \check{D}_a, \qquad
  \mode^{(2)i}_a = s_a{}^i, \qquad
  \mode^{(3)i}_a = a_a{}^{bc} s_b{}^i \check{D}_c.
\end{equation}
Here, $s_a{}^i$ are the components of the $sp_1^C$ Killing vectors, see \cite{Nilsson:1983ru}.%
\footnote{Note that the index $i$ in this section refers to $sp_1$ and is not the same as the $h$-index $i$ in \cref{sec:review_and_method}.}
We normalise them such that $[s^i, s^j] = \epsilon^{ijk} s_k$, where $s^i = s_a{}^i \partial^a$, which one can check explicitly using that the only non-vanishing components are
\begin{equation}
    s_{\hat i}{}^j = \frac{1}{\sqrt{5}} \delta_i^j,
\end{equation}
where we have used the index split $a=(\hat{i},0,i)$ as in \cref{app:octonions} and \cite{Karlsson:2021oxd}.
We will need the $H$-covariant derivative of the Killing vector components \cite{Nilsson:1983ru} (see \cref{app:octonions}),
\begin{equation}
\label{eq:modes:1-form:Ds}
    \check{D}_a s_b{}^i = -3 \epsilon^i{}_{jk} s_a{}^j s_b{}^k.
\end{equation}

A general one-form in the $(p,q;r)$ isometry irrep can now be written as a linear combination of $P_r \mode^{(1)} \phi$, $P_r \mode^{(2)i} \phi$ and $P_r \mode^{(3)i} \phi$,
where $P_r$ projects onto the $sp_1^C$-irrep $(r)_C$.
Note that since $P_r$ acts on the $sp_1^C$ indices of the differential operators and the (suppressed) isometry irrep index of $\phi$, it clearly commutes with $\Delta$, which is also obvious from the fact that the eigenmodes of $\Delta$ fall into isometry irreps.

The projection operator $P_r$ can also be implemented as a differential operator.
As a first step in this direction, note that $P_p \mode^i \phi \propto \mode^j s_j s^i \phi$ since the latter clearly only contains a $r = p$ piece.
More generally, the projector is given by
\begin{equation}
\label{eq:modes:1-form:projector}
    P_r = \prod_{r' \neq r} \frac{C_{\smash[t]{sp_1^C}} - C_{\smash[t]{sp_1^C}}(r')}{C_{\smash[t]{sp_1^C}}(r) - C_{\smash[t]{sp_1^C}}(r')},
\end{equation}
where $r'$ runs over all $sp_1^C$-irreps except $(r)_C$, since the right-hand side clearly vanishes on any other irrep than $(r)_C$ and becomes the identity when restricted to this $sp_1^C$-irrep.%
\footnote{Note that the projector is unique as long as the $sp_1^C$-irreps are non-degenerate. This is true in our setting since $P_r$ will act on $\mode \phi$ and the tensor product of two $sp_1^C$-irreps has a non-degenerate irrep decomposition.}
In this equation, $C_{\smash[t]{sp_1^C}}(r)$ is just the eigenvalue of the Casimir on the $(r)_C$ irrep but to use it in practice, we also need the action of the Casimir operator on $\mode \phi$.
As an example, when the differential operator is a $\vec{3}_{sp_1^C}$, i.e., $\mode^i$, we have (irrespective of the suppressed $so(7)$ index on $\mode^i$)
\begin{equation}
\label{eq:modes:1-form:Casimir_action}
    C_{\smash[t]{sp_1^C}} \mode^i \phi
    = -(T_k T^k \mode^i) \phi - \mode^i (T_k T^k \phi) - 2 (T_k \mode^i) (T^k \phi)
    = \bigl(C_{\smash[t]{sp_1^C}}^\phi + 2\bigr) \mode^i \phi - 2 \epsilon^i{}_{jk} \mode^j s^k \phi,
\end{equation}
where $T_k$ are the $sp_1^C$ generators and the $2$ comes from $C_{\smash[t]{sp_1^C}}(\vec{3}) = 2$.%
\footnote{We write $C_{\smash[t]{sp_1^C}}^\phi = C_{\smash[t]{sp_1^C}}(p)$ and $C_g^\phi = C_g(p,q;p)$ for the eigenvalues of the Casimirs $C_{\smash[t]{sp_1^C}}$ and $C_g$, respectively, on the irrep $(p,q;p)$ of $\phi$.}

The next step is to compute the action of $\Delta$ using \cref{eq:modes:1-form:masterdelta}.
To this end, we first compute
\begin{align}
\label{eq:modes:1-form:calculation:first}
  & a_a{}^{bc} \check{D}_c (\mode^{(1)}_b \phi)
  = \frac{3}{\sqrt{5}} \check{D}_a \phi
  = \frac{3}{\sqrt{5}} \mode^{(1)}_a \phi,
  \\
  &a_a{}^{bc} \check{D}_c (\mode^{(2)i}_b \phi)
  = \frac{6}{\sqrt{5}} s_a{}^i \phi + a_a{}^{bc} s_b{}^i \check{D}_c \phi
  = \frac{6}{\sqrt{5}} \mode^{(2)i}_a \phi + \mode^{(3)i}_a \phi,
  \\
  &\begin{aligned}[b]
    a_a{}^{bc} \check{D}_c (\mode^{(3)i}_b \phi)
    &= \check{D}_a s^i \phi
    - s_a{}^i \check{\Box} \phi - 6 \epsilon^i{}_{jk} s_a{}^j s^k \phi
    - \frac{7}{\sqrt{5}} a_a{}^{bc} s_b{}^i \check{D}_c \phi \\
    &= \mode^{(1)} s^i \phi
    + (C_g - 6) \mode^{(2)i} \phi
    - \frac{7}{\sqrt{5}} \mode^{(3)i} \phi,
  \end{aligned}
\label{eq:modes:1-form:calculation:last}
\end{align}
where, in the last step, we used that $\check{\Box}\phi = - C_g^\phi \phi$, \cref{eq:method:C_g} and \cref{eq:modes:1-form:Casimir_action}.
Putting \cref{eq:modes:1-form:masterdelta} and \crefrange{eq:modes:1-form:calculation:first}{eq:modes:1-form:calculation:last} together we get
\begingroup
\setlength{\arraycolsep}{4pt} %
\begin{equation}
\label{eq:modes:1-form:Delta_result}
  \Delta P_r
  \left(
    \begin{array}{cc}
        \mode^{(1)}_a & \mode^{(2\text{--}3)i}_a
    \end{array}
  \right)
  \phi
  =
  P_r
  \left(
    \begin{array}{cc}
        \mode^{(1)}_a & \mode^{(2\text{--}3)i}_a
    \end{array}
  \right)
  \left(
    \renewcommand*{\arraystretch}{1.2} %
    \begin{array}{c|cc}
      C_g & 0 & -\frac{1}{\sqrt{5}} s^i \\[3pt] \hline
      0 & C_g - \frac{3}{5} & -\frac{1}{\sqrt{5}} (C_g - 6) \\
      0 & -\frac{1}{\sqrt{5}} & C_g + 2
    \end{array}
  \right)
  \phi,
\end{equation}
\endgroup
where we have indicated the $sp_1^C$-irrep block structure of the matrix and $\mode^{(2\text{--}3)i}_a$ should be understood as the block $(\mode^{(2)i}_a\ \mode^{(3)i}_a)$ so that the matrix multiplication between the row vector containing the differential operators $\mode$ and the matrix on the right-hand side makes sense.
Note that the lower left zero block occurs since no $sp_1^C$ index can appear when computing $\Delta \mode \phi$ for an $sp_1^C$ singlet $\mode$.
This block triangular structure is a generic feature that will reappear in the spinor and two-form analyses below.
The above matrix can easily be diagonalised to obtain the eigenmodes and their corresponding eigenvalues.

Here, we are only interested in the transverse eigenmodes.
From \cite{Nilsson:2018lof}, we know that imposing transversality reduces the one-form isometry irrep content from one $\vec{1}_{sp_1^C}$ and two $\vec{3}_{sp_1^C}$ to just two $\vec{3}_{sp_1^C}$.
Computing the divergences we find
\begin{align}
  & D^a \mode^{(1)}_a \phi = \check{\Box} \phi, \\
  & D^a \mode^{(2)i}_a \phi = s^i \phi, \\
  & D^a \mode^{(3)i}_a \phi = - \frac{3}{\sqrt{5}} s^i \phi,
\end{align}
which may be written, using that $C_g^\phi P_r \mode^{(1)}_a s^i \phi = C_g P_r \mode^{(1)}_a s^i \phi$ since $P_{r\neq p} \mode^{(1)}_a s^i \phi = 0$, as
\begingroup
\setlength{\arraycolsep}{4pt}
\begin{equation}
\label{eq:modes:1-form:divergences}
  D^a P_r
  \left(
    \begin{array}{cc}
      \mode^{(1)}_a &
      \mode^{(2\text{--}3)i}_a
    \end{array}
  \right)
  \phi
  =
  P_r
  \left(
    \begin{array}{c|cc}
        - C_g &
      s^i &
      -\frac{3}{\sqrt{5}} s^i
    \end{array}
  \right)
  \phi,
\end{equation}
\endgroup
where the vertical bar on the right-hand side indicate the $sp_1^C$ block structure that is manifest on the left-hand side.
From this, we see that the expectation from \cite{Nilsson:2018lof} is borne out; $\mode^{(1)}_a \phi$ is not transverse but can be combined with $\mode^{(2,3)i}_a \phi$ to make them transverse.

Diagonalising \cref{eq:modes:1-form:Delta_result} and then using \cref{eq:modes:1-form:divergences}, we find the transverse eigenmodes $P_r \tilde{\mode}^{(1)_\pm i}_a \phi$ of $\Delta_1$ and their corresponding eigenvalues $\Delta_1^{(1)_\pm}$, where
\begin{align}
\label{eq:modes:1-form:eigenmodes_eigenvalues:first}
    & \tilde{\mode}^{(1)_\pm i}_a
    = 2 \sqrt{5} \mode^{(1)}_a s^i
    + \frac{1}{\sqrt{5}} \Bigl( 10 C_g - 21 \mp 3 \sqrtsmash[C_G]{20 C_g + 49} \Bigr) \mode^{(2)i}_a
    - \Bigl(7 \pm \sqrtsmash[C_G]{20 C_g + 49}\Bigr) \mode^{(3)i}_a ,
    \\
    & \Delta_1^{(1)_\pm}
    = \frac{m^2}{9} \Bigl(20 C_g + 14 \pm 2 \sqrtsmash[C_G]{20 C_g + 49}\Bigr),
\label{eq:modes:1-form:eigenmodes_eigenvalues:last}
\end{align}
which agrees with the eigenvalues in \cref{eq:eigs:summary:Delta_1:1}.

As mentioned above, there are some special cases of isometry irreps  that can occur when either $p$ or $q$ is small \cite{Nilsson:2018lof}.
To see for exactly which $(p,q;r)$ the eigenmodes and their associated eigenvalues exist, we compute their norms
\begin{equation}
\label{eq:modes:1-form:norm_formula}
  \| P_r \tilde\mode^{(1)_\pm} \phi \|^2
  = \int \mathrm{vol}\, \bigl( \tilde\mode^{(1)_\pm i}_a \phi \bigr) P_r \bigl( \tilde\mode^{(1)_\pm a}_i \phi \bigr),
\end{equation}
where the integral  is over the squashed seven-sphere.
Note that either all or none of the eigenmodes in each copy of an irrep $(p,q;r)$ vanish since the representation is irreducible, hence we have contracted the
$sp_1^C$ indices in the above equation without loosing information about which eigenmodes exist.

To proceed, we use \cref{eq:modes:1-form:projector} to write the projector in terms of the Casimir operator.
In the case at hand, we are dealing with $\vec{3}_{sp_1^C}$ operators, so $r = p,\; p\pm 2$ and $r'$ takes the two values different from $r$ out of these three.
Hence, we have to act with the Casimir $C_{\smash[t]{sp_1^C}}$ twice on the modes.
We already saw in \cref{eq:modes:1-form:Casimir_action} how to act once.
When acting with the Casimir operator again, we also get a term
\begin{equation}
\label{eq:modes:1-form:es_es}
  \epsilon^i{}_{jk} (\epsilon^j{}_{mn} \mode^m s^n) s^k
  = - \mode^i s^2 + \frac{1}{2} \epsilon^i{}_{jk} \mode^j s^k + \mode_j s^{ij},
\end{equation}
where $s^{ij} = s^{(i} s^{j)}$.
Thus, the squared norms $\| P_r \tilde\mode^{(1)_\pm} \phi \|^2$ can be written as linear combinations of
\begin{equation}
  \langle \mode^{(m)i} \phi,\, \delta_{ij} \mode^{(n)j} \phi \rangle,\qquad
  \langle \mode^{(m)i} \phi,\, \epsilon_{ijk} \mode^{(n)j} s^k \phi \rangle,\qquad
  \langle \mode^{(m)i} \phi,\, \mode^{(n)j} s_{ij} \phi \rangle,
\end{equation}
where $m,n = 1,2,3$ are labels (not $sp_1^C$ indices) and $\mode^{(1)i} \equiv \mode^{(1)} s^i$.

To compute the norms of the eigenmodes we note that it suffices to compute the one-form scalar products
\begin{gather}
\label{eq:modes:1-form:scalar_products:first}
    \begin{gathered}
        \langle \mode^{(1)} \phi, \mode^{(1)} \phi \rangle
        = C_g^\phi \| \phi \|^2,\\
        \langle \mode^{(1)} \phi, \mode^{(2)i} s_i \phi \rangle
        = C_{\smash[t]{sp_1^C}}^\phi \| \phi \|^2, \qquad
        \langle \mode^{(1)} \phi, \mode^{(3)i} s_i \phi \rangle
        = - \frac{3}{\sqrt{5}} C_{\smash[t]{sp_1^C}}^\phi \| \phi \|^2,
    \end{gathered}
    \\
    \langle \mode^{(2)i} \phi, \mode^{(2)j} \varphi \rangle
    = \frac{1}{5} \langle \phi, \delta^{ij} \varphi \rangle,\qquad
    \langle \mode^{(2)i} \phi, \mode^{(3)j} \varphi \rangle
    = \frac{1}{\sqrt{5}} \langle \phi, \epsilon^{ijk} s_k \varphi \rangle,\\
\label{eq:modes:1-form:scalar_products:last}
     \langle \mode^{(3)i} \phi, \mode^{(3)j} \varphi \rangle
    = \frac{1}{5} C_g^\phi \langle \phi, \delta^{ij} \varphi \rangle
    - \frac{9}{10} \langle \phi, \epsilon^{ijk} s_k \varphi \rangle
    + \langle \phi, s^{ij} \varphi \rangle,
\end{gather}
by using that $\langle \mode s^i \phi, \mode' \phi \rangle = - \langle \mode \phi, \mode' s^i \phi \rangle$,
which follows from the $sp_1^C$-irrep of $\phi$ being unitarity,
where $\mode, \mode'$ are differential operators (that may have suppressed $sp_1^C$ indices).
Apart from these, we have to compute the zero-form scalar products that appear when putting the above together, e.g.,
\begin{equation}
\label{eq:modes:1-form:0-form_scalar_product}
  \langle \phi, s^{ij} s_{ij} \phi \rangle
  = C_{\smash[t]{sp_1^C}}^\phi \biggl(C_{\smash[t]{sp_1^C}}^\phi - \frac{1}{2}\biggr) \| \phi \|^2.
\end{equation}

\begin{figure}[H]
  \centering
  \begingroup
  \renewcommand{\scale}{0.50cm}  %
  \footnotesize
  \newcommand{\figIppII}{%
      \begin{tikzpicture}[x=\scale, y=\scale]
          \crossdiagramlayout{p+2}
          \foreach \p in {0,...,\range} {
              \foreach \q in {1,...,\range} {
                  \node at (\p,\q) {$\times$};
              }
          }
      \end{tikzpicture}
  }
  \newcommand{\figIp}{%
      \begin{tikzpicture}[x=\scale, y=\scale]
          \crossdiagramlayout{p}
          \foreach \p in {1,...,\range} {
              \foreach \q in {0,...,\range} {
                  \node at (\p,\q) {$\times$};
              }
          }
      \end{tikzpicture}
  }
  \newcommand{\figIpmII}{%
      \begin{tikzpicture}[x=\scale, y=\scale]
          \crossdiagramlayout{p-2}
          \foreach \p in {2,...,\range} {
              \foreach \q in {0,...,\range} {
                  \node at (\p,\q) {$\times$};
              }
          }
      \end{tikzpicture}
  }
  \newcommand{\figIIppII}{%
      \begin{tikzpicture}[x=\scale, y=\scale]
          \crossdiagramlayout{p+2}
          \foreach \p in {0,...,\range} {
              \foreach \q in {0,...,\range} {
                  \node at (\p,\q) {$\times$};
              }
          }
      \end{tikzpicture}
  }
  \newcommand{\figIIp}{%
      \begin{tikzpicture}[x=\scale, y=\scale]
          \crossdiagramlayout{p}
          \foreach \p in {1,...,\range} {
              \foreach \q in {1,...,\range} {
                  \node at (\p,\q) {$\times$};
              }
          }
      \end{tikzpicture}
  }
  \newcommand{\figIIpmII}{%
      \begin{tikzpicture}[x=\scale, y=\scale]
          \crossdiagramlayout{p-2}
          \foreach \p in {2,...,\range} {
              \foreach \q in {0,...,\range} {
                  \node at (\p,\q) {$\times$};
              }
          }
      \end{tikzpicture}
  }
  \newcommand*{\eig}[1]{{\normalsize$\Delta_1^{#1}$}}%
  \begin{tabular}{cc}
      \eig{(1)_-} & \eig{(1)_+}   \\
      \figIIppII  & \figIppII     \\
      \figIIp     & \figIp        \\
      \figIIpmII  & \figIpmII
  \end{tabular}
  \endgroup
  \caption{Transverse one-form cross diagrams. Each cross corresponds to an isometry irrep $(p,q;r)$ of transverse eigenmodes, listed in \cref{eq:modes:1-form:eigenmodes_eigenvalues:first}, of $\Delta_1$ with an associated eigenvalue given in \cref{eq:modes:1-form:eigenmodes_eigenvalues:last}. Both columns have a unified $\vec{3}_{sp_1^C}$ one-form differential operator generating the eigenmodes and a unified expression for the eigenvalues.}
  \label{fig:modes:1-form:cross_diagrams}
\end{figure}
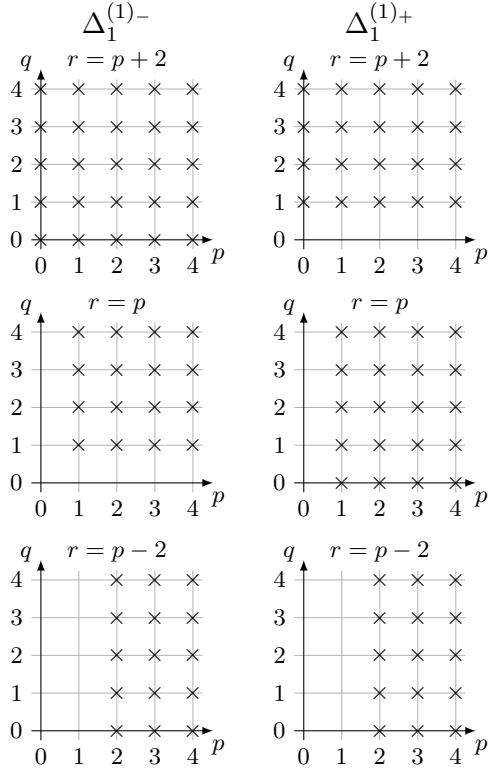

Putting everything together to compute the norms in \cref{eq:modes:1-form:norm_formula}, i.e.,
the eigenmodes from \cref{eq:modes:1-form:eigenmodes_eigenvalues:first},
the projector from \cref{eq:modes:1-form:projector},
the action of the Casimir from \cref{eq:modes:1-form:Casimir_action} and \cref{eq:modes:1-form:es_es},
the one-form scalar products from \crefrange{eq:modes:1-form:scalar_products:first}{eq:modes:1-form:scalar_products:last} and the zero-form scalar products like \cref{eq:modes:1-form:0-form_scalar_product},
we get
\begingroup%
\footnotesize%
\begin{align}
\label{eq:eigenmodes:1-form:norms:first}
  &%
  \begin{aligned}[b]
    \| P_{p+2} \tilde\mode^{(1)_\pm} \phi \|^2
    \propto
    \frac{1}{25} (p+2) (p+3) \Bigl(
      \bigl(20 q (p+q+3) + (5 p + 6)^2\bigr) \bigl(20 q (p+q+3) + (5 p + 13)^2 \bigr)
    \\
    \mp \bigl(20 q (p+q+3) (5 p-1) + (5 p+6)^2 (5 p+13) \bigr) \sqrt{20 q (p+q+3) + (5 p+13)^2}
    \mathrlap{\Bigr),}
  \end{aligned}
  \\ &%
  \begin{aligned}[b]
    \| P_{p} \tilde\mode^{(1)_\pm} \phi \|^2
    \propto
    \frac{2}{25} p (p+2) & \Bigl(
    \bigl(5 q (p+q+3) + (5p + 14)\bigr) \bigl(20 q (p+q+3) + (5 p+7)^2 \bigr)
    \\
    & \pm \bigl(55 q (p+q+3) + (5p + 14)(5p + 7) \bigr) \sqrt{20 q (p+q+3) + (5 p+7)^2}
    \Bigr),
  \end{aligned}
  \\ &%
  \begin{aligned}[b]
    \| P_{p-2} \tilde\mode^{(1)_\pm} \phi \|^2
    \propto
    \frac{1}{25} p (p-1) \Bigl(
      \bigl(20 q (p+q+3) + 5p (5p + 16) + 56\bigr) \bigl(20 q (p+q+3) + 5p (5p + 2) + 49\bigr)
    \\
    \pm \bigl(20 q (p+q+3) (5p + 11) + 25 p^2 (5 p + 13) + 8 (75 p + 49)\bigr) \sqrt{20 q (p+q+3) + 5p (5p + 2) + 49}
    \mathrlap{\Bigr),}
  \end{aligned}
\label{eq:eigenmodes:1-form:norms:last}
\end{align}
\endgroup
where we have dropped the denominators in \cref{eq:modes:1-form:projector} except for the sign to make sure that the proportionality coefficients are positive.
Using the same equations, we have also verified that the eigenmodes are orthogonal, as a consistency check.
Examining for which $(p,q)$ they vanish, we arrive at the cross diagrams in \cref{fig:modes:1-form:cross_diagrams}, which are consistent with \cite{Nilsson:2018lof}.
Note that one immediately sees why there are columns without crosses (i.e., constant $p$) from the representation theory.
For instance, if $\phi$ is an $sp_1^C$ singlet, $\mode^i \phi$ transforms in the $\vec{3} \otimes (0) = (2)$ of $sp_1^C$.
This $(2)$ corresponds to the $p=0$ columns in the $r = p+2$ diagrams and the fact that $\vec{3} \otimes (0)$ does not contain a $(0)$ when decomposed into irreps corresponds to the fact that there are no $p = 0$ crosses in the $r = p$ diagrams.
Note also that it is easy to see why there are columns without crosses from the prefactors in \crefrange{eq:eigenmodes:1-form:norms:first}{eq:eigenmodes:1-form:norms:last}.

In the formalism we use here, there is however no manifest reason for why there are rows without crosses (i.e., constant $q$).
Again looking at \crefrange{eq:eigenmodes:1-form:norms:first}{eq:eigenmodes:1-form:norms:last}, we see that this requires that the expressions under the square roots become perfect squares for particular values of $q$, namely $q = 0$ in the examples at hand, and non-trivial cancellations between the terms.

\subsubsection{Spin-1/2 modes}
Turning now to the spinors, we want to find the eigenmodes of $i\slashed{D}_{1/2}$.
Since this is a linear operator, we will not use \cref{masterdelta} here but instead just act with $i\slashed{D}_{1/2}$ directly.
The structure of the analysis is very similar to the one-form case presented above so
we will just present the results, skipping most  of the calculations.

The spinor isometry irreps from \cite{Nilsson:2018lof} consists of two $\vec{1}_{sp_1^C}$ and two $\vec{3}_{sp_1^C}$.
Following \cite{Nilsson:1983ru}, we first define
\begin{equation}
    \xi^i = i\Gamma^a \eta\, s_a{}^i,
\end{equation}
where $\eta$ is the $G_2$-invariant spinor as above, and then let
\begin{equation}
    \mode^{(1)} = \eta, \qquad
    \mode^{(2)} = i \Gamma^a \eta\, \check{D}_a, \qquad
    \mode^{(3)i} = \xi^i, \qquad
    \mode^{(4)i} = i \Gamma^a \xi^i\, \check{D}_a.
\end{equation}
The $H$-covariant derivatives of $\eta$ and $\xi^i$ are \cite{Nilsson:1983ru} (cf.\ \cref{eq:octonions:eta_def_properties})
\begin{equation}
    \check{D}_a \eta = 0, \qquad\quad
    \check{D}_a \xi^i = -3 \epsilon^i{}_{jk} s_a{}^j \xi^k,
\end{equation}
where the former follows from $\eta$ being $G_2$-invariant and $H \subset G_2$ and the latter follows from the first one and \cref{eq:modes:1-form:Ds}.

Next, we act with $i\slashed{D}_{1/2}$ to obtain
\begingroup
\setlength{\arraycolsep}{4pt} %
\begin{equation}
    i\slashed{D} P_r
    \left(
        \begin{array}{cc}
            \mode^{(1\text{--}2)} &
            \mode^{(3\text{--}4)i}
        \end{array}
    \right)
    \phi
    =
    P_r
    \left(
        \begin{array}{cc}
            \mode^{(1\text{--}2)} &
            \mode^{(3\text{--}4)i}
        \end{array}
    \right)
    \left(
        \renewcommand*{\arraystretch}{1.2} %
        \begin{array}{cc|cc}
            \frac{21}{4\sqrt{5}} & C_g & 0 & -\frac{2}{\sqrt{5}} s^i \\
            1 & -\frac{3\sqrt{5}}{4} & 0 & 0 \\ \hline
            0 & 0 & -\frac{27}{4\sqrt{5}} & C_g - 6 \\
            0 & 0 & 1 & \frac{5\sqrt{5}}{4}
        \end{array}
    \right)
    \phi,
\end{equation}
\endgroup
where we have used the same tricks as in the one-form case and the notation with the $sp_1^C$ block structure is also the same.
Diagonalising this, we find the differential operators generating the eigenmodes and their associated eigenvalues,
which are consistent with \crefrange{eq:eigs:summary:D_1/2:1}{eq:eigs:summary:D_1/2:2} above,
\begin{align}
\label{eq:modes:spinor:eigenmodes:first}
    & \tilde{\mode}^{(1)_\pm}
    = \Bigl(9 \pm \sqrtsmash[C_G]{20 C_g + 81}\Bigr) \mode^{(1)}
    + 2 \sqrt{5} \mode^{(2)},
    \\
    &\begin{aligned}[b]
        \tilde{\mode}^{(2)_\pm i}
        &= \Bigl(7 \pm \sqrtsmash[C_G]{20 C_g + 49}\Bigr) \mode^{(1)} s^i
        + 2 \sqrt{5} \mode^{(2)} s^i + {}
        \\
        &\phantom{{}={}}
        + \frac{1}{\sqrt{5}} \Bigl( 10 C_g - 21 \mp 3 \sqrtsmash[C_G]{20 C_g + 49} \Bigr) \mode^{(3) i}
        + \Bigl(7 \pm \sqrtsmash[C_G]{20 C_g + 49}\Bigr) \mode^{(4) i},
    \end{aligned}
\label{eq:modes:spinor:eigenmodes:last}
    \\
\label{eq:modes:spinor:eigenvalues:first}
    & i\slashed{D}_{1/2}^{(1)_\pm}
    = \frac{m}{3} \biggl(\frac{3}{2} \pm \sqrtsmash[C_G]{20 C_g + 81}\biggr),
    \\
    & i\slashed{D}_{1/2}^{(2)_\pm}
    = \frac{m}{3} \biggl(-\frac{1}{2} \pm \sqrtsmash[C_G]{20 C_g + 49} \biggr).
\label{eq:modes:spinor:eigenvalues:last}
\end{align}

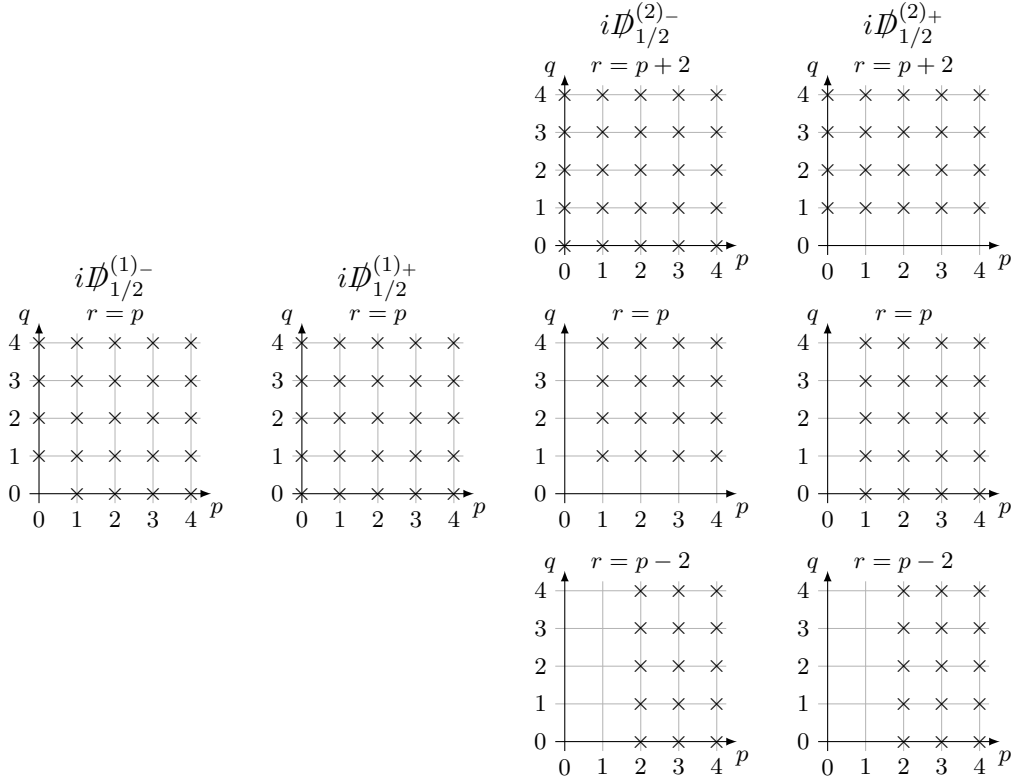
\begin{figure}[H]
  \centering
  \begingroup
  \renewcommand{\scale}{0.50cm}  %
  \footnotesize
  \newcommand{\figI}{%
      \begin{tikzpicture}[x=\scale, y=\scale]
          \crossdiagramlayout{p}
          \foreach \p in {0,...,\range} {
              \foreach \q in {0,...,\range} {
                  \node at (\p,\q) {$\times$};
              }
          }
      \end{tikzpicture}
  }
  \newcommand{\figII}{%
      \begin{tikzpicture}[x=\scale, y=\scale]
          \crossdiagramlayout{p}
          \foreach \p in {1,...,\range} {
              \foreach \q in {0,...,\range} {
                  \node at (\p,\q) {$\times$};
              }
          }
          \foreach \q in {1,...,\range} {
              \node at (0,\q) {$\times$};
          }
      \end{tikzpicture}
  }
  \newcommand{\figIIIppII}{%
      \begin{tikzpicture}[x=\scale, y=\scale]
          \crossdiagramlayout{p+2}
          \foreach \p in {0,...,\range} {
              \foreach \q in {1,...,\range} {
                  \node at (\p,\q) {$\times$};
              }
          }
      \end{tikzpicture}
  }
  \newcommand{\figIIIp}{%
      \begin{tikzpicture}[x=\scale, y=\scale]
          \crossdiagramlayout{p}
          \foreach \p in {1,...,\range} {
              \foreach \q in {0,...,\range} {
                  \node at (\p,\q) {$\times$};
              }
          }
      \end{tikzpicture}
  }
  \newcommand{\figIIIpmII}{%
      \begin{tikzpicture}[x=\scale, y=\scale]
          \crossdiagramlayout{p-2}
          \foreach \p in {2,...,\range} {
              \foreach \q in {0,...,\range} {
                  \node at (\p,\q) {$\times$};
              }
          }
      \end{tikzpicture}
  }
  \newcommand{\figIVppII}{%
      \begin{tikzpicture}[x=\scale, y=\scale]
          \crossdiagramlayout{p+2}
          \foreach \p in {0,...,\range} {
              \foreach \q in {0,...,\range} {
                  \node at (\p,\q) {$\times$};
              }
          }
      \end{tikzpicture}
  }
  \newcommand{\figIVp}{%
      \begin{tikzpicture}[x=\scale, y=\scale]
          \crossdiagramlayout{p}
          \foreach \p in {1,...,\range} {
              \foreach \q in {1,...,\range} {
                  \node at (\p,\q) {$\times$};
              }
          }
      \end{tikzpicture}
  }
  \newcommand{\figIVpmII}{%
      \begin{tikzpicture}[x=\scale, y=\scale]
          \crossdiagramlayout{p-2}
          \foreach \p in {2,...,\range} {
              \foreach \q in {0,...,\range} {
                  \node at (\p,\q) {$\times$};
              }
          }
      \end{tikzpicture}
  }
  \newcommand*{\eig}[1]{{\normalsize$i\slashed{D}_{1/2}^{#1}$}}%
  \begin{tabular}{cccc}
                  &             & \eig{(2)_-}   & \eig{(2)_+}   \\
      \eig{(1)_-} & \eig{(1)_+} & \figIVppII    & \figIIIppII   \\
      \figII      & \figI       & \figIVp       & \figIIIp      \\
                  &             & \figIVpmII    & \figIIIpmII
  \end{tabular}
  \endgroup
  \caption{Spinor cross diagrams. The corresponding eigenmodes of $i\slashed{D}_{1/2}$ are listed in \crefrange{eq:modes:spinor:eigenmodes:first}{eq:modes:spinor:eigenmodes:last} and the associated eigenvalues in \crefrange{eq:modes:spinor:eigenvalues:first}{eq:modes:spinor:eigenvalues:last}.}
  \label{fig:modes:spinor:cross_diagrams}
\end{figure}

Lastly, we compute the scalar products
\begin{gather}
    \begin{alignedat}{2}
        & \langle \mode^{(1)}\phi, \mode^{(1)}\phi \rangle = \|\phi\|^2,\qquad
        && \langle \mode^{(1)}\phi, \mode^{(2)}\phi \rangle = 0,\\
        & \langle \mode^{(1)} \phi, \mode^{(3)i} s_i \phi \rangle = 0,\qquad
        && \langle \mode^{(1)} \phi, \mode^{(4)i} s_i \phi \rangle = C_{\smash[t]{sp_1^C}}^\phi \|\phi\|^2,
    \end{alignedat}
    \\
    \begin{gathered}
        \langle \mode^{(2)}\phi, \mode^{(2)}\phi \rangle = C_g^\phi  \|\phi\|^2, \\
        \langle \mode^{(2)} \phi, \mode^{(3)i} s_i \phi \rangle = C_{\smash[t]{sp_1^C}}^\phi \|\phi\|^2,\qquad
        \langle \mode^{(2)} \phi, \mode^{(4)i} s_i \phi \rangle = \frac{3}{\sqrt{5}} C_{\smash[t]{sp_1^C}}^\phi \|\phi\|^2,
    \end{gathered}
    \\
    \langle \mode^{(3)i} \phi, \mode^{(3)j} \varphi \rangle = \frac{1}{5} \langle \phi, \delta^{ij} \varphi \rangle,\qquad
    \langle \mode^{(3)i} \phi, \mode^{(4)j} \varphi \rangle = - \frac{1}{\sqrt{5}} \langle \phi, \epsilon^{ijk} s_k \varphi \rangle, \\
    \langle \mode^{(4)i} \phi, \mode^{(4)j} \varphi \rangle = \frac{1}{5} C_g^\phi \langle \phi, \delta^{ij} \varphi \rangle - \frac{7}{5} \langle \phi, \epsilon^{ijk} s_k \varphi \rangle,
\end{gather}
to investigate when the eigenmodes above exist and find the results presented in \cref{fig:modes:spinor:cross_diagrams}, which is consistent with \cite{Nilsson:2018lof}.
\subsection{Two-form modes}%
\label{sec:eigenmodes:two-forms}
When we now set out to find explicit expressions for the transverse ($D^a Y_{ab} = 0$) two-form eigenmodes $Y_{ab}$ of $\Delta_2$,
we will face several technical challenges beyond those found in the one-form and spinor analyses.
In order to only have to act with one derivative, we will again make use of \cref{masterdelta}, this time  written as
\begin{equation}
\label{eq:modes:2-form:masterdelta}
    \Delta_2
    = C_g + 3 P_7 - \frac{2}{\sqrt{5}} \check{\mD}_{[2]},
\end{equation}
where $\check{\mD}_{[2]} Y_{ab} = a_{[a}{}^{cd} \check{D}_{|d} Y_{c|b]}$, as before.

As in the analyses above, the first step is to construct a basis for the two-forms in a generic isometry irrep $(p,q;r)$.
Before implementing transversality, the two-form isometry irreps consists of one $\vec{1}_{sp_1^C}$, five $\vec{3}_{sp_1^C}$ and one $\vec{5}_{sp_1^C}$, according to \cite{Nilsson:2018lof}.
The appearance of a $\vec{5}_{sp_1^C}$ here is new compared to the previous cases and is, in fact, the largest $sp_1^C$-irrep needed for all of the $so(7)$-irreps relevant to the supergravity compactification \cite{Nilsson:2018lof}.
This is one of the technical complications referred to above.

After some  algebra we arrive at the following set of differential operators to generate the two-form eigenmodes
\begin{gather}
\label{eq:modes:2-form:building-modes:first}
    \mode^{(1)}_{ab} = a_{ab}{}^c \check{D}_c,\\[2pt]
    \begin{alignedat}{2}
        &\mode^{(2)i}_{ab} = a_{ab}{}^c s_c{}^i,\qquad
        &&\mode^{(3)i}_{ab} = \epsilon^i{}_{jk} s_a{}^j s_b{}^k,\\
        &\mode^{(4)i}_{ab} = s_{[a}{}^i \check{D}_{b]},\qquad
        &&\mode^{(5)i}_{ab} = c_{ab}{}^{cd} s_c{}^i \check{D}_d,\\
        &\mode^{(6)i}_{ab} = a_{[a|}{}^{cd} s_{c}{}^i \check{D}_{d|b]},
        &&
    \end{alignedat}
    \\[4pt]
    \mode^{(7)ij}_{ab} = s_{[a}{}^{\{i|} a_{b]}{}^{cd} s_c{}^{|j\}} \check{D}_d,
\label{eq:modes:2-form:building-modes:last}
\end{gather}
where the braces around $i, j$ on the last line indicate the symmetric traceless part, i.e., the $\vec{5}_{sp_1^C}$ in $\vec{3}_{sp_1^C} \otimes \vec{3}_{sp_1^C}$.
Note the appearance of the differential operator
\begin{equation}
\check{D}_{ab} = \check{D}_{(a} \check{D}_{b)},
\end{equation}
 which  is the symmetrised composition of two derivatives,
in the expression for  $\mode^{(6)i}_{ab}$.  That the corresponding mode is necessary  is a fundamental result of the analysis presented here.
To establish that this mode is needed one can compute $\check{\mD}_{[2]}$ on the more standard mode functions obtained from $\mode^{(4)i}_{ab} $ and verify that the mode function
$\mode^{(6)i}_{ab}\phi$ gets generated, as we will see below.
The acceptance that the mode functions  $\mode^{(6)i}_{ab}\phi$ are unavoidable is  the key  that makes  it possible to complete the analysis of the two-form
modes.%
\footnote{Mode functions can also be constructed using $\Gamma^a$ matrices but these must then be decomposed into octonionic quantities and $Sp_1^C$ Killing vector components
in order to check that different mode functions are really independent. One example is $Y^{ij}_{ab}= i\bar\xi^{\{i}\Gamma_{abc}\xi^{j\}}\check{D}^c\phi$ where the right-hand side can be
rewritten using the identity  $i \bar\xi^{\{i}\Gamma_{abc}\xi^{j\}}=6 s_{[a}{}^{\{i|} a_{bc]}{}^d s_d{}^{|j\}}$.}

Of course,  one has to make sure that the mode functions constructed here are independent.  One such concern could involve
\begin{equation}
    \epsilon^i{}_{jk} s_{[a}{}^j a_{b]}{}^{cd} s_c{}^k \check{D}_d
\end{equation}
that seems to give an additional mode function independent of the above ones.
However, the identity
\begin{equation}
    \epsilon^i{}_{jk} a_{[a}{}^{cd} s_{b]}{}^j s_c{}^k
    = \frac{1}{2} \epsilon^i{}_{jk} a_{ab}{}^c s_c{}^j s^{dk}
    + \frac{1}{2\sqrt{5}} c_{ab}{}^{cd} s_c{}^i
\end{equation}
shows that the above expression is in fact a linear combination of $\epsilon^i{}_{jk} \mode^{(2)j}_{ab} s^k$ and $\mode^{(5)i}_{ab}$.

Next, we compute the action of $\check{\mD}_{[2]}$ and $P_7$ and put everything together using \cref{eq:modes:2-form:masterdelta} to obtain
\begingroup
\setlength{\arraycolsep}{4pt} %
\begin{equation}
    \Delta P_r
    \left(
        \begin{array}{c c c}
            \mode^{(1)}_{ab} &
            \mode^{(2\text{--}6)i}_{ab} &
            \mode^{(7)jk}_{ab}
        \end{array}
    \right)
    \phi
    =
    P_r
    \left(
        \begin{array}{c c c}
            \mode^{(1)}_{ab} &
            \mode^{(2\text{--}6)i}_{ab} &
            \mode^{(7)jk}_{ab}
        \end{array}
    \right)
    \mathrm{M}_{\Delta_2}
    \phi,
\end{equation}
\endgroup
where
\begingroup
\small
\setlength{\arraycolsep}{3pt}  %
\renewcommand*{\arraystretch}{1.5} %
\begin{align}
\label{eq:modes:2-form:Delta_matrix}
    & \text{\normalsize$\mathrm{M}_{\Delta_2} =$}
    \\ \nonumber &%
    \left(
    \begin{array}{c|ccccc|c}
        C_g + \frac{18}{5}  & 0 & 0 & 0 & \frac{2}{\sqrt{5}} s^i & s^i & 0
        \\[2pt] \hline
        0 & C_g + \frac{27}{5} & \frac{1}{2\sqrt{5}} (2 - \Delta C) &
        \frac{3}{4\sqrt{5}} \Delta C & \frac{2}{\sqrt{5}} C_g^\phi &
        \frac{1}{40} (44 C_g^\phi - 87 \Delta C) &
        \frac{7}{10} \delta_i^{\{j} s^{k\}}
        \\
        0 & -\frac{18}{\sqrt{5}} & C_g - \frac{6}{5} & 0 & 0 &
        -\frac{3}{2\sqrt{5}} (2 C_g^\phi - 3 \Delta C) &
        -\frac{3}{\sqrt{5}} \delta_i^{\{j} s^{k\}}
        \\
        0 & -\frac{2}{\sqrt{5}} & 0 & C_g - \frac{7}{10} & \frac{6}{5} &
        -\frac{1}{20\sqrt{5}} (20 C_g^\phi + 30 \Delta C - 33) &
        -\frac{1}{\sqrt{5}} \delta_i^{\{j} s^{k\}}
        \\
        0 & \frac{2}{\sqrt{5}} & \frac{1}{5} & \frac{1}{10} & C_g + \frac{12}{5} &
        -\frac{3}{20 \sqrt{5}} (10 \Delta C - 7) & 0
        \\
        0 & 0 & 0 & -\frac{1}{\sqrt{5}} & -\frac{4}{\sqrt{5}} & C_g - \frac{1}{10} & 0
        \\[2pt] \hline
        0 & 0 & 0 & 0 & 0 & 0 & C_g
    \end{array}
    \right)
\end{align}
\endgroup
and we have defined $\Delta C \equiv C_{\smash[t]{sp_1^C}} - C_{\smash[t]{sp_1^C}}^\phi - 2$.
Note the appearance of $C_g$, $C_g^\phi$ and $\Delta C$ in this expression, which makes it structurally more complicated than the one-form and spin-1/2 cases above where only $C_g$ enters in the analogous matrices.
In fact, one might be worried that all three of them will also enter in the two-form eigenvalues, since we know from \cref{sec:eigenvalues} that the eigenvalues should not contain either $C_g^\phi$ or $\Delta C$.
As we will see below, the eigenvalues do match what we found in \cref{sec:eigenvalues} but some eigenmodes contain $C_g^\phi$ and $\Delta C$.%
\footnote{Note that $C_g = C_g^\phi + 3 \Delta C + 6$, which follows from \cref{eq:method:C_g}, enables terms containing $C_g^\phi$ and $\Delta C$ to cancel.}

One complication in this calculation is the appearance of the Weyl tensor so we should recall that
\cref{masterweyl} relates the Weyl tensor acting on two-forms to the Casimirs of $g_2$ and $h$.
However, while $C_h$ only has one eigenvalue on $\vec{7}_{g_2}$, it has several when acting  on  $\vec{14}_{g_2}$ in $\vec{21}_{so(7)}$, and hence one
has to construct the projectors onto the various $h$-irreps using $s_a{}^i$, $a_{abc}$ and $c_{abcd}$.
These projectors and some identities that facilitate the calculation can be found in \cref{app:octonions,app:gtwoprojectors}.

Since we are only  interested in the transverse eigenmodes, we also compute the divergences
\begingroup
\setlength{\arraycolsep}{4pt} %
\begin{equation}
    \begin{aligned}[b]
        & D^b P_r
        \left(
            \begin{array}{ccc}
                \mode^{(1)}_{ba} &
                \mode^{(2\text{--}6)i}_{ba} &
                \mode^{(7)jk}_{ba}
            \end{array}
        \right)
        \phi
        = P_r
        \left(
        \begin{array}{cc}
            \mode^{(1)}_a & \mode^{(2\text{--}3)i}_a
        \end{array}
        \right)
        \\
        &\quad \times
        \left(
            \renewcommand*{\arraystretch}{1.5} %
            \begin{array}{c|ccccc|c}
                0 & 0 & 0 & \frac{1}{2} s^i & 0 & -\frac{7}{4\sqrt{5}} s^i & 0
                \\[3pt] \hline
                0 & \frac{3}{\sqrt{5}} & 1 + \frac{1}{2} \Delta C & \frac{1}{2} C_g^\phi &
                3 \Delta C & -\frac{1}{4\sqrt{5}} C_g^\phi + \frac{3 \sqrt{5}}{4} \Delta C &
                \frac{3}{2 \sqrt{5}} \delta_i^{\{j} s^{k\}}
                \\
                0 & 1 & 0 & 0 & -2 \sqrt{5} & \frac{1}{2} C_g^\phi + \frac{3}{4} \Delta C -1 &
                \frac{1}{2} \delta_i^{\{j} s^{k\}}
            \end{array}
        \right)
        \phi,
    \end{aligned}
\end{equation}
\endgroup
where $\mode^{(1)}_a$ and $\mode^{(2\text{--}3)i}_a$ are the one-form differential operators from \cref{eq:modes:1-form:basis}.

The eigenmodes of $\Delta_2$ are found by diagonalising \cref{eq:modes:2-form:Delta_matrix}.
We then use the above equation to check which of these are transverse.
This results in eigenmodes $P_r \tilde{\mode}_{ab} \phi$ and associated eigenvalues given by
\begin{align}
\label{eq:modes:2-form:eigenmodes:first}
    & \tilde{\mode}^{(1)}_{ab} = \mode^{(1)}_{ab},
    \\
    &\begin{aligned}[b]
        \tilde{\mode}^{(2)_\pm i}_{ab}
        &= \frac{1}{10} \Bigl(-5 C_g - 21 \mp 3\sqrtsmash[C_G]{20 C_g + 49}\Bigr) \mode^{(1)}_{ab} s^i
        \\ &\phantom{{}={}}
        - \frac{1}{4} C_g \Bigl(C_g + C_g^\phi + 20 \pm 2 \sqrtsmash[C_G]{20 C_g + 49} \Bigr) \mode^{(2)i}_{ab}
        \\ &\phantom{{}={}}
        + \frac{3}{2 \sqrt{5}} C_g \Bigl(13 \pm \sqrtsmash[C_G]{20 C_g + 49}\Bigr) \mode^{(3)i}_{ab}
        + \frac{7}{2 \sqrt{5}} C_g \mode^{(4)i}_{ab}
        \\ &\phantom{{}={}}
        - \frac{1}{4\sqrt{5}} C_g \Bigl(15 \pm \sqrtsmash[C_G]{20 C_g + 49}\Bigr) \mode^{(5)i}_{ab}
        + C_g \mode^{(6)i}_{ab},
    \end{aligned}
    \\
    &\begin{aligned}[b]
        \tilde{\mode}^{(3)i}_{ab}
        & = \frac{1}{36} C_g (5 \Delta C + 16) \mode^{(1)}_{ab} s^i
        \\ &\phantom{{}={}}
        + \frac{1}{4} \bigl(2 {C_g}^2 - C_g (7 \Delta C + 16) - 3 (\Delta C + 2) (3 \Delta C - 4) \bigr) \mode^{(2)i}_{ab}
        \\ &\phantom{{}={}}
        - \frac{\sqrt{5}}{2} \bigl(2 {C_g}^2 - 3 C_g (3 \Delta C + 8) + 36 (\Delta C + 2)\bigr) \mode^{(3)i}_{ab}
        \\ &\phantom{{}={}}
        + \frac{1}{2 \sqrt{5}} \bigl(C_g (10 \Delta C + 13) - 57 (\Delta C + 2)\bigr) \mode^{(4)i}_{ab}
        \\ &\phantom{{}={}}
        - \frac{1}{4 \sqrt{5}} \bigl(C_g (5 \Delta C + 6) - 30 (\Delta C + 2)\bigr) \mode^{(5)i}_{ab}
        - \bigl(C_g + 3 (\Delta C + 2)\bigr) \mode^{(6)i}_{ab},
    \end{aligned}
    \\
    &\begin{aligned}[b]
        \tilde{\mode}^{(3)' ij}_{ab}
        & = \frac{5}{36} (\Delta C + 2) \mode^{(1)}_{ab} s^{\{ij\}}
        + \frac{1}{2} \bigl(C_g - 2 (\Delta C + 2) \bigr) \mode^{(2)\{i }_{ab} s^{j\}}
        \\ &\phantom{{}={}}
        - \frac{\sqrt{5}}{2} \bigl(2 C_g - 9 (\Delta C + 2) \bigr) \mode^{(3)\{i }_{ab} s^{j\}}
        + \sqrt{5} (\Delta C + 2) \mode^{(4)\{i }_{ab} s^{j\}}
        \\ &\phantom{{}={}}
        - \frac{\sqrt{5}}{4} (\Delta C + 2) \mode^{(5)\{i }_{ab} s^{j\}}
        - \bigl(C_g + 3 (\Delta C + 2) \bigr) \mode^{(7)ij}_{ab},
    \end{aligned}
\label{eq:modes:2-form:eigenmodes:last}
    \\
\label{eq:modes:2-form:eigenvalues:first}
    & \Delta_2^{(1)} = \frac{m^2}{9} \bigl(20 C_g + 72\bigr),
    \\
    & \Delta_2^{(2)_\pm} = \frac{m^2}{9} \Bigl(20 C_g + 44 \pm 4 \sqrtsmash[C_G]{20 C_g + 49} \Bigr),
    \\
    & \Delta_2^{(3),(3)'} = \frac{m^2}{9} 20 C_g,
\label{eq:modes:2-form:eigenvalues:last}
\end{align}
which is consistent with \crefrange{eq:eigs:summary:Delta_2:1}{eq:eigs:summary:Delta_2:3}.

Lastly, we compute the norms of these eigenmodes to see for which $(p,q;r)$ they exist.
We employ the same strategy as in the one-form case above, using \cref{eq:modes:1-form:projector} to implement the projectors $P_r$.
The calculation is a lot more involved, particularly because of the symmetrised derivatives $\check{D}_{ab}$ in $\mode^{(6)i}_{ab}$ and the $\vec{5}_{sp_1^C}$ mode operator $\mode^{(7)ij}_{ab}$.
The latter implies that there, in some cases, are four factors in \cref{eq:modes:1-form:projector}, so we need relations like \cref{eq:modes:1-form:es_es} for differential operators transforming in $\vec{5}_{sp_1^C}$ with up to four $s^i$ instead of just two.

\begin{figure}[H]
  \centering
  \begingroup %
  \renewcommand{\scale}{0.41cm} %
  \footnotesize %
  \newcommand{\figIp}{%
      \begin{tikzpicture}[x=\scale, y=\scale]
          \crossdiagramlayout{p}
          \foreach \p in {1,...,\range} {
              \foreach \q in {0,...,\range} {
                  \node at (\p,\q) {$\times$};
              }
          }
          \foreach \q in {1,...,\range} {
              \node at (0,\q) {$\times$};
          }
      \end{tikzpicture}%
  }%
  \newcommand{\figIIppII}{%
      \begin{tikzpicture}[x=\scale, y=\scale]
          \crossdiagramlayout{p+2}
          \foreach \p in {1,...,\range} {
              \foreach \q in {0,...,\range} {
                  \node at (\p,\q) {$\times$};
              }
          }
          \foreach \q in {2,...,\range} {
              \node at (0,\q) {$\times$};
          }
      \end{tikzpicture}%
  }%
  \newcommand{\figIIp}{%
      \begin{tikzpicture}[x=\scale, y=\scale]
          \crossdiagramlayout{p}
          \foreach \p in {2,...,\range} {
              \foreach \q in {0,...,\range} {
                  \node at (\p,\q) {$\times$};
              }
          }
          \foreach \q in {1,...,\range} {
              \node at (1,\q) {$\times$};
          }
      \end{tikzpicture}%
  }%
  \newcommand{\figIIpmII}{%
      \begin{tikzpicture}[x=\scale, y=\scale]
          \crossdiagramlayout{p-2}
          \foreach \p in {3,...,\range} {
              \foreach \q in {0,...,\range} {
                  \node at (\p,\q) {$\times$};
              }
          }
          \foreach \q in {1,...,\range} {
              \node at (2,\q) {$\times$};
          }
      \end{tikzpicture}%
  }%
  \newcommand{\figIIIppII}{%
      \begin{tikzpicture}[x=\scale, y=\scale]
          \crossdiagramlayout{p+2}
          \foreach \p in {0,...,\range} {
              \foreach \q in {1,...,\range} {
                  \node at (\p,\q) {$\times$};
              }
          }
      \end{tikzpicture}%
  }%
  \newcommand{\figIIIp}{%
      \begin{tikzpicture}[x=\scale, y=\scale]
          \crossdiagramlayout{p}
          \foreach \p in {1,...,\range} {
              \foreach \q in {0,...,\range} {
                  \node at (\p,\q) {$\times$};
              }
          }
      \end{tikzpicture}%
  }%
  \newcommand{\figIIIpmII}{%
      \begin{tikzpicture}[x=\scale, y=\scale]
          \crossdiagramlayout{p-2}
          \foreach \p in {2,...,\range} {
              \foreach \q in {0,...,\range} {
                  \node at (\p,\q) {$\times$};
              }
          }
      \end{tikzpicture}%
  }%
  \newcommand{\figIVppII}{%
      \begin{tikzpicture}[x=\scale, y=\scale]
          \crossdiagramlayout{p+2}
          \foreach \p in {0,...,\range} {
              \foreach \q in {0,...,\range} {
                  \node at (\p,\q) {$\times$};
              }
          }
      \end{tikzpicture}%
  }%
  \newcommand{\figIVp}{%
      \begin{tikzpicture}[x=\scale, y=\scale]
          \crossdiagramlayout{p}
          \foreach \p in {1,...,\range} {
              \foreach \q in {1,...,\range} {
                  \node at (\p,\q) {$\times$};
              }
          }
      \end{tikzpicture}%
  }%
  \newcommand{\figIVpmII}{%
      \begin{tikzpicture}[x=\scale, y=\scale]
          \crossdiagramlayout{p-2}
          \foreach \p in {2,...,\range} {
              \foreach \q in {0,...,\range} {
                  \node at (\p,\q) {$\times$};
              }
          }
      \end{tikzpicture}%
  }%
  \newcommand{\figVppIV}{%
      \begin{tikzpicture}[x=\scale, y=\scale]
          \crossdiagramlayout{p+4}
          \foreach \p in {0,...,\range} {
              \foreach \q in {1,...,\range} {
                  \node at (\p,\q) {$\times$};
              }
          }
      \end{tikzpicture}%
  }%
  \newcommand{\figVppII}{%
      \begin{tikzpicture}[x=\scale, y=\scale]
          \crossdiagramlayout{p+2}
          \foreach \p in {1,...,\range} {
              \foreach \q in {2,...,\range} {
                  \node at (\p,\q) {$\times$};
              }
          }
      \end{tikzpicture}%
  }%
  \newcommand{\figVp}{%
      \begin{tikzpicture}[x=\scale, y=\scale]
          \crossdiagramlayout{p}
          \foreach \p in {2,...,\range} {
              \foreach \q in {1,...,\range} {
                  \node at (\p,\q) {$\times$};
              }
          }
      \end{tikzpicture}%
  }%
  \newcommand{\figVpmII}{%
      \begin{tikzpicture}[x=\scale, y=\scale]
          \crossdiagramlayout{p-2}
          \foreach \p in {3,...,\range} {
              \foreach \q in {1,...,\range} {
                  \node at (\p,\q) {$\times$};
              }
          }
      \end{tikzpicture}%
  }%
  \newcommand{\figVpmIV}{%
      \begin{tikzpicture}[x=\scale, y=\scale]
          \crossdiagramlayout{p-4}
          \foreach \p in {4,...,\range} {
              \foreach \q in {0,...,\range} {
                  \node at (\p,\q) {$\times$};
              }
          }
      \end{tikzpicture}%
  }%
  \newcommand*{\eig}[1]{{\normalsize$\Delta_{2}^{#1}$}}%
  \begin{tabular}{ccccc}%
                &               &               &                 & \eig{(3)'}    \\
                & \eig{(2)_-}   & \eig{(2)_+}   & \eig{(3)}       & \figVppIV     \\
      \eig{(1)} & \figIIIppII   & \figIVppII    & \figIIppII      & \figVppII     \\
      \figIp    & \figIIIp      & \figIVp       & \figIIp         & \figVp        \\
                & \figIIIpmII   & \figIVpmII    & \figIIpmII      & \figVpmII     \\
                &               &               &                 & \figVpmIV
  \end{tabular}%
  \endgroup
  \caption{Transverse two-form cross diagrams. The corresponding eigenmodes are listed in \crefrange{eq:modes:2-form:eigenmodes:first}{eq:modes:2-form:eigenmodes:last} and the associated eigenvalues in \crefrange{eq:modes:2-form:eigenvalues:first}{eq:modes:2-form:eigenvalues:last}. Note that the last column is associated to a linear combination of the eigenmodes formed from $\tilde{\mode}^{(3) i}_{ab}$ and $\tilde{\mode}^{(3)' ij}_{ab}$ which is orthogonal to $P_r \tilde{\mode}^{(3) i}_{ab} \phi$.}
  \label{fig:modes:2-form:cross_diagrams}
\end{figure}
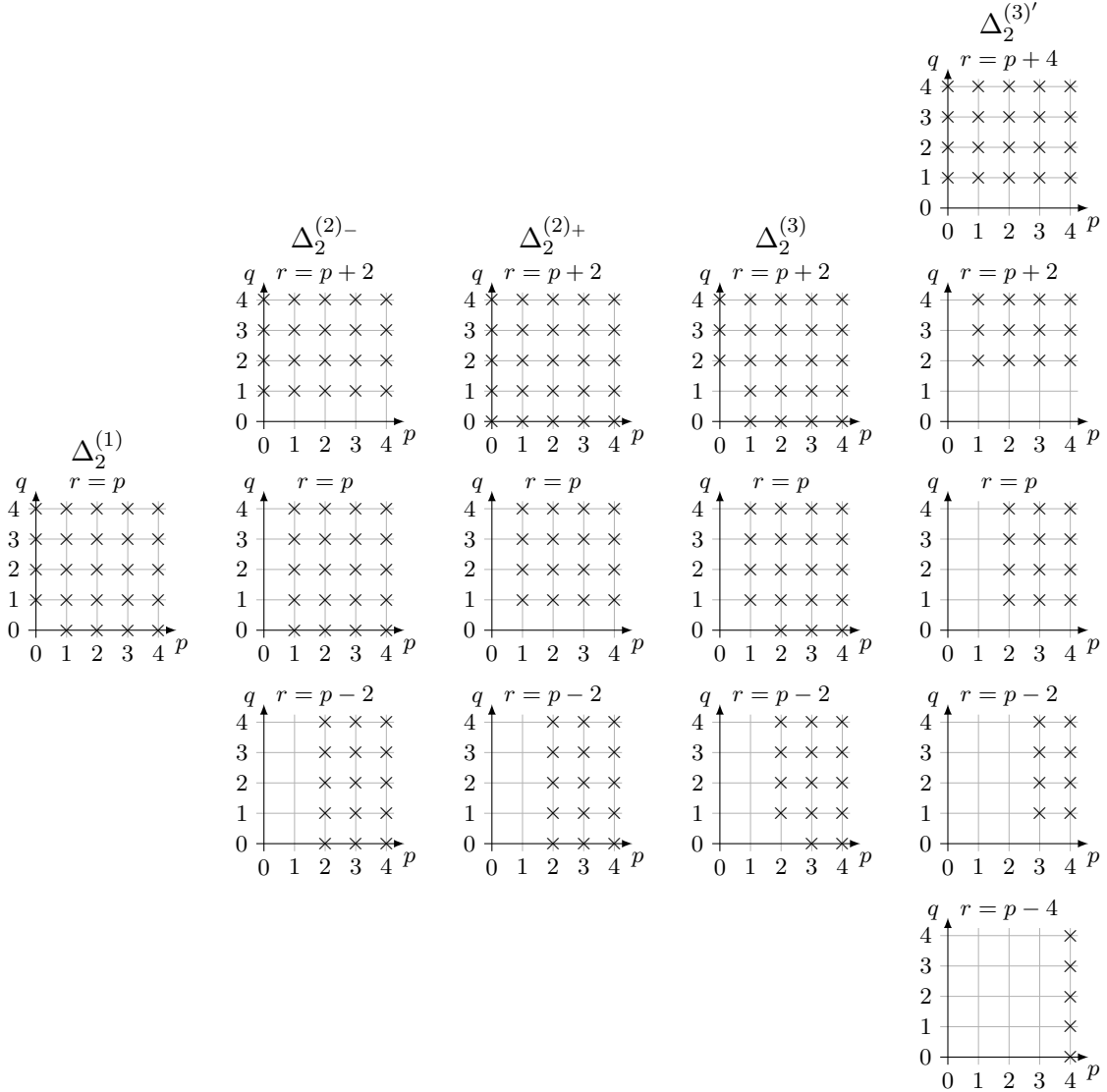

Using the results of the scalar product computations (see \cref{app:non_transv_and_2-form_scalar_prods}), we derive for which isometry irreps the eigenvalues exist.
The result, which is consistent with \cite{Nilsson:2018lof}, is presented in \cref{fig:modes:2-form:cross_diagrams}.

It turns out that $P_r \tilde{\mode}^{(3)i}_{ab} \phi$ and $P_r \tilde{\mode}^{(3)' ij}_{ab} \phi$ are not orthogonal for $r = p,\, p \pm 2$.
To arrive at the correct cross diagrams, we compute the norm of a linear combination of $P_r \tilde{\mode}^{(3)i}_{ab} \phi$ and $P_r \tilde{\mode}^{(3)' ij}_{ab} \phi$ orthogonal to the former.
This results in some arbitrariness in the split of the $\vec{3}$ and $\vec{5}$ cross diagrams corresponding to the same eigenvalue, $\Delta_2^{(3)} = \Delta_2^{(3)'}$.
Specifically, it is arbitrary whether the $r=p-2,\; q=0$ crosses, $r=p,\; q=0$ crosses and $r=p+2,\; q\leq 1$ crosses in the three $\Delta_2^{(3)}$ cross diagrams are placed where they are or at the corresponding places in the five $\Delta_2^{(3)'}$ cross diagrams since there exists a differential operator transforming in $\vec{5}_{sp_1^C}$ with such cross diagrams.
Note, however, that for instance the $p=1$ crosses in the $r=p$ diagram of $\Delta_2^{(3)}$ cannot be moved to $\Delta_2^{(3)'}$ since $P_r \mode_{ab} \phi$ vanishes for $r=p=1$ for any $\mode$ transforming in $\vec{5}_{sp_1^C}$.

\subsection{Isometry singlet modes}%
\label{sec:eigenmodes:isometry_singlets}

We end this section by giving the singlet mode functions, constructed recently in \cite{Nilsson:2023ctq}, in the notation of this paper.
The  singlet modes that have not yet been constructed  in this paper belong to the spectra of $ i\slashed{D}_{3/2}$, $\Delta_L$ and $Q$.
These singlet modes will be further discussed in the next section.

To construct the isometry singlet modes, we use that isometry irrep singlet fields are constructed from isotropy singlet tensors in a one-to-one fashion.
Thus, the $G$-singlets are obtained  by decomposing the $Spin(7)$-irrep of the field into $H$-pieces and then using the resulting $H$-singlets.
The relevant decompositions can be found in, e.g., \cite{Nilsson:2018lof}, from which we see that the only $H$-singlets are one in $\vec{1}_{g_2}$ and one in $\vec{27}_{g_2}$.
Hence, all $G$-singlets can be constructed using the $G_2$ singlets and the additional $H$ singlet in $\vec{27}_{g_2}$, which is the traceless part of $s_a{}^i s_{bi}$.
Using this method, we must check transversality of the singlets explicitly.

Explicitly, the scalar singlet is obviously a constant, $\phi = 1$, with eigenvalue
\begin{equation}
    \Delta_0\+ 1 = 0,
\end{equation}
while the spinor is the $G_2$-invariant $\eta$ with
\begin{equation}
    i \slashed{D}_{1/2}\+ \eta = \frac{7 m}{2} \eta.
\end{equation}
There are no one-form or two-form singlets since neither $\vec{7}_{g_2}$ nor $\vec{14}_{g_2}$ contain any $H$-singlets.
Turning to $\vec{27}$, there is one singlet given by
\begin{equation}
    h_{ab} \coloneqq s_{\{a}{}^i s_{b\}i}
    = s_a{}^i s_{bi} - \frac{3}{35} \delta_{ab},
\end{equation}
as mentioned above.
A short calculation shows that this is transverse and, using \cref{masterdelta}, that
\begin{equation}
    \Delta_L h_{ab} = \frac{28 m^2}{9} h_{ab}.
\end{equation}

Turning to three-forms, there is one singlet, $a_{abc}$, appearing already when decomposing $\vec{35}$ under $Spin(7) \to G_2$.
From $\check{D}_a a_{bcd} = 0$, it easy to see that $d a = 4m\+ c$ and hence
\begin{equation}
    Q a = 4m\+ a.
\end{equation}
There is another three-form singlet since there is a $\vec{27}_{g_2}$ in the $G_2$ decomposition of $\vec{35}_{so(7)}$.
Using what we saw in \cref{sec:eigenvalues}, this can be written as $a_{[a b}{}^d h_{c]d}$, where $h_{ab}$ is the $\vec{27}$ singlet from above, and a calculation gives
\begin{equation}
    Q (a_{[a b}{}^d h_{c]d}) = - \frac{2 m}{3} a_{[a b}{}^d h_{c]d}.
\end{equation}
Lastly, there is one vector-spinor singlet since the decomposition of $\vec{48}_{so(7)}$ contains $\vec{27}_{g_2}$.
Again using the decomposition presented in \cref{sec:eigenvalues}, we can write it as $i\Gamma^b h_{ba} \eta$ and compute the eigenvalue
\begin{equation}
    i\slashed{D}_{3/2} (i\Gamma^b h_{ba} \eta) = - \frac{m}{6}\+ i\Gamma^b h_{ba} \eta.
\end{equation}
These results agree with \cite{Nilsson:2023ctq}.

\section{Implications for the \texorpdfstring{$\mN = 1$}{N = 1} and \texorpdfstring{$\mN = 0$}{N = 0} vacua}%
\label{sec:supermultiplet}

The full content of isometry irreps of the squashed seven-sphere compactification, including the structure of the supermultiplets in the left-squashed case listed below, was derived in \cite{Nilsson:2018lof}
and presented there in the form of cross diagrams, like those in \cref{sec:eigenmodes} (see the appendix of \cite{Nilsson:2018lof}).
While this reference  gives no information about the possible values of $E_0$, with the results obtained in \cite{Ekhammar:2021gsg} and in the previous sections of this paper we can finally
associate these supermultiplets  with their proper mass and energy values  $E_0$. There are general formulas for these values covering infinite sets
of supermultiplets but some Wess--Zumino multiplets  with a low value of the Casimir $C_g$  require a separate scrutiny since they may be associated with
some unusual sign choices. In the right-squashed vacuum,
the fields in these multiplets also give rise to ambiguities concerning their boundary conditions as was demonstrated recently for the $C_g=0$
Wess--Zumino multiplets in  \cite{Nilsson:2023ctq}. This last issue appears also in the left-squashed supersymmetric vacuum but in a more ordered fashion as we will see below
when we study the Wess--Zumino multiplets with $C_g>0$.

We present the results for the general cases in the tables in \cref{spectrum_summary}, which ends with a discussion of  the special low $C_g$ cases.
How to arrive at these results are explained in \cref{sec:supermultiplets:details}.
The spectrum of  the right-squashed non-supersymmetric $AdS_4$ theory obtained by skew-whiffing, or flipping the orientation flipping of $S^7$, is discussed in \cref{sec:supermultiplets:skew-whiffing}.
In \cref{sec:marginal_ops_in_N=0} we  analyse  some implications for the boundary theory, in particular the potential presence of marginal operators relevant for the issue of stability of the skew-whiffed theory.

\subsection{Supermultiplets in the left-squashed vacuum}
\label{spectrum_summary}

In terms of the $Spin(2,3)$-irreps $D(E_0,s)$, the $\mN =1$ supermultiplets that appear in eleven-dimensional supergravity compactified on the left-squashed seven-sphere are \cite{Nilsson:2018lof}:%
\footnote{Note the order of the $E_0$ values, i.e., $(E_0, E_0-\frac{1}{2}, E_0+\frac{1}{2}, E_0)$ chosen so that short gauge supermultiplets contain only the first two entires.}
\begin{alignat}{4}
\label{spintwo}
    1 &\times \bigl(D(E_0,2^+) &&\oplus D(E_0-\tfrac{1}{2}, \tfrac{3}{2}) &&\oplus D(E_0+\tfrac{1}{2}, \tfrac{3}{2}) &&\oplus D(E_0, 1^+)\bigr), \\
    6 &\times \bigl(D(E_0,\tfrac{3}{2}) &&\oplus D(E_0-\tfrac{1}{2}, 1^{\pm}) &&\oplus D(E_0+\tfrac{1}{2}, 1^{\mp}) &&\oplus D(E_0, \tfrac{1}{2})\bigr), \\
    6 &\times \bigl(D(E_0,1^-) &&\oplus D(E_0-\tfrac{1}{2}, \tfrac{1}{2}) &&\oplus D(E_0+\tfrac{1}{2}, \tfrac{1}{2}) &&\oplus D(E_0, 0^-)\bigr), \\
    8 &\times \bigl(D(E_0,1^+) &&\oplus D(E_0-\tfrac{1}{2}, \tfrac{1}{2}) &&\oplus D(E_0+\tfrac{1}{2}, \tfrac{1}{2}) &&\oplus D(E_0, 0^+)\bigr), \\
\label{wz}
    14 &\times \bigl(D(E_0, \tfrac{1}{2}) &&\oplus D(E_0-\tfrac{1}{2}, 0^{\pm}) &&\oplus D(E_0+\tfrac{1}{2}, 0^{\mp}) \mathrlap{\bigr),}
\end{alignat}

As explained in \cite{Nilsson:2018lof} the multiplicities (numbers in front of the multiplets) refer to the number of cross diagrams
connected to these multiplets as given in the appendix of \cite{Nilsson:2018lof}. In the following we will refer to these Heidenreich  supermultiplets  \cite{Heidenreich:1982rz} by
their highest spin component, except for the last case, the Wess--Zumino multiplets.

Using the results of this paper, we can associate masses and energy values $E_0$ to the supermultiplets listed above, see \cref{sec:supermultiplets:details} for details on how this can be done.
The results are given in  two tables,  \cref{spinonetotwomultiplets} for those with maximum spin $2^+$, $3/2$, $1^-$ and $1^+$, and \cref{wesszuminopositivecgmultiplets}, for the Wess--Zumino supermultiplets with $C_g>0$.
The two special cases, the Wess--Zumino multiplets with $C_g=0$, are  given separately after these tables, in \cref{wesszuminocgequaltozero}.
These latter multiplets were also discussed recently in \cite{Nilsson:2023ctq}.

\begin{table}[ht]
    \centering
    \setlength{\defaultaddspace}{4pt}
    \begin{tabular}{lllllcl}
        \toprule
        $s^p$ & $E_0$ & $E_0-\frac{1}{2}$ & $E_0+\frac{1}{2}$ & $E_0$  & $\mathclap{Sp_1^C}$ & $E_0$ values\\ \midrule
        $2^+$ & $2^+(\Delta_0^{(1)})$ & $\frac{3}{2}(i\slashed{D}_{1/2}^{(1)_+})$ & $\frac{3}{2}(i\slashed{D}_{1/2}^{(1)_-})$ &$1^+(\Delta_2^{(1)})$ & $\bf{1}$ & $\frac{3}{2}+\frac{1}{6}\sqrtsmash[C_G]{20C_g+81}$\\ \addlinespace
        $\frac{3}{2}_1$ & $\frac{3}{2}(i\slashed{D}_{1/2}^{(2)_-})$ & $1^+(\Delta_2^{(2)_+})$ &$ 1_{\scriptscriptstyle (+)}^-(\Delta_1^{(1)_-}) $& $\frac{1}{2}(i\slashed{D}_{3/2}^{(3)_+})$ &$\bf{3}$ &  $\frac{3}{2}+\frac{5}{6}+\frac{1}{6}\sqrtsmash[C_G]{20C_g+49}$\\ \addlinespace
        $\frac{3}{2}_2$ & $\frac{3}{2}(i\slashed{D}_{1/2}^{(2)_+})$ &  $1_{\scriptscriptstyle (-)}^-(\Delta_1^{(1)_+}) $ & $1^+(\Delta_2^{(2)_-})$ & $\frac{1}{2}(i\slashed{D}_{3/2}^{(3)_-})$ &$\bf{3}$  &  $\frac{3}{2}-\frac{5}{6}+\frac{1}{6}\sqrtsmash[C_G]{20C_g+49}$\\ \addlinespace
        $1^-_1$ & $1_{\scriptscriptstyle (+)}^-(\Delta_1^{(1)_+})$ & $\frac{1}{2}(i\slashed{D}_{3/2}^{(4)_+})$ & $\frac{1}{2}(i\slashed{D}_{1/2}^{(2)_+})$ & $0^-(Q^{(2)_+})$&$\bf{3}$  &  $\frac{3}{2}+\frac{5}{3}+\frac{1}{6}\sqrtsmash[C_G]{20C_g+49}$\\ \addlinespace
        $1^-_2$ & $1_{\scriptscriptstyle (-)}^-(\Delta_1^{(1)_-})$ & $\frac{1}{2}(i\slashed{D}_{1/2}^{(2)_-})$ & $\frac{1}{2}(i\slashed{D}_{3/2}^{(4)_-})$ & $0^-(Q^{(2)_-})$&$\bf{3}$  &  $\frac{3}{2}-\frac{5}{3}+\frac{1}{6}\sqrtsmash[C_G]{20C_g+49}$\\ \addlinespace
        $1^+_1$ & $1^+(\Delta_2^{(3)})$ &  $\frac{1}{2}(i\slashed{D}_{3/2}^{(2)_-})$ &  $\frac{1}{2}(i\slashed{D}_{3/2}^{(2)_+})$ &$0^+(\Delta_L^{(1)})$ &$\bf{3}$  &  $\frac{3}{2}+\frac{1}{6}\sqrtsmash[C_G]{20C_g+9}$\\ \addlinespace
        $1^+_2$ & $1^+(\Delta_2^{(3)'})$ &  $\frac{1}{2}(i\slashed{D}_{3/2}^{(2)'_-})$ & $\frac{1}{2}(i\slashed{D}_{3/2}^{(2)'_+})$ &$0^+(\Delta_L^{(1)'})$ &$\bf{5}$  &  $\frac{3}{2}+\frac{1}{6}\sqrtsmash[C_G]{20C_g+9}$
        \\ \bottomrule
    \end{tabular}
    \caption{Supermultiplets with maximum spin $s = 2,\; 3/2,\; 1$ and parity $p$. Here, each entry,  represented by  $s(\text{operator}_{\text{mode type}}^{\text{eigenvalue}})$,
    corresponds to a specific  spin component of a supermultiplet.
    The notation indicates also the relevant  cross diagrams of $\Delta_p$, for $p=0,1,2$, and $i\slashed{D}_{1/2}$  as given
    in \cref{sec:eigenmodes}. The sign subscript on $1^-_{\scriptscriptstyle (\pm)}$ shows which branch of the mass formula for spin $1^-$ in \cref{table:massop} that is being  used.  All operator eigenvalues are listed in \cref{sec:eigenvalues:summary}.
    The $Sp_1^C$ entries specify the number of cross diagrams belonging to the supermultiplet.
    The short, i.e., massless, supermultiplets are $2^+$ for $(p,q;r)=(0,0;0)$ and $1^-_2$ for $(2,0;0)$ and $(0,0;2)$.
    }
    \label{spinonetotwomultiplets}
\end{table}

The irreps in the  $Sp_1^C$   column in \cref{spinonetotwomultiplets} correspond to the sets of cross diagrams (see \cref{sec:eigenmodes} or the appendix of \cite{Nilsson:2018lof}) relevant   for each supermultiplet:
That is, $\bf 1$  means that there
is just one cross diagram with $r=p$, $\bf 3 $ that there are three cross diagrams with $r=p$, $r=p\pm 2$  and finally $\bf 5 $  implies five cross diagrams with $r=p$, $r=p\pm 2$ and $r=p\pm 4$.
More precisely, it corresponds to the $sp_1^C$-irrep of the differential operators generating the modes.
In the process of finding which cross diagrams belong to each of these sets of supermultiplets they are also given a specific operator eigenvalue. How to obtain this information is explained in \cref{sec:supermultiplets:details}.

The last two  supermultiplets $1^+_1$ and $1^+_2$  in \cref{spinonetotwomultiplets} contain  fields with identical  isometry irreps (for $r=p$ and $r=p\pm 2$) and with the same  values of $E_0$.  However, as shown in \cref{sec:eigenmodes},
the eigenmodes of $\Delta_2$ in the two cases are nevertheless different, as indicated by the superscripts on $\Delta_2^{(3)}$ and $\Delta_2^{(3)'}$,  but with the  same  eigenvalues (see \crefrange{eq:modes:2-form:eigenmodes:first}{eq:modes:2-form:eigenvalues:last} and  \cref{fig:modes:2-form:cross_diagrams}). This phenomenon  appears twice also  in the table for the
Wess--Zumino multiplets  with $C_g > 0$, \cref{wesszuminopositivecgmultiplets}. It is tempting to try to tie this to the appearance of $\Delta_L$ in all these supermultiplets but we leave  this  issue for future work.
We also note that none of the scalar and Dirac fields
in  \cref{spinonetotwomultiplets} have Neumann boundary conditions since they all have $E_0\ge 3/2$ (all irreps in the spin $1^-$ cross diagrams have $C_g\ge 19/4$).
In fact, it is easy to see from the supermultiplet structure and the unitarity bounds that higher-spin supermultiplets never admit Neumann boundary conditions.%
\footnote{For massless vector supermultiplets, the spinor has degenerate boundary conditions ($E_0 = 3/2$).}

\begin{table}[H]
    \centering
    \setlength{\defaultaddspace}{4pt}
    \begin{tabular}{lll @{$\mskip\thinmuskip$}c@{$\mskip\thickmuskip$} lcl}
        \toprule
        Mult. & $E_0$ &$E_0-\frac{1}{2}$ && $E_0+\frac{1}{2}$ &   $\mathclap{Sp_1^C}$& $E_0$ values\\ \midrule
        WZ1 &$\frac{1}{2}(i\slashed{D}^{(1)_+}_{1/2})$ & $0^-(Q^{(3)_+})$ && $0_{\scriptscriptstyle (+)}^+(\Delta_0^{(1)})$ & $\bf{1}$ & $\frac{3}{2}+\frac{5}{2}+\frac{1}{6}\sqrtsmash[C_G]{20C_g+81}$\\ \addlinespace
        WZ2 & $\frac{1}{2}(i\slashed{D}^{(1)_-}_{1/2})$ & $0^+_{\scriptscriptstyle (-)}(\Delta_0^{(1)})$ & $\leftrightarrow$ & $0^-(Q^{(3)_-})$ & $\bf{1}$ & $\frac{3}{2}\pm \bigl(-\frac{5}{2}+\frac{1}{6}\sqrtsmash[C_G]{20C_g+81}\bigr)$\\ \addlinespace
        WZ3 &$\frac{1}{2}(i\slashed{D}_{3/2}^{(1)_+})$ & $0^+(\Delta_L^{(2)_+})$ && $0^-(Q^{(1)_+})$  & $\bf{1}$  & $\frac{3}{2}+\frac{5}{6}+\frac{1}{6}\sqrtsmash[C_G]{20C_g+1}$\\ \addlinespace
        WZ4 &$\frac{1}{2}(i\slashed{D}_{3/2}^{(1)'_+})$ & $0^+(\Delta_L^{(2)'_+})$ && $0^-(Q^{(1)'_+})$  & $\bf{5}$  & $\frac{3}{2}+\frac{5}{6}+\frac{1}{6}\sqrtsmash[C_G]{20C_g+1}$\\ \addlinespace
        WZ5 & $\frac{1}{2}(i\slashed{D}_{3/2}^{(1)_-})$ &$0^-(Q^{(1)_-})$ && $0^+(\Delta_L^{(2)_-})$ & $\bf{1}$  &  $\frac{3}{2}-\frac{5}{6}+\frac{1}{6}\sqrtsmash[C_G]{20C_g+1}$\\ \addlinespace
        WZ6 & $\frac{1}{2}(i\slashed{D}_{3/2}^{(1)'_-})$ &$0^-(Q^{(1)'_-})$ && $0^+(\Delta_L^{(2)'_-})$ & $\bf{5}$  &  $\frac{3}{2}-\frac{5}{6}+\frac{1}{6}\sqrtsmash[C_G]{20C_g+1}$
        \\ \bottomrule
    \end{tabular}
    \caption{Wess--Zumino supermultiplets for $C_g>0$. See \cref{spinonetotwomultiplets} for definitions.
    The upper sign in the WZ2 $E_0$ value corresponds to $(D,D,D)$ boundary conditions for $p+q \geq 3$, $(D,N,D)$ for $p+q=2$ and $(N,N,D)$ for $p+q=1$.
    The scalar $0^+$ and pseudo-scalar $0^-$ should change places in the WZ2 supermultiplet if the lower sign in the $E_0$ formula is used, as indicated by the arrow.
    The lower sign is only valid for $p+q = 2$ and $p+q = 1$ and then corresponds to $(N,N,D)$ and $(D,N,D)$ boundary conditions, respectively.
    The subscript on $0^+_{\scriptscriptstyle (\pm)}$
    indicates which branch of $M^2(\Delta_0)$ is used.}
    \label{wesszuminopositivecgmultiplets}
\end{table}

We now turn to the six infinite sequences of Wess--Zumino multiplets with $C_g>0$ in \cref{wesszuminopositivecgmultiplets}.
The cross diagrams and operator eigenvalues for the first two Wess--Zumino multiplets in this list are specified by the spin-1/2 entries,
see \cref{fig:modes:spinor:cross_diagrams} and \crefrange{eq:modes:spinor:eigenvalues:first}{eq:modes:spinor:eigenvalues:last}.
Note that the isometry singlet mode of $\Delta_0$ should be removed from $0^+_{\scriptscriptstyle (-)}$, but not from $0^+_{\scriptscriptstyle (+)}$, as explained in \cite{Duff:1986hr}.
Looking at WZ1--2, this is consistent with the fact that the $(0,0;0)$ eigenmode of $i\slashed{D}_{1/2}$ has an eigenvalue given by $i\slashed{D}_{1/2}^{(1)_+}$ and there is no $(0,0;0)$ eigenmode of $i\slashed{D}_{1/2}$ with eigenvalue $i\slashed{D}_{1/2}^{(1)_-}$, see \cref{fig:modes:spinor:cross_diagrams}.

\begin{figure}[H]
    \centering
    \begingroup
    \renewcommand{\scale}{0.41cm}
    \footnotesize
    \newcommand{\figIppIV}{%
        \begin{tikzpicture}[x=\scale, y=\scale]
            \crossdiagramlayout{p+4}
            \foreach \p in {0,...,\range} {
                \foreach \q in {2,...,\range} {
                    \node at (\p,\q) {$\times$};
                }
            }
        \end{tikzpicture}
    }
    \newcommand{\figIppII}{%
        \begin{tikzpicture}[x=\scale, y=\scale]
            \crossdiagramlayout{p+2}
            \foreach \p in {1,...,\range} {
                \foreach \q in {1,...,\range} {
                    \node at (\p,\q) {$\times$};
                }
            }
        \end{tikzpicture}
    }
    \newcommand{\figIp}{%
        \begin{tikzpicture}[x=\scale, y=\scale]
            \crossdiagramlayout{p}
            \foreach \p in {0,...,1} {
                \foreach \q in {1,...,\range} {
                    \node at (\p,\q) {$\times$};
                }
            }
            \foreach \p in {2,...,\range} {
                \node at (\p,0) {$\times$};
            }
            \foreach \p in {2,...,\range} {
                \foreach \q in {1,...,\range} {
                    \node at (\p,\q) {$\boxtimes$};
                }
            }
        \end{tikzpicture}
    }
    \newcommand{\figIpmII}{%
        \begin{tikzpicture}[x=\scale, y=\scale]
            \crossdiagramlayout{p-2}
            \foreach \p in {3,...,\range} {
                \foreach \q in {0,...,\range} {
                    \node at (\p,\q) {$\times$};
                }
            }
        \end{tikzpicture}
    }
    \newcommand{\figIpmIV}{%
        \begin{tikzpicture}[x=\scale, y=\scale]
            \crossdiagramlayout{p-4}
            \foreach \p in {4,...,\range} {
                \foreach \q in {0,...,\range} {
                    \node at (\p,\q) {$\times$};
                }
            }
        \end{tikzpicture}
    }
    \newcommand{\figIIppIV}{%
        \begin{tikzpicture}[x=\scale, y=\scale]
            \crossdiagramlayout{p+4}
            \foreach \p in {0,...,\range} {
                \foreach \q in {0,...,\range} {
                    \node at (\p,\q) {$\times$};
                }
            }
        \end{tikzpicture}
    }
    \newcommand{\figIIppII}{%
        \begin{tikzpicture}[x=\scale, y=\scale]
            \crossdiagramlayout{p+2}
            \foreach \p in {1,...,\range} {
                \foreach \q in {1,...,\range} {
                    \node at (\p,\q) {$\times$};
                }
            }
        \end{tikzpicture}
    }
    \newcommand{\figIIp}{%
        \begin{tikzpicture}[x=\scale, y=\scale]
            \crossdiagramlayout{p}
            \node at (0,0) {$\otimes$};
            \foreach \p in {0,...,1} {
                \foreach \q in {2,...,\range} {
                    \node at (\p,\q) {$\times$};
                }
            }
            \foreach \p in {2,...,\range} {
                \node at (\p,0) {$\times$};
                \node at (\p,1) {$\times$};
            }
            \foreach \p in {2,...,\range} {
                \foreach \q in {2,...,\range} {
                    \node at (\p,\q) {$\boxtimes$};
                }
            }
        \end{tikzpicture}
    }
    \newcommand{\figIIpmII}{%
        \begin{tikzpicture}[x=\scale, y=\scale]
            \crossdiagramlayout{p-2}
            \foreach \p in {3,...,\range} {
                \foreach \q in {1,...,\range} {
                    \node at (\p,\q) {$\times$};
                }
            }
        \end{tikzpicture}
    }
    \newcommand{\figIIpmIV}{%
        \begin{tikzpicture}[x=\scale, y=\scale]
            \crossdiagramlayout{p-4}
            \foreach \p in {4,...,\range} {
                \foreach \q in {0,...,\range} {
                    \node at (\p,\q) {$\times$};
                }
            }
        \end{tikzpicture}
    }
    \begin{tabular}{ccccc}
        & & {\normalsize WZ3--4} & & \\
        \figIpmIV  & \figIpmII  & \figIp  & \figIppII  & \figIppIV  \\[4pt]
        & & {\normalsize WZ5--6} & & \\
        \figIIpmIV & \figIIpmII & \figIIp & \figIIppII & \figIIppIV
    \end{tabular}
    \endgroup
    \caption{%
        The cross diagrams in the appendix of \cite{Nilsson:2018lof} that remain when all other supermultiplets with their respective cross diagrams have been removed.
        These twelve diagrams correspond to  two sets of $\vec{1}\oplus \vec{5}$.
        The displayed division into two different sets, see \cref{wesszuminopositivecgmultiplets}, relies on the results  of \cite{Duboeuf:2022mam}  for small $p$ or $q$.
        The multiplicities of the isometry irreps are one for crosses and two for boxes with crosses.
        The single encircled cross, in the irrep $(0,0;0)$, is the Page supermultiplet (see below) containing the mode ``coming from nowhere'' in the analysis of \cite{Nilsson:2018lof, Nilsson:2023ctq}.
    }
    \label{fig:colour-split}
\end{figure}
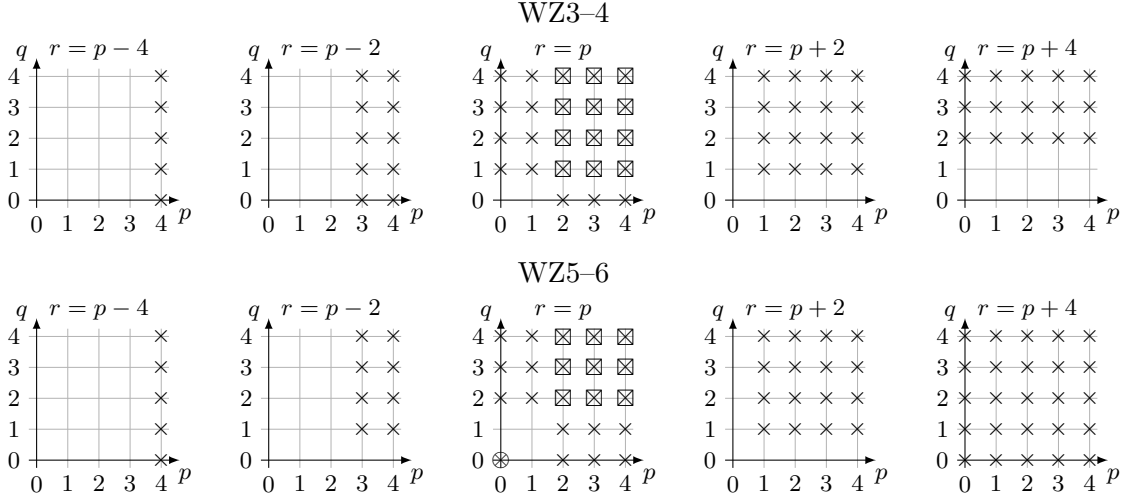

For the Wess--Zumino multiplets numbered 3 to 6, on the other hand,
it is not possible, with the information obtained in this paper, to group the remaining twelve cross diagrams, for each of the operators $i\slashed{D}_{3/2}$, $\Delta_L$ and $Q$, in the appendix of  \cite{Nilsson:2018lof} into two separate sets of ${\vec 1}+{\vec 5}$.
However, by invoking also results from \cite{Duboeuf:2022mam} this can be done.%
\footnote{Note that the $\Delta$ of \cite{Duboeuf:2022mam} is the lowest conformal dimension of a given supermultiplet and, hence, shifted compared to the $E_0$-value we use here, $\Delta = E_0 - 1/2$, see \cref{footnote:E0}.}
The result is presented in \cref{fig:colour-split}.
If one constructs the corresponding eigenmodes, we expect that the crosses below the boxed crosses cannot uniquely be assigned to the $\vec{1}$ or the $\vec{5}$, similar to what we found in \cref{sec:eigenmodes:two-forms}.

From the general structure of a Wess--Zumino multiplet $D(\frac{1}{2}, E_0) \oplus D(0^\pm, E_0 - \frac{1}{2}) \oplus D(0^\mp, E_0 + \frac{1}{2})$, we see that there can be no Neumann boundary conditions if $E_0 \geq 2$.
For $\frac{3}{2} < E_0 < 2$, the boundary conditions are $(D,N,D)$ and if all boundary conditions are flipped the fields fit into a supermultiplet $(\frac{1}{2}, 0^\mp, 0^\pm)$ where the scalar and pseudo-scalar have switched places and the boundary conditions are $(N, N, D)$.%
\footnote{The case $E_0 = 3/2$ is identical to the $3/2 < E_0 < 2 $ one except that the boundary conditions for the spinor now stay the same if the boundary conditions for the scalar and pseudo-scalar are flipped.}

That only irreps with $C_g \geq 4$ occur in \cref{fig:colour-split}, except for the isometry singlet with $C_g = 0$, implies that Neumann boundary conditions are never possible in WZ3--6, only in WZ2.
This happens for the  five  irreps with $p+q \leq 2$  and $C_g$ values given in \cref{wzcgvalueswithdandnbc}.

Specifically, the $(1/2, 0^+, 0^-)$ WZ2 supermultiplet as given in \cref{wesszuminopositivecgmultiplets} have $(D,D,D)$ boundary conditions for $p+q \geq 3$, $(D,N,D)$ for $p+q = 2$  and $(N,N,D)$ for $p+q = 1$.
For $p+q=2$ (last three entries in \cref{wzcgvalueswithdandnbc}) and $p+q=1$ (first two entries in \cref{wzcgvalueswithdandnbc}) the boundary conditions of all three fields can be flipped and they then fit in a $(1/2, 0^-, 0^+)$ supermultiplet with boundary conditions $(N,N,D)$ and $(D,N,D)$, respectively.
Note that the $E_0$ value of the supermultiplet given in the table should be replaced by $3 - E_0$ in these cases.

Also for the $C_g$ values in the range
$12<C_g<25$ ($p+q=3$) both Dirichlet and Neumann boundary conditions  are possible in WZ2 but now only for the scalars $0^+$.
A complete list of fields with $E_0$ in this window is provided in \cref{tab:D_or_N:scalars} below.
After skew-whiffing there seems to be no reason for choosing one or the other of the boundary conditions for these $0^+$ fields.
The supermultiplets in the left-squashed case  containing these scalar fields can, however, only be of the kind $(D,D,D)$.

\begin{table}[H]
    \centering
    \begin{tabular}{l|lllll}
        \toprule
        $(p,q;r)$  &   $(0,1;0)$ & $(1,0;1)$  & $(1,1;1)$  & $(0,2;0)$  &  $(2,0;2)$  \\
        $(d_2, d_1)$  &  $({\bf 5}, {\bf 1})$  & $({\bf 4}, {\bf 2})$  &  $({\bf 8}, {\bf 2})$ & $({\bf 14}, {\bf 1})$ & $({\bf 10}, {\bf 3})$ \\
        $C_g$ & 4  & 19/4  &   39/4 &  10&   12
        \\ \bottomrule
    \end{tabular}
    \caption{Irreps with $r=p$ and $C_g>0$ for which Wess--Zumino multiplets in WZ2 admit both Dirichlet and Neumann boundary conditions.
    The implications of this are explained in the text. Here, $(d_2, d_1)$ denotes the dimensions of the $sp_2$- and $sp_1^C$-irreps.}
    \label{wzcgvalueswithdandnbc}
\end{table}
It would be interesting  to find a way to discriminate between the different boundary conditions respecting supersymmetry.
This could potentially be achieved by relating the squashed vacuum to the round one with $\mN = 8$.
In this context, we note that if one squashes the seven-sphere continuously and follows the spinors in the isometry irrep $(0,0;2)$ of $G = Sp_2 \times Sp_1^C$, which comes from the $Spin(8)$-irrep $(0,0,1,0)$ on the round sphere (see \cref{decomp}), the relevant eigenvalue is \cite{Nilsson:1983ru}
\begin{equation}
    i\slashed{D}_{1/2}(0,0;2)
    = - \frac{\sqrt{7}}{10} \frac{10\lambda^2 + 25}{\sqrt{1 + 8\lambda^2 -2\lambda^4}} m(\lambda),
\end{equation}
where
\begin{equation}
    m^2(\lambda) = \frac{1 + 8\lambda^2 -2\lambda^4}{28\lambda^2}
\end{equation}
comes from demanding that $R = 42m^2$ for all values of the squashing parameter $\lambda$.
From this, we see that the associated $AdS_4$ field starts out as a fermionic singleton, with $i\slashed{D}_{1/2}(0,0;2) = \frac{-7m}{2}$ and $E_0 = 1$, in the round sphere vacuum ($\lambda^2 = 1$) and ends up as the massless spin-1/2 companion of the gauge field in the short massless supermultiplet related to the $Sp_1^C$ isometry after squashing ($\lambda^2 = 1/5$), with $i\slashed{D}_{1/2}^{(2)_-}(0,0;2) = \frac{-9m}{2}$ and $E_0 = \frac{3}{2}$.
Hence, this is one example of Higgsing involving a singleton and, thus, also a change of boundary conditions triggered by squashing.%
\footnote{The language used here describes what happens when going from round to squashed, for consistency with \cite{Nilsson:2023ctq,Nilsson:2018lof}, even though the round vacuum has to be in the IR of the squashed one for any potential RG-flow connecting the two, see \cref{sec:conclusions}.}

Another relevant point is that there can be exactly marginal singlet operators that can be used to deform the dual CFT, as discussed in, e.g., \cite{Witten:2001ua}.
We will return to the issue  of marginal operators in \cref{sec:marginal_ops_in_N=0} where we give the full content of marginal single- and multi-trace operators for the theory discussed in this paper.
A perhaps relevant fact here is that the right-squashed spectrum does not allow for any single- or double-trace marginal operators.
If these considerations are applied to the left-squashed case, such deformations will break supersymmetry.

As an example where different boundary conditions are possible, we consider the irrep $(1,0 ;1)$ (with dimension $(\vec{4}, \vec{2})$) in the Einstein-squashed spectrum.
The scalar $0^+$ in this multiplet has $M^2(0^+_{\scriptscriptstyle (-)}) = 95m^2/9$, from which we see that
\begin{equation}
    0^+_{\scriptscriptstyle (-)}(\Delta_0^{(1)})_{(1,0;1)}\colon\quad
    E_0=\frac{3}{2}\pm\frac{1}{6} \bigl(18-4\sqrt{11}\+ \bigr) > \frac{1}{2},
\end{equation}
which is  compatible with both Dirichlet ($E_0 > 3/2$) and Neumann ($1/2 < E_0 < 3/2$) boundary conditions. For the other two fields in this Wess--Zumino multiplet we have
\begin{alignat}{2}
    &\frac{1}{2}(i\slashed{D}_{1/2}^{(1)_-})_{(1,0;1)}\colon\quad
    && E_0=\frac{3}{2}\pm\frac{1}{6} \bigl(15-4\sqrt{11}\+ \bigr) > 1,
    \\[2pt]
    & 0^-(Q^{(3)_-})_{(1,0;1)}\colon\quad
    && E_0=\frac{3}{2}\pm\frac{1}{6}\bigl(4\sqrt{11}-12\bigr) > \frac{1}{2},
\end{alignat}
which also allow for both types of boundary conditions.  One should note here that the three fields in a supermultiplet must have the same sign of the square root term
in $E_0$, due to supersymmetry, which implies that the supermultiplets will involve fields with different boundary conditions as described above.

Remarkably, it is not consistent to assume that all fields will retain their boundary conditions when
squashing the seven-sphere from the round $S^7$ to the Einstein-squashed one, even though the spectrum can be understood in this way through a Higgs/de-Higgs mechanism \cite{Nilsson:2018lof,Nilsson:2023ctq}. This fact is clear from looking at the irrep $(0,2;0)$ for which all three fields come from fields with Dirichlet boundary conditions in the round sphere
vacuum which leads to an impossible set of boundary conditions (see \cref{wzcgvalueswithdandnbc}) in the left-squashed case if supersymmetry is kept
intact. In fact, also for the irrep $(1,0;1)$ it is clear that one cannot keep the boundary
conditions when squashing since the spin-1/2  field comes from the massless field $(1,0,1,0)$ with $E_0=3/2$ in the round spectrum.
This field has neither Dirichlet nor Neumann boundary conditions, as discussed in, e.g., \cite{Witten:2001ua}.

The last supermultiplets to account for are the  Wess--Zumino ones with $C_g=0$.
There are only two such multiplets, see \cref{wesszuminocgequaltozero}, which is clear from the cross diagrams in \cite{Nilsson:2018lof}.
We have listed the two Wess--Zumino multiplets that exist for $C_g=0$ separately since all the relevant mode functions can be constructed explicitly as
done recently in \cite{Nilsson:2023ctq} and reproduced in \cref{sec:eigenmodes:isometry_singlets} of this paper.
The results obtained there are summarised in \cref{tab:supermultiplets:singlet-summary}.

\begin{table}[H]
    \centering
    \setlength{\defaultaddspace}{4pt}
    \begin{tabular}{llll}
        \toprule
        Spin & Eigenvalue & Mass & $E_0$ \\ \midrule
        $\frac{1}{2}$ & $i\slashed{D}_{1/2}^{(1)_+} = \frac{7m}{2}$ & $M = -8m$ & $\frac{11}{2}$ \\ \addlinespace
        $\frac{1}{2}$ & $i\slashed{D}_{3/2}^{(1)_-} = - \frac{m}{6}$ & $M = \frac{4m}{3}$ & $\frac{13}{6}$ \\ \addlinespace
        $0^+$ & $\Delta_0^{(1)} = 0$ & $M^2_{\scriptscriptstyle (+)} = 80m^2$ & $6$ \\ \addlinespace
        $0^+$ & $\Delta_L^{(2)_-} = \frac{28 m^2}{9}$ & $M^2 = - \frac{8 m^2}{9}$ & $\frac{3}{2} \pm \frac{1}{6}$ \\ \addlinespace
        $0^-$ & $Q^{(3)_+} = 4m$ & $M^2 = 48 m^2$ & $5$ \\ \addlinespace
        $0^-$ & $Q^{(1)_-} = - \frac{2m}{3}$ & $M^2 = \frac{40 m^2}{9}$ & $\frac{8}{3}$
        \\ \bottomrule
    \end{tabular}
    \caption{$AdS_4$ spin, operator eigenvalues, masses and $E_0$ values of isometry singlet ($C_g = 0$) modes and the associated $AdS_4$ fields. The plus subscript on $M^2$ indicates the branch of the mass formula in \cref{table:massop}. The $\pm$ in the $E_0$ column corresponds to different boundary conditions. Below, we will see that supersymmetry fixes the plus sign.}
    \label{tab:supermultiplets:singlet-summary}
\end{table}

The fields corresponding to isometry singlet modes (see \cref{tab:supermultiplets:singlet-summary}) fit into two Wess--Zumino supermultiplets uniquely, given in \cref{wesszuminocgequaltozero}.
All three fields in both  these supermultiplets have  Dirichlet boundary conditions, which in the case of the Page multiplet is  a consequence of supersymmetry.
This is most easily seen by noting that the spin-1/2 fields have $E_0 > 2$.

After skew-whiffing
the situation is a bit more complicated since there is  no supersymmetry in the right-squashed vacuum and several  choices of boundary conditions become possible \cite{Nilsson:2023ctq}.
We shall return to this issue below.

\begin{table}[H]
    \centering
    \setlength{\defaultaddspace}{4pt}
    \begin{tabular}{llllcl}
        \toprule
        Mult. & $E_0$ & $E_0-\frac{1}{2}$ &$E_0+\frac{1}{2}$ &   $\mathclap{Sp^C_1}$& $E_0$ values\\ \midrule
        $(\text{WZ1})_0$ &$\frac{1}{2}({i\slashed{D}}_{1/2}^{(1)_+})$ & $0^-(Q^{(3)_+})$& $0_{\scriptscriptstyle (+)}^+(\Delta_0^{(1)})$  & $\bf{1}$ & $\frac{3}{2}+4=\frac{11}{2}$\\ \addlinespace
        Page &$\frac{1}{2}(i\slashed{D}_{3/2}^{(1)_-})$ & $0^+(\Delta_L^{(2)_-})$ & $0^-(Q^{(1)_-})$ & $\bf{1}$  & $\frac{3}{2}+\frac{2}{3}=\frac{13}{6}$
        \\ \bottomrule
    \end{tabular}
    \caption{Wess--Zumino supermultiplets with $C_g = 0$.}
    \label{wesszuminocgequaltozero}
\end{table}

While the $(\text{WZ1})_0$ value $E_0=\frac{3}{2}+4$ fits into the general
$E_0$ formula for WZ1 this is not the case for the Page multiplet which does not fit into any  of the $E_0$ formulas  in \cref{wesszuminopositivecgmultiplets}.
However, it does nevertheless belong to WZ5--6 as we will now explain.
The reason for the odd behaviour of the Page multiplet is that it utilises a special property that arises for $C_g=0$ in the $E_0$ formulas for all three fields in the multiplet,
similar to the five cases in \cref{wzcgvalueswithdandnbc}.
These formulas are of the form $\frac{3}{2} \pm |\sqrtsmash[C_G]{20 C_g + 1} - X|$ where the argument of the absolute value is negative only for the singlet isometry irrep.
This leads to the scalar and pseudo-scalar of the Page multiplet changing places relative to WZ5--6 and the $E_0$ value being reflected in $3/2$ relative to the formula in \cref{wesszuminopositivecgmultiplets}.

Note that if the $E_0$ value of the spinor in the Page multiplet would have been less than $2$, the boundary conditions of the fields could have been flipped and the multiplet would have fitted exactly into WZ5--6 as given in the table.
Attempting this with the actual values of $E_0$ leads to a violation of the unitarity bounds.

The Page WZ-multiplet contains the  scalar that was discussed by Page in \cite{Page:1983mke}. This scalar is the squashing mode of the  $S^7$ compactification
and was shown by Page to have $M^2=-\frac{8}{9} m^2$ (just above the BF bound) in the Einstein-squashed case.

Clearly there are no singletons in the spectrum of the left-squashed seven-sphere \cite{Nilsson:2018lof}.
The singleton that exists after skew-whiffing to the right-squashed vacuum \cite{Nilsson:2018lof}   will be discussed below.

\subsection{Boundary conditions and marginal operators in the \texorpdfstring{$\mN \!=\! 1$}{N = 1} and \texorpdfstring{$\mN \!=\! 0$}{N = 0} vacua}
\label{sec:supermultiplets:skew-whiffing}
\settocdepth{subsubsection}
\setsecnumdepth{subsubsection}

One reason for constructing the entire operator spectrum of the left-squashed seven-sphere is that
one can use supersymmetry as a consistency check of the results and then flip its orientation, i.e., perform a so called skew-whiffing \cite{Duff:1986hr}, to produce
the right-squashed compactification with no supersymmetry.
This solution is known to be BF-stable \cite{Duff:1984sv} despite having no supersymmetry and there should therefore exist, in view of the swampland $AdS$ swampland conjecture first discussed in \cite{Ooguri:2016pdq},  some other decay mode with respect to which the right-squashed vacuum is unstable.

One possible such decay mode is related to  special composite operators  in the boundary theory of the $AdS_4$ gravity theory and condensation of scalar composites in the bulk. As discussed in the Introduction (and in the references cited there), if marginal operators exist ($\Delta=3$), either single, double- or multi-trace,
that may signal an instability of the $AdS_4$ theory.  For the gravity theory in the bulk, this corresponds to looking for composite scalar gauge singlet states with total $E_0=3$.

Such trace operators can be of only a few kinds, namely, (schematically) $\lambda^2 \phi^n$ for $n=0,1,2$ and $\phi^n$ for $n=1,\hdots,6$, with the spinor $\lambda$ and scalar $\phi$ representing both the bulk fields and their dual single-trace operators (with $\Delta=E_0$).
Note that in the last cases in these lists the fields have to be singletons and scalar singletons are known to exist only in the round sphere case \cite{Nilsson:2018lof}.
There is also a spinorial singleton with $E_0=1$ in the right-squashed spectrum \cite{Nilsson:2018lof} which could  play a role here.
In fact, in the recent analysis of the singlet sector of the right-squashed spectrum in \cite{Nilsson:2023ctq} the fields in this sector were found to either give rise to two kinds of marginal triple-trace
operators, of which one involved the fermionic singleton mentioned above,  or to give rise to no marginal operators depending on
which  of two possible choices of boundary conditions one imposes on the fields.

Here, we will generalise  this analysis to all fields by including also the ones   coming from modes with $C_g> 0$, i.e., non-singlet modes.
These fields have non-trivial properties even in the left-squashed supersymmetric case which were analysed in \cref{spectrum_summary}.
In \cref{sec:boundary-conditions}, we will therefore perform a complete analysis of all fields which can have Neumann boundary conditions  in both the left- and  right-squashed vacuum, after discussing skew-whiffing and the right-squashed spectrum in \cref{sec:skew-whiffing}.
Marginal operators will be discussed in detail in \cref{sec:marginal_ops_in_N=0}.
There we will find that, since there exist several fields in the squashed vacua  with  $E_0\le \frac{3}{2}$, bound states
containing different fields may give rise to large  numbers of marginal operators depending on the boundary conditions. In fact, one such mixed example was discussed recently in \cite{Nilsson:2023ctq}.

\subsubsection{Skew-whiffing and the spectrum of the right-squashed vacuum}
\label{sec:skew-whiffing}
To see  the effects on the spectrum of an orientation flip,  we should  recall how it is implemented, see for instance  \cite{Duff:1986hr}.
The definition of the linear operators $Q$, $i\slashed{D}_{1/2}$ and $i\slashed{D}_{3/2}$ are sensitive to the orientation and, as explained in \cite{Duff:1986hr}, the sign of their spectra are flipped when the orientation is changed.%
\footnote{This way to implement skew-whiffing is designed to keep the $AdS_4$ the same when flipping, including the flux determined by $m$. One way to do this is to leave the vielbein, the
epsilon-tensor and the gamma-matrices intact on $S^7$ and produce the sign flip in the volume form by letting the coordinates $y^i$ go to $-y^i$.
The same effect can be obtained by changing $m\rightarrow -m$ and keeping the $S^7$ intact.}
Thus this has implications only for the $AdS_4$ fields with spin 3/2, 1/2 and the pseudo-scalars $0^-$.
In practice, we can implement this conveniently by only referring to the operators, eigenvalues and cross diagrams of the left-squashed seven-sphere and flip the signs of the linear operators in the mass formulas as follows (cf.\ \cref{table:massop}):
\begin{alignat}{2}
    & \tfrac{3}{2}_{\scriptscriptstyle (\pm)}(i\slashed{D}_{1/2})\colon\qquad
    && M_R = i\slashed{D}_{1/2} + \frac{7m}{2}, \\[2pt]
    & \tfrac{1}{2}_{\scriptscriptstyle (\pm)}(i\slashed{D}_{1/2})\colon\qquad
    && M_R = i\slashed{D}_{1/2} - \frac{9m}{2}, \\[2pt]
    & \tfrac{1}{2}_{\scriptscriptstyle (\pm)}(i\slashed{D}_{1/2})\colon\qquad
    && M_R = -i\slashed{D}_{3/2} + \frac{3m}{2}, \\[2pt]
    & 0^-_{\scriptscriptstyle (\pm)}(Q)\colon\qquad
    && M_R^2 = (Q-3m)^2 - m^2,
\end{alignat}
where we have   introduced the following notation for masses of  scalar and spin-1/2 fields in the left- and right-squashed vacua:  We keep the standard notation $M^2$ and $M$
in the cases where the orientation does not matter, while, when it does matter, we from now on use the notation
$M^2_L$ or $M_L$ and $M^2_R$ or $M_R$ for the left- and right-squashed cases, respectively.
The right-squashed masses have several interesting consequences. The first one is that the spinor $\eta$, satisfying $i\slashed{D}_{1/2}\eta=\frac{7m}{2}\eta$,
 is no longer a Killing spinor and hence  does not lead to a massless spin-3/2 field in $AdS_4$
gravity theory and supersymmetry is therefore lost. Furthermore, when using this spinor for the spin-1/2 fields one finds a field with mass $M=-m$ which, if given Neumann boundary conditions, will have
$E_0=1$. This spinorial singleton is known from \cite{Nilsson:2018lof} to exist in the right-squashed case (see also \cite{Nilsson:2023ctq}).

\subsubsection{Possible boundary conditions for the fields in both vacua}
\label{sec:boundary-conditions}
We will here tabulate the   scalar fields in  $AdS_4$ in the mass range $-m^2 \le M^2 \le 3m^2$ and the fermions having  $|M| \le m$.  These are the only masses  for which
fields with both Dirichlet and Neumann boundary conditions are possible.
Since the boundary conditions affect the conformal dimension of the dual operator, this is crucial for the analysis of marginal multi-trace operators in \cref{sec:marginal_ops_in_N=0}.

To list these mass values we need the
$C_g$ Casimir
\begin{equation}
    C_g(p,q;r)=\frac{1}{2}(p^2+2q^2+4p+6q+2pq)+\frac{3}{4}r(r+2).
\end{equation}
We will specify the irreps using the Dynkin labels $(p,q;r)$ as above.
The dimensions of the $sp_2$- and $sp_1$-irreps can be obtained from these as
\begin{align}
    & d_2(p,q) = \frac{1}{6}(p+2q+3)(p+q+2)(p+1)(q+1),\\
    & d_1(r) = r+1,
\end{align}
respectively.

The  issue of stability, and  as we saw above its connection to the existence of fields with  $E_0\le \frac{3}{2}$, is
thus related to  when Neumann boundary conditions are possible on $AdS_4$ for scalar and
Dirac fields. This  is clear
from  their energy formulas
\begin{equation}
    E_0(s=0^{\pm})=\frac{3}{2}\pm \frac{1}{2} \sqrt{\frac{M^2}{m^2}+1}\ge \frac{1}{2},\qquad
    E_0\biggl(s=\frac{1}{2}\biggr)=\frac{3}{2}\pm \frac{1}{2} \biggl|\frac{M}{m}\biggr|\ge 1,
\end{equation}
where the bounds are due to unitarity. Note also the immediate implication that  for scalars $M^2\ge -m^2$ which is the BF bound. This implies that Neumann boundary conditions
are possible, implementing also unitarity,  provided that
\begin{equation}
    \text{Scalars: } -m^2 < M^2 \leq 3m^2,\qquad
    \text{Spinors: } -m \leq M \leq m.
\end{equation}

Inserting the various operator spectra into the formulas for the mass matrices we can check if there are modes that satisfy these bounds. We might suspect from experience with the round $S^7$ spectrum
that such states are quite rare. In fact, for the round $S^7$ the only such fields are the  $(2,0,0,0)$ (i.e., ${\bf 35}_v$  of $so(8)$) scalars   in the $Spin(2,3)$-irrep $D(1,0^+)$ and
the supersingleton\footnote{Singletons belong to the lower (minus) branch of the $E_0$  formulas for scalars and spin-1/2 fermions but have a more intricate behaviour
near the $AdS_4$ boundary than ordinary fields with Neumann boundary conditions. Often, singletons are considered as fields confined to the boundary.}
in $D(\frac{1}{2}, 0)\oplus D(1, \frac{1}{2})$ in the $so(8)$-irreps $(1,0,0,0)$ and $ (0,0, 1,0)$ (i.e., ${\bf 8}_v$ and ${\bf 8}_c$) as shown in \cite{Nilsson:2018lof}. There are also some fields in the round spectrum
with $E_0=\frac{3}{2}$ that hence have neither Dirichlet nor Neumann boundary conditions. These are scalars  from the first massive level in the irrep $(3,0,0,0)$  (or ${\bf 112}_v$) and spin-1/2 fields in the irrep $(1,0,1,0)$ (or ${\bf 56}_s$).
There are no pseudo-scalars of any similar kind.
However, we find here (see below) that in the $AdS_4$ spectra coming from the squashed sphere,  fields of these kinds are quite abundant. The exception here is the singletons which
exist only in the form of a singlet in the right-squashed vacuum \cite{Nilsson:2018lof, Nilsson:2023ctq}.

We list, for the squashed sphere vacua, the scalar, pseudo-scalar and spinor fields in the mass ranges specified above, that is, the isometry irreps for which both Dirichlet and Neumann boundary conditions are possible, in \cref{tab:D_or_N:scalars,tab:D_or_N:pseudo-scalars,tab:D_or_N:spinors}, respectively.
The analysis in the left-squashed case will show that whenever there are
two branches (either in the operator eigenvalues or in $M^2$) only the minus branch will have any irreps in these ranges.
Similarly, only the plus branches of eigenvalues appear in the right-squashed part of the pseudo-scalar and spinor tables.
The scalar fields closest to the BF bound are the ones in the $(p,q;r)=(2,0;2)$ isometry irrep of $0^+_{\scriptscriptstyle (-)}(\Delta_0^{(1)})$, both in the left- and right-squashed case.
These have $M^2 = -(12\sqrt{321}-212) m^2/3$, which is less than one part per mille above the BF bound $-m^2$.

Here we note that the singlet spin-3/2 mode, which belong to the spectrum of $i\slashed{D}_{3/2}^{(1)_-}$, is not present in neither the left- nor right-squashed part of \cref{tab:D_or_N:spinors} due to it having $E_0 = 13/6 > 2$ and $E_0 = 7/3 > 2$, respectively.
Even though the $E_0$ formula corresponding to this eigenvalue in the left-squashed case is $\frac{3}{2} \pm \frac{1}{6} |-5 + \sqrtsmash[C_G]{20 C_g + 1}|$, where the square root could a-priori make Neumann boundary conditions possible, Neumann boundary conditions are in fact never allowed by unitarity due to which values of $C_g$ occur in the cross diagram associated to this eigenvalue.

In the right-squashed vacuum we find the
members of the $(\text{WZ1})_0$ multiplet that were discussed above and in \cite{Nilsson:2023ctq}, namely the singlet ones in $1/2(i\slashed{D}_{1/2}^{(1)_+})$ and $0^-(Q^{(3)_+})$.
The issue of instability due to possible marginal operators constructed from these two singlet modes was discussed  in \cite{Nilsson:2023ctq} with the conclusion that marginal operators can either appear or not depending on the boundary conditions of the corresponding fields. We will discuss this further in \cref{sec:marginal_ops_in_N=0}.

We might add here that in general  it is not possible  to demand that a given eigenmode of
an operator should generate fields in the left- and right-squashed vacua with the same boundary conditions.
In the left-squashed vacuum, we know from \cref{spectrum_summary} that either the spinors associated with $i\slashed{D}_{1/2}^{(1)_-}$ or the pseudo-scalars associated with $Q^{(3)_-}$ need to have Neumann boundary conditions for the isometry irreps $(0,1;0)$ and $(1,0;1)$ to respect supersymmetry.
However, we see from \cref{tab:D_or_N:pseudo-scalars,tab:D_or_N:spinors} that the fields associated with these modes in the right-squashed vacuum can only have Dirichlet boundary conditions.
Some fields must be allowed to change boundary conditions also when squashing, as was mentioned already in \cref{spectrum_summary} where the
example of the WZ2 supermultiplet in the irrep $(0,2;0)$ was given.

The spin-3/2 fields are also affected by skew-whiffing but, in this case, only what happens to the Killing spinor mode leads to anything interesting and this was already discussed above.

The scalar fields with $\frac{5}{2} < E_0 \le 3 $ and the spinor fields with $2 < E_0 \leq 3$ might also be of interest since they too are dual to relevant single-trace operators. These have masses satisfying $3m^2 < M^2 \le 8m^2$ and $m < |M| \leq 3m$, respectively, and can thus only  have  Dirichlet
boundary conditions.  We list these in \cref{app:fields_in_ezero_range}.

The implications for the issue of possible instabilities due to marginal operator appearing in the right-squashed vacuum are discussed in \cref{sec:marginal_ops_in_N=0}. Having tabulated all fields
with small masses we note that some of them have indeed negative values which is in accord with the assertion of  \cite{Andriot:2022brg}.

\begin{table}[H]
    \centering
    \setlength{\defaultaddspace}{3pt}
    \setlength{\tabcolsep}{10pt}
    \begin{tabular}{lllll}
        \toprule
        Field & $(p,q;r)$ & $C_g$ & $M^2/m^2$ & $E_0$
        \\ \midrule
        $0^+_{\scriptscriptstyle (-)}(\Delta_0^{(1)})$
        & $(0,1;0)$ & $4$ & $\frac{1}{9} \bigl(476 - 36\sqrt{161}\+\bigr)$ & $\frac{3}{2} \pm \frac{1}{6} \bigl(18-\sqrt{161}\+\bigr)$ \\ \addlinespace
        & $(1,0;1)$ & $19/4$ & $\frac{1}{9} \bigl(491 - 144\sqrt{11}\+\bigr)$ & $\frac{3}{2} \pm \frac{1}{3} \bigl(9-2 \sqrt{11}\+\bigr)$ \\ \addlinespace
        & $(1,1;1)$ & $39/4$ & $-\frac{1}{3} \bigl(24 \sqrt{69} - 197\bigr)$ & $\frac{3}{2} \pm \frac{1}{3} \bigl(9-\sqrt{69}\+\bigr)$ \\ \addlinespace
        & $(0,2;0)$ & $10$ & $-\frac{1}{9} \bigl(36 \sqrt{281} - 596\bigr)$ & $\frac{3}{2} \pm \frac{1}{6} \bigl(18-\sqrt{281}\+\bigr)$ \\ \addlinespace
        & $(2,0;2)$ & $12$ & $-\frac{1}{3} \bigl(12 \sqrt{321} - 212\bigr)$ & $\frac{3}{2} \pm \frac{1}{6} \bigl(18-\sqrt{321}\+\bigr)$ \\ \addlinespace
        & $(1,2;1)$ & $67/4$ & $-\frac{1}{9} \bigl(144 \sqrt{26} - 731\bigr)$ & $\frac{3}{2} \pm \frac{1}{3} \bigl(2 \sqrt{26}-9\bigr)$ \\ \addlinespace
        & $(0,3;0)$ & $18$ & $0$ & $\frac{3}{2} \pm \frac{1}{2}$ \\ \addlinespace
        & $(2,1;2)$ & $18$ & $0$ & $\frac{3}{2} \pm \frac{1}{2}$ \\ \addlinespace
        & $(3,0;3)$ & $87/4$ & $\frac{1}{3} \bigl(277 - 24\sqrt{129}\+\bigr)$ & $\frac{3}{2} \pm \frac{1}{3} \bigl(\sqrt{129}-9\bigr)$ \\ \addlinespace
        $0^+(\Delta_L^{(2)_-})$
        & $(0,0;0)$ & $0$ & $-\frac{8}{9}$ & $\frac{3}{2} \pm \frac{1}{6}$
        \\ \addlinespace[\aboverulesep] \bottomrule
    \end{tabular}
    \caption{Scalars that can have both Dirichlet and Neumann boundary conditions. The listed fields and their properties
    are the same in the   left- and the right-squashed vacuum. Note that only the five first rows can respect supersymmetry with Neumann boundary conditions in the left-squashed vacuum, see \cref{spectrum_summary}. All brackets are positive.}
    \label{tab:D_or_N:scalars}
\end{table}

\begin{table}[H]
    \centering
    \setlength{\defaultaddspace}{3pt}
    \setlength{\tabcolsep}{10pt}
    \begin{tabular}{lllll}
        \toprule
        Field ($LS^7$) & $(p,q;r)$ & $C_g$ & $M^2_L/m^2$ & $E_0$
        \\ \midrule
        $0^-(Q^{(2)_-})$
        & $(1,1;1)$ & $39/4$ & $\frac{5}{9} \bigl(67 - 8 \sqrt{61}\+\bigr)$ & $\frac{3}{2} \pm \frac{1}{3} \bigl(\sqrt{61}-5\bigr)$ \\ \addlinespace
        & $(0,1;2)$ & $10$ & $\frac{20}{9} \bigl(17 - \sqrt{249}\+\bigr)$ & $\frac{3}{2} \pm \frac{1}{6} \bigl(\sqrt{249}-10\bigr)$ \\ \addlinespace
        $0^-(Q^{(3)_-})$
        & $(0,1;0)$ & $4$ & $-\frac{8}{9} \bigl(3 \sqrt{161} - 37\bigr)$ & $\frac{3}{2} \pm \frac{1}{6} \bigl(\sqrt{161}-12\bigr)$ \\ \addlinespace
        & $(1,0;1)$ & $19/4$ & $-\frac{1}{9} \bigl(96 \sqrt{11} - 311\bigr)$ & $\frac{3}{2} \pm \frac{2}{3} \bigl(\sqrt{11}-3\bigr)$ \\ \addlinespace
        & $(1,1;1)$ & $39/4$ & $\frac{1}{3} \bigl(137-16 \sqrt{69}\+\bigr)$ & $\frac{3}{2} \pm \frac{1}{3} \bigl(\sqrt{69}-6\bigr)$ \\ \addlinespace
        & $(0,2;0)$ & $10$ & $\frac{8}{9} \bigl(52-3 \sqrt{281}\+\bigr)$ & $\frac{3}{2} \pm \frac{1}{6} \bigl(\sqrt{281}-12\bigr)$ \\ \addlinespace
        & $(2,0;2)$ & $12$ & $\frac{8}{3} \bigl(19 - \sqrt{321}\+\bigr)$ & $\frac{3}{2} \pm \frac{1}{6} \bigl(\sqrt{321}-12\bigr)$
        \\ \addlinespace[\aboverulesep] \midrule
        Field ($RS^7$) & $(p,q;r)$ & $C_g$ & $M^2_R/m^2$ & $E_0$
        \\ \midrule
        $0^-(Q^{(1)_+})$
        & $(0,1;0)$ & $4$ & $-\frac{8}{9}$ & $\frac{3}{2} \pm \frac{1}{6}$ \\ \addlinespace
        & $(1,1;1)$ & $39/4$ & $\frac{7}{9}$ & $\frac{3}{2} \pm \frac{2}{3}$ \\ \addlinespace
        & $(0,2;0)$ & $10$ & $\frac{4}{9} \bigl(73 - 5 \sqrt{201}\+\bigr)$ & $\frac{3}{2} \pm \frac{1}{6} \bigl(\sqrt{201}-10\bigr)$ \\ \addlinespace
        & $(2,0;2)$ & $12$ & $\frac{4}{9} \bigl(83 - 5 \sqrt{241}\+\bigr)$ & $\frac{3}{2} \pm \frac{1}{6} \bigl(\sqrt{241}-10\bigr)$ \\ \addlinespace
        & $(3,0;1)$ & $51/4$ & $3$ & $\frac{3}{2} + 1$ \\ \addlinespace
        $0^-(Q^{(2)_+})$
        & $(1,0;1)$ & $19/4$ & $\frac{7}{9}$ & $\frac{3}{2} \pm \frac{2}{3}$ \\ \addlinespace
        & $(2,0;0)$ & $6$ & $\frac{16}{9}$ & $\frac{3}{2} \pm \frac{5}{6}$ \\ \addlinespace
        $0^-(Q^{(3)_+})$
        & $(0,0;0)$ & $0$ & $0$ & $\frac{3}{2} \pm \frac{1}{2}$
        \\ \addlinespace[\aboverulesep] \bottomrule
    \end{tabular}
    \caption{Pseudo-scalars that can have both Dirichlet and Neumann boundary conditions. The list is divided into the ones occurring in the left-squashed vacuum ($LS^7$) and
    those belonging to the right-squashed vacuum ($RS^7$).
    Some details  about which $(p,q;r)$ belong to which branch of  $0^-(Q^{(1)_{\pm}})$  have been collected from \cite{Duboeuf:2022mam} and used here.
    Note that the last entry for $0^-(Q^{(1)_+})$ is special since the option with a minus sign in $E_0$, corresponding to a  singleton, is not possible in the $RS^7$ vacuum. This follows from the results in
    \cite{Nilsson:2018lof}. Note also that Neumann boundary conditions for $0^-(Q^{(2)_-})$ do not respect supersymmetry in the left-squashed vacuum, see \cref{spectrum_summary}. All brackets are positive.}
    \label{tab:D_or_N:pseudo-scalars}
\end{table}

\begin{table}[H]
    \centering
    \setlength{\defaultaddspace}{3pt}
    \setlength{\tabcolsep}{10pt}
    \begin{tabular}{lllll}
        \toprule
        Field ($LS^7$) & $(p,q;r)$ & $C_g$ & $M_L/m$ & $E_0$
        \\ \midrule
        $\frac{1}{2}(i\slashed{D}_{1/2}^{(1)_-})$
        & $(0,1;0)$ & $4$ & $-\frac{1}{3}\bigl(15 - \sqrt{161}\+\bigr)$ & $\frac{3}{2} \pm \frac{1}{6} \bigl(15-\sqrt{161}\+\bigr)$ \\ \addlinespace
        & $(1,0;1)$ & $19/4$ & $-\frac{1}{3}\bigl(15 - 4\sqrt{11}\+\bigr)$ & $\frac{3}{2} \pm \frac{1}{6} \bigl(15-4 \sqrt{11}\+\bigr)$ \\ \addlinespace
        & $(1,1;1)$ & $39/4$ & $\frac{1}{3} \bigl(2\sqrt{69} - 15\bigr)$ & $\frac{3}{2} \pm \frac{1}{6} \bigl(2 \sqrt{69}-15\bigr)$ \\ \addlinespace
        & $(0,2;0)$ & $10$ & $\frac{1}{3} \bigl(\sqrt{281} - 15\bigr)$ & $\frac{3}{2} \pm \frac{1}{6} \bigl(\sqrt{281}-15\bigr)$ \\ \addlinespace
        & $(2,0;2)$ & $12$ & $\frac{1}{3} \bigl(\sqrt{321} - 15\bigr)$ & $\frac{3}{2} \pm \frac{1}{6} \bigl(\sqrt{321}-15\bigr)$ \\ \addlinespace
        $\frac{1}{2}(i\slashed{D}_{1/2}^{(2)_-})$
        & $(2,0;0)$ & $6$ & $0$ & $\frac{3}{2}$ \\ \addlinespace
        & $(0,0;2)$ & $6$ & $0$ & $\frac{3}{2}$ \\ \addlinespace
        & $(1,1;1)$ & $39/4$ & $\frac{1}{3} \bigl(2 \sqrt{61}-13\bigr)$ & $\frac{3}{2} \pm \frac{1}{6} \bigl(2 \sqrt{61}-13\bigr)$ \\ \addlinespace
        & $(0,1;2)$ & $10$ & $\frac{1}{3} \bigl(\sqrt{249} - 13\bigr)$ & $\frac{3}{2} \pm \frac{1}{6} \bigl(\sqrt{249}-13\bigr)$
        \\ \addlinespace[\aboverulesep] \midrule
        Field ($RS^7$) & $(p,q;r)$ & $C_g$ & $M_R/m$ & $E_0$
        \\ \midrule
        $\frac{1}{2}(i\slashed{D}_{1/2}^{(1)_+})$
        & $(0,0;0)$ & $0$ & $-1$ & $\frac{3}{2} - \frac{1}{2}$ \\ \addlinespace
        & $(0,1;0)$ & $4$ & $\frac{1}{3}\bigl(\sqrt{161} - 12\bigr)$ & $\frac{3}{2} \pm \frac{1}{6} \bigl(\sqrt{161}-12\bigr)$ \\ \addlinespace
        & $(1,0;1)$ & $19/4$ & $\frac{4}{3} \bigl(\sqrt{11}-3\bigr)$ & $\frac{3}{2} \pm \frac{2}{3} \bigl(\sqrt{11}-3\bigr)$ \\ \addlinespace
        $\frac{1}{2}(i\slashed{D}_{1/2}^{(2)_+})$
        & $(1,0;1)$ & $19/4$ & $-\frac{2}{3}$ & $\frac{3}{2} \pm \frac{1}{3}$ \\ \addlinespace
        & $(2,0;0)$ & $6$ & $-\frac{1}{3}$ & $\frac{3}{2} \pm \frac{1}{6}$ \\ \addlinespace
        & $(1,1;1)$ & $39/4$ & $\frac{2}{3} \bigl(\sqrt{61}-7\bigr)$ & $\frac{3}{2} \pm \frac{1}{3} \bigl(\sqrt{61}-7\bigr)$ \\ \addlinespace
        & $(0,1;2)$ & $10$ & $\frac{1}{3} \bigl(\sqrt{249} - 14\bigr)$ & $\frac{3}{2} \pm \frac{1}{6} \bigl(\sqrt{249}-14\bigr)$ \\ \addlinespace
        & $(2,0;2)$ & $12$ & $1$ & $\frac{3}{2} + \frac{1}{2}$ \\ \addlinespace
        & $(2,1;0)$ & $12$ & $1$ & $\frac{3}{2} + \frac{1}{2}$
        \\ \addlinespace[\aboverulesep] \bottomrule
    \end{tabular}
    \caption{Spinors that can have both Dirichlet and Neumann boundary conditions.
    The special entries are of two kinds: 1) The one in the  $LS^7$  part of the table with $E_0 = \frac{3}{2}$, which have neither Dirichlet nor Neumann boundary conditions
    (see \cite{Breitenlohner:1982jf}). 2) The first and last entries in the $RS^7$ part of the table where only one sign in $E_0$ is given as
    dictated by the fact that the only singleton that exists
    belongs to the   $RS^7$ vacuum and is a singlet \cite{Nilsson:2018lof}.
    Note that Neumann boundary conditions for $\frac{1}{2}(i\slashed{D}_{1/2}^{(2)_-})$ do not respect supersymmetry in the left-squashed vacuum, see \cref{spectrum_summary}.
    All brackets are positive.}
    \label{tab:D_or_N:spinors}
\end{table}

\subsubsection{Marginal operators in the dual CFTs}%
\label{sec:marginal_ops_in_N=0}

Having obtained a detailed understanding of the spectrum of the non-supersymmetric right-squashed vacuum in the previous subsection, including
the ambiguities due to the many possible choices of boundary conditions, we can now analyse the question of marginal operators.

As a preamble we consider the left-squashed spectrum of fields and list the possible marginal operators. It should be noted that these operators cannot, of course,
lead to instabilities since stability is in this case guaranteed by supersymmetry. However, despite being part of supermultiplets,
the  boundary conditions of the fields involved, and hence their $E_0$ values, are not uniquely determined as we saw in the previous subsections.

The scalar $sp_2 \times sp_1^C$-singlet marginal operators, depending on the boundary conditions and even for boundary conditions that do not respect supersymmetry, in the left-squashed case are given in \cref{tab:LS7:marginals}.
With boundary conditions preserving supersymmetry, the only scalar singlet marginal operators are those guaranteed by supersymmetry, formed from the spinors with $E_0 = 3/2$ from the massless vector supermultiplets.

\begin{table}[H]
    \centering
    \begin{tabular}{llll}
        \toprule
        \#$\vec{1}_g^{p}$ & Field 1: $E_0$ & \dots &
        \\ \midrule
        $3^+$
        & $0^+_{\scriptscriptstyle (-)} (\Delta_0^{(1)})_{(2,1;2)}^- \colon\; 1$
        & $0^+_{\scriptscriptstyle (-)} (\Delta_0^{(1)})_{(2,1;2)}^- \colon\; 1$
        & $0^+_{\scriptscriptstyle (-)} (\Delta_0^{(1)})_{(2,1;2)}^- \colon\; 1$
        \\ \addlinespace
        $1^+$
        & $0^+_{\scriptscriptstyle (-)} (\Delta_0^{(1)})_{(0,3;0)}^- \colon\; 1$
        & $0^+_{\scriptscriptstyle (-)} (\Delta_0^{(1)})_{(2,1;2)}^- \colon\; 1$
        & $0^+_{\scriptscriptstyle (-)} (\Delta_0^{(1)})_{(2,1;2)}^- \colon\; 1$
        \\ \addlinespace
        $2^-$
        & $0^+_{\scriptscriptstyle (-)} (\Delta_0^{(1)})_{(2,1;2)}^- \colon\; 1$
        & $0^- (Q^{(3)_-})_{(1,1;1)}^- \colon\; \frac{7}{2} - \sqrt{\frac{23}{3}}$
        & $0^+_{\scriptscriptstyle (-)} (\Delta_0^{(1)})_{(1,1;1)}^- \colon\; \sqrt{\frac{23}{3}} - \frac{3}{2}$
        \\ \addlinespace
        $1^-$
        & $0^+_{\scriptscriptstyle (-)} (\Delta_0^{(1)})_{(0,3;0)}^- \colon\; 1$
        & $0^- (Q^{(3)_-})_{(1,1;1)}^- \colon\; \frac{7}{2} - \sqrt{\frac{23}{3}}$
        & $0^+_{\scriptscriptstyle (-)} (\Delta_0^{(1)})_{(1,1;1)}^- \colon\; \sqrt{\frac{23}{3}} - \frac{3}{2}$
        \\ \addlinespace
        $1^-$
        & $0^+_{\scriptscriptstyle (-)} (\Delta_0^{(1)})_{(2,1;2)}^- \colon\; 1$
        & $0^- (Q^{(3)_-})_{(2,0;2)}^- \colon\; \frac{7}{2} - \frac{1}{6}  \sqrt{321}$
        & $0^+_{\scriptscriptstyle (-)} (\Delta_0^{(1)})_{(2,0;2)}^- \colon\; \frac{1}{6} \sqrt{321} - \frac{3}{2}$
        \\ \addlinespace
        $1^+$
        & $\frac{1}{2} (i\slashed{D}_{1/2}^{(2)_-})_{(2,0;0)} \colon\; \frac{3}{2}$
        & $\frac{1}{2} (i\slashed{D}_{1/2}^{(2)_-})_{(2,0;0)} \colon\; \frac{3}{2}$ &
        \\ \addlinespace
        $1^+$
        & $\frac{1}{2} (i\slashed{D}_{1/2}^{(2)_-})_{(0,0;2)} \colon\; \frac{3}{2}$
        & $\frac{1}{2} (i\slashed{D}_{1/2}^{(2)_-})_{(0,0;2)} \colon\; \frac{3}{2}$ &
        \\ \addlinespace[\aboverulesep]
        \bottomrule
    \end{tabular}
    \caption{Possible scalar $g=sp_2\times sp_1^C$ singlet marginal operators with parity $p$ in $LS^7$ depending on the boundary conditions.
    The subscripts after the eigenvalues indicate the isometry irreps $(p,q;r)$ while the superscript indicates
    the boundary conditions when applicable ($+$ for Dirichlet, $-$ for Neumann). The leftmost column gives the number of $g$-singlets in the
    composite. Note that only the ones containing two spinors are compatible  with the boundary conditions that preserve $\mN = 1$ supersymmetry.}
    \label{tab:LS7:marginals}
\end{table}

Turning to the more interesting case of the right-squashed non-supersymmetric vacuum, there is a lot more freedom in choosing
boundary conditions leading to a much longer list of marginal operators. We present these in \cref{tab:RS7:marginals}.
There are several features of this table that deserve a comment. We use here the language of the boundary $CFT_3$ theory although the notation is from the bulk theory.

First, we note that in the right-squashed vacuum there are no marginal single- or double-trace
operators,%
\footnote{This fact might be of some importance, see for instance \cite{Giombi:2017mxl}  for the role of such operators.}
while  both marginal triple- and quadruple-trace operators occur. Secondly, as observed in \cite{Nilsson:2023ctq},
in the singlet sector the two possible boundary conditions for the pseudo-scalar generate
either two marginal operators or none  (see \cref{tab:RS7:marginals}). This phenomenon generalises to the entire spectrum. In other words, instead of choosing the Neumann boundary conditions on which almost all of
\cref{tab:RS7:marginals} is based, we can use Dirichlet boundary conditions when possible and thereby eliminate all of the marginal operators.
With such boundary conditions,
the theory cannot be afflicted by any instability issues related to scalar singlet marginal operators.

Although they seem possible, it remains to be seen if such choices of boundary conditions can be made freely
or if they are constrained in some way. In any case, although this may seem like a step towards proving stability
it is just a small one since there may exist a number of other decay modes we are still ignorant about.

\begin{table}[H]
    \centering
    \begin{tabular}{llll}
        \toprule
        \#$\vec{1}_g^{p}$ & Field 1: $E_0$ & \dots &
        \\ \midrule
        $1^+$
        & $0^- (Q^{(2)_+})_{(2,0;0)}^- \colon\; \frac{2}{3}$
        & $0^- (Q^{(2)_+})_{(2,0;0)}^- \colon\; \frac{2}{3}$
        & $0^+ (\Delta_L^{(2)_-})_{(0,0;0)}^+ \colon\; \frac{5}{3}$
        \\ \addlinespace
        $1^-$
        & $0^- (Q^{(2)_+})_{(2,0;0)}^- \colon\; \frac{2}{3}$
        & $0^- (Q^{(2)_+})_{(2,0;0)}^- \colon\; \frac{2}{3}$
        & $0^- (Q^{(1)_+})_{(0,1;0)}^+ \colon\; \frac{5}{3}$
        \\ \addlinespace
        $1^+$
        & $0^- (Q^{(1)_+})_{(1,1;1)}^- \colon\; \frac{5}{6}$
        & $0^- (Q^{(1)_+})_{(1,1;1)}^- \colon\; \frac{5}{6}$
        & $0^+ (\Delta_L^{(2)_-})_{(0,0;0)}^- \colon\; \frac{4}{3}$
        \\ \addlinespace
        $1^-$
        & $0^- (Q^{(1)_+})_{(1,1;1)}^- \colon\; \frac{5}{6}$
        & $0^- (Q^{(1)_+})_{(1,1;1)}^- \colon\; \frac{5}{6}$
        & $0^- (Q^{(1)_+})_{(0,1;0)}^- \colon\; \frac{4}{3}$
        \\ \addlinespace
        $1^-$
        & $0^- (Q^{(1)_+})_{(1,0;1)}^- \colon\; \frac{5}{6}$
        & $0^- (Q^{(1)_+})_{(1,1;1)}^- \colon\; \frac{5}{6}$
        & $0^- (Q^{(1)_+})_{(0,1;0)}^- \colon\; \frac{4}{3}$
        \\ \addlinespace
        $1^+$
        & $0^- (Q^{(1)_+})_{(1,0;1)}^- \colon\; \frac{5}{6}$
        & $0^- (Q^{(1)_+})_{(1,0;1)}^- \colon\; \frac{5}{6}$
        & $0^+ (\Delta_L^{(2)_-})_{(0,0;0)}^- \colon\; \frac{4}{3}$
        \\ \addlinespace
        $1^-$
        & $0^- (Q^{(1)_+})_{(1,0;1)}^- \colon\; \frac{5}{6}$
        & $0^- (Q^{(1)_+})_{(1,0;1)}^- \colon\; \frac{5}{6}$
        & $0^- (Q^{(1)_+})_{(0,1;0)}^- \colon\; \frac{4}{3}$
        \\ \addlinespace
        $1^-$
        & $0^- (Q^{(3)_+})_{(0,0;0)}^- \colon\; 1$
        & $0^- (Q^{(3)_+})_{(0,0;0)}^- \colon\; 1$
        & $0^- (Q^{(3)_+})_{(0,0;0)}^- \colon\; 1$
        \\ \addlinespace
        $1^-$
        & $0^- (Q^{(3)_+})_{(0,0;0)}^- \colon\; 1$
        & $0^+_{\scriptscriptstyle (-)} (\Delta_0^{(1)})_{(0,3;0)}^- \colon\; 1$
        & $0^+_{\scriptscriptstyle (-)} (\Delta_0^{(1)})_{(0,3;0)}^- \colon\; 1$
        \\ \addlinespace
        $1^-$
        & $0^- (Q^{(3)_+})_{(0,0;0)}^- \colon\; 1$
        & $0^+_{\scriptscriptstyle (-)} (\Delta_0^{(1)})_{(2,1;2)}^- \colon\; 1$
        & $0^+_{\scriptscriptstyle (-)} (\Delta_0^{(1)})_{(2,1;2)}^- \colon\; 1$
        \\ \addlinespace
        $1^+$
        & $0^+_{\scriptscriptstyle (-)} (\Delta_0^{(1)})_{(0,3;0)}^- \colon\; 1$
        & $0^+_{\scriptscriptstyle (-)} (\Delta_0^{(1)})_{(2,1;2)}^- \colon\; 1$
        & $0^+_{\scriptscriptstyle (-)} (\Delta_0^{(1)})_{(2,1;2)}^- \colon\; 1$
        \\ \addlinespace
        $3^+$
        & $0^+_{\scriptscriptstyle (-)} (\Delta_0^{(1)})_{(2,1;2)}^- \colon\; 1$
        & $0^+_{\scriptscriptstyle (-)} (\Delta_0^{(1)})_{(2,1;2)}^- \colon\; 1$
        & $0^+_{\scriptscriptstyle (-)} (\Delta_0^{(1)})_{(2,1;2)}^- \colon\; 1$
        \\ \addlinespace
        $1^+$
        & $0^- (Q^{(2)_+})_{(2,0;0)}^- \colon\; \frac{2}{3}$
        & $0^- (Q^{(2)_+})_{(2,0;0)}^- \colon\; \frac{2}{3}$
        & \\
        & $0^- (Q^{(2)_+})_{(2,0;0)}^- \colon\; \frac{2}{3}$
        & $0^- (Q^{(3)_+})_{(0,0;0)}^- \colon\; 1$
        \\ \addlinespace
        $1^-$
        & $0^- (Q^{(2)_+})_{(2,0;0)}^- \colon\; \frac{2}{3}$
        & $0^- (Q^{(2)_+})_{(2,0;0)}^- \colon\; \frac{2}{3}$
        & \\
        & $0^- (Q^{(2)_+})_{(2,0;0)}^- \colon\; \frac{2}{3}$
        & $0^+_{\scriptscriptstyle (-)} (\Delta_0^{(1)})_{(0,3;0)}^- \colon\; 1$
        \\ \addlinespace
        $8^+$
        & $0^- (Q^{(2)_+})_{(2,0;0)}^- \colon\; \frac{2}{3}$
        & $0^- (Q^{(2)_+})_{(2,0;0)}^- \colon\; \frac{2}{3}$
        & \\
        & $0^- (Q^{(1)_+})_{(1,1;1)}^- \colon\; \frac{5}{6}$
        & $0^- (Q^{(1)_+})_{(1,1;1)}^- \colon\; \frac{5}{6}$
        \\ \addlinespace
        $4^+$
        & $0^- (Q^{(2)_+})_{(2,0;0)}^- \colon\; \frac{2}{3}$
        & $0^- (Q^{(2)_+})_{(2,0;0)}^- \colon\; \frac{2}{3}$
        & \\
        & $0^- (Q^{(1)_+})_{(1,1;1)}^- \colon\; \frac{5}{6}$
        & $0^- (Q^{(2)_+})_{(1,0;1)}^- \colon\; \frac{5}{6}$
        \\ \addlinespace
       $ 3^+$
        & $0^- (Q^{(2)_+})_{(2,0;0)}^- \colon\; \frac{2}{3}$
        & $0^- (Q^{(2)_+})_{(2,0;0)}^- \colon\; \frac{2}{3}$
        & \\
        & $0^- (Q^{(2)_+})_{(1,0;1)}^- \colon\; \frac{5}{6}$
        & $0^- (Q^{(2)_+})_{(1,0;1)}^- \colon\; \frac{5}{6}$
        \\ \addlinespace
        $1^-$
        & $0^- (Q^{(3)_+})_{(0,0;0)}^- \colon\; 1$
        & $\frac{1}{2} (i\slashed{D}_{1/2}^{(1)_+})_{(0,0;0)}^- \colon\; 1$
        & $\frac{1}{2} (i\slashed{D}_{1/2}^{(1)_+})_{(0,0;0)}^- \colon\; 1$
        \\ \addlinespace[\aboverulesep]
        \bottomrule
    \end{tabular}
    \caption{Possible scalar $g=sp_2 \times sp_1^C$ singlet marginal operators with parity $p$ in $RS^7$ depending on the boundary conditions, with notation as in \cref{tab:LS7:marginals}. Note that the lines without a number in the leftmost column are continuation lines.}
    \label{tab:RS7:marginals}
\end{table}

\resettocdepth
\resetsecnumdepth

\section{Conclusions and comments}%
\label{sec:conclusions}

In this paper, we have reported on technical progress on the computation of the eigenvalue spectra of operators
on the squashed seven-sphere relevant in the compactification of eleven-dimensional supergravity.
The full isometry irrep spectrum was derived in \cite{Nilsson:2018lof}, including the structure of  the $AdS_4$ supermultiplet spectrum for the left-squashed vacuum.
With the results of \cite{Karlsson:2021oxd} and this paper  on the solution of the spin-3/2 equation,
these supermultiplets can be associated with  values of $E_0$, pinpointing the $Spin(2,3)$-irrep $D(E_0, s)$, thus providing a full understanding of the left-squashed spectrum.%
\footnote{Note that, as explained in \cref{spectrum_summary}, we use results of \cite{Duboeuf:2022mam} to resolve some ambiguities that would otherwise appear for the supermultiplets denoted WZ3--6 for small $p$ or $q$.}
This is the case also after skew-whiffing to the right-squashed non-supersymmetric vacuum,
whose spectrum becomes fully determined up to a number of ambiguities connected to the possibility to  choose  either Dirichlet or Neumann   boundary conditions
for a number of fields. Some ambiguities of this kind exist also in the supersymmetric left-squashed case.
These results are summarised in \crefrange{spinonetotwomultiplets}{tab:RS7:marginals} of \cref{sec:supermultiplet}.

This progress was made possible by deriving more powerful versions of some key formulas used in solving operator equations on coset manifolds, see \cite{Karlsson:2021oxd} and \cref{sec:review_and_method},
here the squashed $S^7$ represented by the  coset $Sp_2\times Sp_1^C/Sp_1 \times Sp_1 $. Using these refined formulas, in particular \cref{masterweyl} and \cref{masterdelta},
we  finally  managed to solve
also the spin-3/2 equation. These results are summarised in \cref{sec:eigenvalues:summary} but they were reported on already in \cite{Karlsson:2021oxd} where some additional details can
be found compared to the presentation
in this paper.

The structure of the results is to a large extent as expected apart from some surprising features: We found degeneracies in  the spectrum of two-forms, in the sense that
different mode functions have identical operator eigenvalues.
We demonstrated this explicitly by constructing all eigenmodes of the operator $\Delta_2$,
providing  a direct means of calculating  the eigenvalues.
These novel results are presented in \cref{sec:eigenmodes}. This kind of degeneracy  appears in (almost) all supermultiplets that involve fields associated to the eigenmodes of the
Lichnerowicz operator on the squashed seven-sphere, as was shown in \cref{sec:supermultiplet}.

We also found  that supersymmetry alone is not enough to determine the boundary conditions for a number of supermultiplets.
This issue was discussed in \cref{sec:supermultiplet} where  the supermultiplet structure was presented in full detail.

The implications for the skew-whiffed  non-supersymmetric right-squashed seven-sphere compactification are discussed at the end of the previous section.
Here, the boundary condition ambiguities  appear for many  fields which implies that one can find a large number of scalar singlet marginal operators
in the boundary $CFT_3$.  This is of interest since
the right-squashed case  is BF-stable and has, in fact, not yet been shown to be unstable, although this is predicted by the swampland
$AdS$ swampland conjecture \cite{Ooguri:2016pdq,Freivogel:2016qwc}. However, in order to determine the fate of the right-squashed vacuum one must analyse the corresponding $\beta$ functions
to determine whether  a marginal operator leads to an instability or not. We leave this for a future study.
Interestingly enough, there are multiple sets of boundary conditions for which no scalar singlet marginal operators exists.
This phenomenon was recently observed in \cite{Nilsson:2023ctq} to
be a property of the singlet sector of the spectrum.%
\footnote{Generally, we believe that Dirichlet boundary conditions can always be chosen in non-supersymmetric vacua such that the only possible scalar singlet marginal operators are single-trace operators or multi-trace operators dual to composites containing a singleton or fields with $E_0 = 3/2$.}
Another property of the space of singlet scalar marginal operators found in \cref{sec:marginal_ops_in_N=0} is that there are none that are either single- or double-trace.
The implications of this fact might be worth investigating.

The squashed spectra clearly also imply that the boundary theories contain a large number of relevant operators which can generate RG-flows to other CFTs.
For instance, a double-trace operator constructed as the square of a single-trace operator dual to a scalar field with Neumann boundary conditions generates a flow that leads, in the infrared, to the dual scalar field having Dirichlet boundary conditions \cite{Witten:2001ua}.
It would be interesting to investigate whether there are RG-flows connecting the boundary conditions preserving supersymmetry in the left-squashed vacuum.
Another interesting potential RG-flow is one that corresponds to a domain-wall interpolating between either of the squashed vacua and the round $S^7$ vacuum in the bulk, as discussed, but not found explicitly, in \cite{Ahn:1999dq,Campos:2000yu} (see also, e.g., \cite{Witten:2001ua,Papadimitriou:2007sj,Giombi:2017mxl}) or one that connects the left- and right-squashed vacua.

The operator corresponding to the field associated with the squashing mode (the $\Delta_L$-row in \cref{tab:supermultiplets:singlet-summary}) is relevant in the squashed vacuum and irrelevant in the round vacuum \cite{Nilsson:2023ctq}.
This indicates that an RG-flow connecting the two would have to start from the squashed vacuum in the UV and flow to the round one in the IR.
Another argument for this is provided by the F-theorem \cite{Jafferis:2010un,Jafferis:2011zi,Closset:2012vg}, which states that the $S^3$ free energy decreases along RG-trajectories.
Holographically, the free energy is given by%
\footnote{It should be understood that this, and what follows, is only valid to leading order in $N$ or $L^2/G_N$.}
\begin{equation}
    \mathcal{F}_{\! S^3} = I_{EAdS} = \frac{\pi L^2}{2 G_N},
\end{equation}
where $I_{EAdS}$ is the regularised on-shell action of Euclidean $AdS$, $L^2$ the $AdS$ curvature radius and $G_N$ the four-dimensional Newton's constant.
By dimensional reduction, we have $G_N \propto \ell_\mathrm{Pl}{}^9 / \operatorname{Vol}(M_7)$ where $\ell_\mathrm{Pl}$ is the eleven-dimensional Planck length.
The flux quantisation condition gives the number of M2-branes as $N \propto \operatorname{Vol}(M_7) / (\ell_\mathrm{Pl}{}^6 L)$.
Putting this together to eliminate $\ell_\mathrm{Pl}$, we get $\mathcal{F}_{\! S^3} \propto N^{3/2} L^{7/2} / \sqrt{\operatorname{Vol}(M_7)}$.
Lastly, computing the volume of the squashed $S^7$ and fixing the proportionality constant through the well-known round $S^7$ answer or by properly keeping track of all numerical factors in the quantisation condition and the dimensional reduction (see, e.g., \cite{Jafferis:2011zi}), we arrive at (c.f.\ \cite{Ahn:1999dq,Page:1983mke,Campos:2000yu})
\begin{equation}
    \mathcal{F}_{\! S^3}^{\+(\mathrm{round})} = \frac{\pi \sqrt{2}}{3} N^{3/2},
    \qquad\quad
    \mathcal{F}_{\! S^3}^{\+(\mathrm{squashed})} = \frac{\pi \sqrt{2}}{3} \sqrt{\frac{5^5}{3^7}}\,  N^{3/2}.
\end{equation}
Note that the latter is about $20\+\%$ larger than the former, consistent with the above assertion that the RG-flow, if it exists, goes from squashed in the UV to round in the IR.

It would also be interesting to investigate if the marginal operators found in this paper can be used to argue holographically for a manifold of non-supersymmetric dual CFTs, in line with \cite{Witten:2001ua}, which requires computing the $\beta$ functions of the marginal operators.
However, no solid conclusions can be drawn without incorporating $1/N$ or finite $N$ effects, which currently seems out of reach.

The methods and results of this paper can also be used to analyse compactifications on $S^7/Z_k$. Some $AdS/CFT$ aspects  in this
context are discussed in, e.g., \cite{Ooguri:2008dk,Murugan:2016aty}.

From the results of this paper it is  clear that  all values of $E_0$ in the squashed vacua are of the form
\begin{equation}
    E_0=\biggl(\frac{3}{2}\pm A\biggr)+\frac{1}{6}\sqrtsmash[C_G]{20C_g+X},\qquad
    X=1,\,9,\,49,\,\text{or}\, 81,
\end{equation}
with a constant $A$ that is also restricted to a small number of special rational values. The simplicity of this result is striking,
especially in view of how the various values of $E_0$ arose, potentially containing three layers of nested square roots,
one each coming from the operator eigenvalues, the mass formulas and finally from the formulas for $E_0(s)$ as functions of $M$ or $M^2$.
Even more intriguing is the fact that also the operator eigenvalue spectra
are given by  similar formulas, either with or without a  square root  term.
From \cref{sec:eigenvalues} we see that the general forms of the eigenvalues of operators quadratic  in derivatives are
\begin{equation}
    \kappa^2=\frac{m^2}{9}(20C_g+B),
    \qquad
    \kappa^2=\frac{m^2}{9} \Bigl(20C_g+\tilde A \pm \tilde B\sqrtsmash[C_G]{20C_g+X}\Bigr),
\end{equation}
while for linear operators, the eigenvalues are all of the form
\begin{equation}
    \kappa=\frac{m}{3} \Bigl(\pm \hat A \pm \sqrtsmash[C_G]{20C_g+X}\Bigr).
\end{equation}
It is interesting to note, however, that all eigenvalues of the form $\kappa^2=\frac{m^2}{9}(20C_g+\tilde A \pm \tilde B\sqrtsmash[C_G]{20C_g+X})$ can be
expressed as a square minus a constant term proportional to $m^2$ which immediately eliminates the square root in the corresponding formula for $M^2$,
as was noticed already in  \cite{Ekhammar:2021gsg}. In fact, looking at the table of mass matrices in \cref{sec:review_and_method}, we realise that the same phenomenon will
appear when using the mass formulas in the expressions for $E_0$. The reason behind the universal simple results for $E_0$ is then clear.
This might nevertheless point towards the existence of a more direct derivation of  $E_0$.

In this context we should mention that the expressions for
$E_0$ for the various supermultiplets was recently derived  by entirely different methods in \cite{Duboeuf:2022mam}, giving in addition a refined version of the
results in terms of a universal formula for $E_0$. Interestingly enough, these methods do not involve solving the operator equations on the squashed seven-sphere but instead rely on ExFT methods.
The results of this paper partially overlap with those of \cite{Duboeuf:2022mam}, in particular, the $E_0$ results that were derived explicitly for finitely many levels in \cite{Duboeuf:2022mam}.
Note also that the many ambiguities in the  choice of Dirichlet or Neumann boundary conditions we have found are not discussed in \cite{Duboeuf:2022mam}.

We end with an observation concerning the degeneracies mentioned above. There are three pairs of identical supermultiplets,
all containing a scalar sector governed by $\Delta_L$, one pair of $1^+$ and two pairs of Wess--Zumino multiplets.
We know from explicit calculations of the  spectrum of $\Delta_2$-eigenmodes in \cref{sec:eigenmodes}, which gives also $E_0$,
that an infinite set of pairs of  different eigenmodes (with $r=p$ or $r=p\pm 2 $) with  identical isometry irreps appear with the same eigenvalues and, thus, also the same $E_0$ for the corresponding fields. We do not have an
explanation for this feature but hope to come back to it in a future publication.

Instead, we point out some facts related to the operator  $\Delta_L$ which is involved in all these cases.
It is known from the work of House and Micu \cite{House:2004pm} that $\Delta_L$ is related to a first order differential operator, here denoted as $Q_L$, by
\begin{equation}
    \Delta_L-3m^2=(Q_L+m)^2,
\end{equation}
which implies
\begin{equation}
    M^2(0^{+(2)})=\Delta_L-4m^4=(Q_L+m)^2-m^2 \ \implies\
    E_0(0^{+(2)},Q_L)=\frac{3}{2}\pm \frac{1}{2} \biggl|\frac{Q_L}{m}+1\biggr|.
\end{equation}
Note that this seems to imply that $\Delta_L \ge 3m^2$ which, however, must be supplemented by a precise statement about which modes of the two
operators the  House--Micu equation  is valid for. This is a rather non-trivial issue as we will indicate below.

One might interpret this to mean that the three  known eigenvalues of $\Delta_L$ could be used twice, that is, that we should be able to use
the positive and negative parts of the spectrum of $Q_L$ when taking  the square root of  the House--Micu formula.
Thus we consider
\begin{align}
\label{eq:conclusions:QL:a}
    \Delta_L^{(1)} = \frac{m^2}{9}(20 C_g+36)
    &\implies
    Q_L=-m\pm\frac{m}{3}\sqrtsmash[C_G]{20C_g+9},
    \\
\label{eq:conclusions:QL:b}
    \Delta_L^{(2)_+} = \frac{m^2}{9} \Bigl(20 C_g+32 + 4\sqrtsmash[C_G]{20C_g+1}\Bigr)
    &\implies
    Q_L=-m \pm \biggl(\frac{2}{3}m + \frac{m}{3}\sqrtsmash[C_G]{20C_g+1} \biggr),
    \\
\label{eq:conclusions:QL:c}
    \Delta_L^{(2)_-} = \frac{m^2}{9} \Bigl(20 C_g+32 - 4\sqrtsmash[C_G]{20C_g+1}\Bigr)
    &\implies
    Q_L=-m \pm \biggl|-\frac{2}{3}m + \frac{m}{3}\sqrtsmash[C_G]{20C_g+1} \biggr|.
\end{align}
However, this interpretation has a number of problems. First, $Q_L=-m\pm\frac{m}{3}\sqrtsmash[C_G]{20C_g+9}$ is not a possible set of
eigenvalues for an operator like $Q$ acting on transverse three-forms as we saw in \cref{sec:eigenvalues}.
One issue here is that although the  irrep ${\vec 27}$ describes both
the symmetric second rank tensors and part of the three-forms,  the transversality conditions act differently on these tensors making the House--Micu formula difficult to apply.
For instance, at least one of the sets of suggested $Q_L$-eigenvalues in \cref{eq:conclusions:QL:a} could instead be relevant for non-transverse three-form modes.
Furthermore, only one of the two signs in \cref{eq:conclusions:QL:b,eq:conclusions:QL:c} give existing $Q$-eigenvalues and these are precisely the eigenvalues of $Q$ that correspond to eigenmodes with only a $\vec{27}$-part and that are used by the pseudo-scalars in the supermultiplets using $\Delta_L^{(2)_\pm}$, i.e., WZ3--6.
So, $Q$ does not resolve the analogous degeneracies of $\Delta_3$ and since these $Q$-eigenvalues, which should also be $Q_L$-eigenvalues, have cross diagrams that already cover all isometry irreps of these supermultiplets, it seems like $Q_L$ cannot resolve the degeneracies either.

\section*{Acknowledgements}
J.K.\ thanks Junho Hong, Emanuel Malek, Henning Samtleben, Thomas Van Riet and, in particular, Nikolay Bobev for useful comments and discussions.
The second author, B.E.W.N., is grateful to Mike Duff and Chris Pope for many discussions and
collaboration at an early stage of this work.
We are also grateful to Igor Klebanov and the referee for discussions and comments on parts of the manuscript.
J.K.\ is supported by the Research Foundation - Flanders (FWO) doctoral fellowship 1171823N and, in part, by the FWO research grants G092617N and G0H9318N.
B.E.W.N.\ is partly supported by the Wilhelm and Martina Lundgren Foundation.

\appendix
\section{Conventions on octonions, gamma-matrices and \texorpdfstring{$Sp_1^C$}{Sp1C} Killing vectors}
\label{app:octonions}

The octonions are elements of a non-associative algebra spanned by a real unit $1$ and octonionic imaginary units $o_a$, where $a=1,\hdots,7$, with multiplication
\begin{equation}
    o_a o_b = - \delta_{ab} + a_{ab}{}^c o_c,
\end{equation}
where $a_{abd}$ (index lowered by $\delta_{cd}$) are the completely antisymmetric octonionic structure constants with independent components
\begin{equation}
    a_{abc}=1 \quad\text{for } abc=123, 257, 365, 437, 761, 642, 415,
\end{equation}
in a particular basis.
By splitting the index as $a=(\hat i, 0, i)$, where $\hat i=(\hat 1, \hat 2, \hat 3)=(1,2,3)$ and $i=(5,6,7)$, the non-zero components of $a_{abc}$ can conveniently be summarised as follows%
\footnote{The indices $i,j,k$ should be shifted to $(1,2,3)$ to make sense of \cref{eq:octonions:a_split}.}
\begin{equation}
\label{eq:octonions:a_split}
    a_{\hat i \hat j \hat k}=\epsilon_{ijk},\qquad
    a_{ij \hat k}=-\epsilon_{ijk},\qquad
    a_{0i \hat j}=-\delta_{ij},
\end{equation}
in terms of the standard totally antisymmetric  $\epsilon_{ijk}$ and the Kronecker delta $\delta_{ij}$.
The Hodge dual of $a_{abc}$, denoted $c_{abcd}$, is
\begin{equation}
    c_{abcd} = \frac{1}{6} \epsilon_{abcdefg}\+ a^{efg},
\end{equation}
where the seven-dimensional epsilon tensor is totally antisymmetric with $\epsilon_{1234567}=1$. This implies
that $c_{abcd}$ has non-zero components
\begin{equation}
    c_{0 i j k }=\epsilon_{ijk},\qquad
    c_{0 \hat i \hat j k }=-\epsilon_{ijk},\qquad
    c_{\hat i \hat j}{}^{kl}=-2\delta_{ij}^{kl}.
\end{equation}

It is sometimes   convenient  to express the seven-dimensional  gamma-matrices $\Gamma_a$
in terms of $a_{abc}$, $c_{abcd}$ and the Killing spinor  of the squashed seven-sphere, $\eta$, satisfying%
\footnote{The octonions can be identified with $Spin(7)$ spinors and the gamma-matrices $\Gamma_a$ realised through multiplication by $o_a$. $G_2$ is the automorphism group of the octonions or, equivalently, the subgroup of $Spin(7)$ leaving the unit octonion $1$, identified with the spinor $\eta$, invariant. See, e.g., \cite{Karlsson:2021oxd} for more details on this perspective and how to derive some of the following relations using it.}
\begin{equation}
\label{eq:octonions:eta_def_properties}
    D_a \eta = -\frac{i}{2}m\+ \Gamma_a \eta,\qquad
    \bar\eta \eta=1.
\end{equation}
The octonionic conventions above lead then to the relations
\begin{equation}
    a_{abc}=i\bar\eta\Gamma_{abc}\eta,\qquad
    c_{abcd}=-\bar\eta\Gamma_{abcd}\eta,
\end{equation}
which are consistent with the gamma-matrices satisfying their defining relations
\begin{equation}
\label{eq:conventions:gamma_def-properties}
    \{\Gamma_a,\Gamma_b\}=2\delta_{ab},\qquad
    \Gamma_{abcdefg}=i\epsilon_{abcdefg}\bf{1}.
\end{equation}
The fact that the squashed seven-sphere is a \emph{weak} $G_2$ manifold is then encoded in the equation
\begin{equation}
    D_a a_{bcd}=m\+ c_{abcd},
\end{equation}
which is a direct consequence of the above relations with the octonionic structure constants and  the Killing spinor equation.

The consistency of these conventions can be verified by explicitly writing out the seven $8\times 8$ gamma-matrices, with spinor indices $A,B,C,\hdots$ taking the eight values  $A=(\hat 0, a)$, as follows:%
\footnote{Note that vector $a,b,c,...$ and spinor $A, B, C, ...$ indices
are raised and lowered by $\delta_{ab}$ and $\delta_{AB}$ (and their inverses), respectively.}
\begin{equation}
    (\Gamma_a)_B{}^C\colon \qquad
    (\Gamma_a)_b{}^c=ia_{ab}{}^{c},\quad
    (\Gamma_a)_b{}^{\hat 0}=-i\delta_{ab}.
\end{equation}
Note that all seven $\Gamma_a$ are purely imaginary and antisymmetric and that they satisfy \cref{eq:conventions:gamma_def-properties}.

To derive and simplify the components of $\Gamma_{ab}$ and $\Gamma_{abc}$ we need some structure constant identities.
We list these and, for the convenience of the reader, some other structure constant identities that are used heavily throughout the paper.
We have
\begin{align}
\label{eq:octonoins:Fierz:first}
    & a_{abc}\+ a^{cde}=2\delta_{ab}^{de}+c_{ab}{}^{de},\\
    & a_{abc}\+ a^{bcd}=6\delta_a^d,\\
    & a_{abc}\+ a^{abc}=42.
\end{align}
The analogous identities involving $c_{abcd}$ are
\begin{align}
    & c_{abcd}\+ c^{defg} = a_{abc}\+ a^{efg}-9c_{[ab}{}^{[ef}\delta_{c]}^{g]}-6\delta_{abc}^{efg},\\
    & c_{abcd}\+ c^{cdef}=8\delta_{ab}^{ef}+2c_{ab}{}^{ef},\\
    & c_{abcd}\+ c^{bcde}=-24\delta_a^e,\\
    & c_{abcd}\+ c^{abcd}=168.
\end{align}
Some identities involving contractions between the $a_{abc}$ and $c_{abcd}$ structure constants are:
\begin{align}
    & c_{abcd}\+ a^{def}=6a_{[ab}{}^{[e}\delta_{c]}^{f]},\\
    & c_{abcd}\+ a^{cde}=4a_{ab}{}^e,\\
    & c_{abcd}\+ a^{bcd}=0.
\end{align}
Two other identities that have been used in this work are
\begin{align}
\label{eq:octonoins:Fierz:last}
    \delta_{a b}
    &= \frac{1}{4!}a_{a}{}^{cd} a_{b}{}^{ef} c_{cdef},\\[2pt]
    a_{[ab}{}^{[d}a_{c]}{}^{ef]}
    &= \frac{1}{3}a_{abc}a^{def}-2c_{[ab}{}^{[de}\delta_{c]}^{f]}.
\end{align}
These and the following identities may be verified directly by using the components in the basis given above. By using the Fierz identity of the $G_2$-invariant $\eta$,
\begin{equation}
    \Gamma_a \eta \bar{\eta} \Gamma^a = 1 - \eta \bar{\eta},
\end{equation}
one can also prove \crefrange{eq:octonoins:Fierz:first}{eq:octonoins:Fierz:last} more elegantly.

Using the octonionic definition of the gamma-matrices
$(\Gamma_{a})_{B}{}^{C}$ above we can express also the antisymmetric matrices $(\Gamma_{ab})_{CD}$ this way. We find
\begin{equation}
    (\Gamma_{ab})_C{}^D\colon\qquad
    (\Gamma_{ab})^{cd}=-c_{ab}{}^{cd}+2\delta_{ab}^{cd},\quad
    (\Gamma_{ab})_c{}^{\hat 0}=-a_{abc}.
\end{equation}
Finally, for the symmetric matrices $(\Gamma_{abc})_{DE}$ we have
\begin{align}
\nonumber
    (\Gamma_{abc})_D{}^E\colon\qquad
    & (\Gamma_{abc})^{de}= -ic_{[ab}{}^{ef}a_{c]f}{}^{d} - ia_{[ab}{}^e \delta_{c]}^{d} - 2ia_{[ab}{}^d \delta_{c]}^{e},\\
\nonumber
    & (\Gamma_{abc})_d{}^{\hat 0}=-ic_{abcd},\\
\label{gammathreeoct}
    & (\Gamma_{abc})_{\hat 0}{}^{\hat 0}=-ia_{abc}.
\end{align}
The first one of these expressions can be simplified further using the identity (note the order of the indices)
\begin{equation}
    c_{[ab|d|}{}^{f}a_{c]f}{}^{e} = 6 \bigl(\delta_{[a}^{[c'}a_{bd]}{}^{e]} \delta_{c c'}\bigr) \big|_{[abc]}.
\end{equation}
Then, the required symmetry property in $de$ becomes clear:
\begin{equation}
    (\Gamma_{abc})^{de}=i\bigl(a_{abc}\delta^{de}-6\delta^{(d}_{[a}a_{bc]}{}^{e)}\bigr).
\end{equation}
Finally, we see that $\operatorname{Tr}(\Gamma^{abc})=0$ as it must.

The relations above show that this representation of the gamma-matrices  is consistent with the definition of the structure constants $a_{abc}$ in
terms of the Killing spinor $\eta$.
To clarify this further  we can solve the Killing spinor equation and get an explicit form of this spinor. We know from the left-squashed $S^7$ compactification that the
holonomy is $G_2$  and hence that the Killing spinor equation has one solution. Normalising it to $\bar \eta \eta=1$
we for now assume that, in the above basis, $\eta$ has only one non-zero real constant component. Then
\begin{align}
\nonumber
    &D_a \eta_B = - \frac{i}{2} m (\Gamma_a)_B{}^C \eta_C\colon \\
\nonumber
    B=b\colon\quad
    &\partial_a \eta_b + \frac{1}{4} \omega_{ade} \bigl((\Gamma^{de})_b{}^c \eta_c + (\Gamma^{de})_b{}^{\hat 0} \eta_{\hat 0}\bigr)
    = -\frac{i}{2}m \bigl((\Gamma_a)_{b}{}^{c} \eta_c + (\Gamma_a)_{b}{}^{{\hat 0}} \eta_{\hat 0}\bigr) \\[2pt]
    B=\hat{0}\colon\quad
    &\partial_a \eta_{\hat 0} + \frac{1}{4}\omega_{ade} (\Gamma^{de})_{\hat 0}{}^c \eta_c = -\frac{i}{2}m (\Gamma_a)_{\hat 0}{}^c \eta_c
\end{align}
It is clear from these two equations that a spinor $\eta$ with only a constant non-zero component $\eta_{\hat 0}$ has a chance to be a solution. Assuming this to be the case, the second equation is solved and the first becomes the single
condition on the spin connection
\begin{equation}
    \frac{1}{4}\omega_a{}^{de}a_{bde} = \frac{m}{2}\delta_{ab}\quad\implies\quad
    \hat\omega_{abc}=\frac{m}{3}a_{abc},
\end{equation}
where the hat indicates that there might be (and are in case of the squashed seven-sphere) other terms that vanish when contracted with $a_{abc}$ this way.
This condition is in fact satisfied but the easiest way to verify the assumed form of the Killing spinor is  instead to consider the
integrability condition
\begin{equation}
\label{eq:octonions:integrability}
    W_{ab}{}^{cd}\Gamma_{cd}\eta = 0,
\end{equation}
where $W_{ab}{}^{cd}$ is the Weyl tensor and the only linearly independent solution is
\begin{equation}
    \eta_A=(1,0,....,0).
\end{equation}
Note that this explicit form of $\eta$ and the gamma-matrices is consistent with $\bar\eta \Gamma_{abc} \eta=-ia_{abc}$.
One can check that this $\eta$ solves not only the integrability condition but also the Killing spinor equation on the left-squashed seven-sphere but not on the right-squashed one \cite{Awada:1982pk,Duff:1983ajq}.
That the integrability condition \cref{eq:octonions:integrability} has exactly one independent solution is equivalent to the fact that the holonomy of the Killing spinor equation is $G_2$, i.e., $\frac{1}{4} W_{ab}{}^{cd} \Gamma_{cd}$ generates the Lie algreba $g_2$.

There are also identities where structure constants are contracted  with the $Sp_1^C$ Killing vector components
\begin{equation}
    s_a{}^j
    =(s_{\hat i}{}^j,\; s_0{}^j,\; s_i{}^j)
    =\biggl( \frac{1}{\sqrt{5}}\delta^j_{i},\; 0,\; 0\biggr).
\end{equation}
The $Sp_1^C$ Killing vectors are used heavily
in the mode calculations in \cref{sec:eigenmodes}. From the explicit realisation of  $a_{abc}$ and $c_{abcd}$ given above one finds the following identities:
\begin{align}
    & a_{a}{}^{bc} s_b{}^i s_c{}^j = \frac{1}{\sqrt{5}} \epsilon^{ij}{}_k s_a{}^k,\\
    & c_{ab}{}^{cd} s_c{}^i s_d{}^j = \frac{1}{\sqrt{5}} \epsilon^{ij}{}_k a_{ab}{}^c s_c{}^k - 2 s_{[a}{}^i s_{b]}{}^j,\\
    & s_{[a|}{}^i s^d{}_{i} a_{d|bc]} = \frac{1}{15} a_{abc} + \frac{2 \sqrt{5}}{3} \epsilon_{ijk} s_a{}^i s_b{}^j s_c{}^k,
\end{align}
In addition, there is an important derivative relation  given in \cite{Nilsson:1983ru} which in our notation  reads
\begin{equation}
    D_a s_b{}^i = \frac{1}{2\sqrt{5}} a_{ab}{}^c s_c{}^i -3 \epsilon^i{}_{jk} s_a{}^j s_b{}^k.
\end{equation}

\section{\texorpdfstring{$G_2$}{G2} and \texorpdfstring{$H$}{H} projection operators and Casimirs}
\label{app:gtwoprojectors}

Projection operators picking out the various $G_2$-irreps occurring in the decompositions of two-forms, three-forms, spinors and vector-spinors on $S^7$ under $Spin(7)\rightarrow G_2$ are, for the convenience of the reader, given below (the other $so(7)$-irreps of interest are also $g_2$-irreps and so do not need to be decomposed).
For spinors, one- and two-forms, we also give the projection operators onto the $H$-irreps occurring when decomposing $Spin(7) \rightarrow G_2 \rightarrow H$.
This is relevant for \cref{sec:eigenmodes}, in particular \cref{sec:eigenmodes:two-forms} since $C_h$ is not proportional to the identity on $\vec{14}_{g_2} \subset \vec{21}_{so(7)}$.
In \cref{app:gtwoprojectors:Casimirs}, we give our conventions for the Casimirs of the various groups.

\subsection{Projection operators}
\subsubsection{One-forms}
The one-form irrep $\vec{7}$ of $so(7)$ does not decompose when breaking $so(7)$ to $g_2$.
However, when breaking to $h = sp_1^A \oplus sp_1^{B+C} \subset g_2$, it decomposes as $\vec{7} \rightarrow (0,2) \oplus (1,1)$.
Since $s_a{}^i s^b{}_i$ is a $H$-singlet and has rank three when seen as an operator acting on one-forms, we have
\begin{equation}
\label{eq:projector:one-form:H_first}
    (P_{(0,2)})_a{}^b = 5 s_a{}^i s^b{}_i,
\end{equation}
where we have fixed the coefficient by demanding $P_{(0,2)}{}^2 = P_{(0,2)}$.
It follows that the projector onto $(1,1)$ is
\begin{equation}
\label{eq:projector:one-form:H_second}
    (P_{(1,1)})_a{}^b = \delta_a^b - 5 s_a{}^i s^b{}_i.
\end{equation}

\subsubsection{Two-forms}
The two-form irrep $\vec{21}$ decomposes into $\vec{7} \oplus \vec{14}$ under $g_2$.
Since $a_a{}^{bc}$ can be viewed as an intertwiner $\vec{21} \to \vec{7}$, we have
\begin{align}
    & (P_7)_{ab}{}^{cd} = \frac{1}{6}a_{ab}{}^e a^{cde} = \frac{1}{6} (2\delta_{ab}^{cd} + c_{ab}{}^{cd}), \\
    & (P_{14})_{ab}{}^{cd} = (1 - P_7)_{ab}{}^{cd} = \frac{1}{6} (4\delta_{ab}^{cd} - c_{ab}{}^{cd}).
\end{align}
From this, we see that
\begin{equation}
    c_{ab}{}^{cd} = (4P_7 - 2P_{14})_{ab}{}^{cd},
\end{equation}
which is often useful in computations.

When restricting to $H$, the $\vec{7}$ decomposes as described in the one-form case above and the $\vec{14}$ decomposes into $(0,2) \oplus (2,0) \oplus (1,3)$.
The projection operators projecting onto the $H$-irreps in $\vec{7}$ can be constructed using $a_{abc}$ and the projectors from the one-form case.
For instance, we see that (c.f.\ \cref{eq:projector:one-form:H_first})
\begin{equation}
    (P_{(0,2)\subset\vec{7}})_{ab}{}^{cd}
    = \frac{5}{6} a_{ab}{}^e s_e{}^i s^f{}_{\! i}\+ a_f{}^{cd}
    = 75 (P_7 \varsigma P_7)_{ab}{}^{cd},
\end{equation}
where
\begin{equation}
    \varsigma_{ab}{}^{cd}
    = s_{[a|}{}^i s^{[c|}{}_i\+  s_{|b]}{}^j s^{|d]}_j
\end{equation}
is a useful building block for constructing the projectors.
One can verify that all four $g_2$-blocks of $\varsigma_{ab}{}^{cd}$, viewed as a linear operator acting on two-forms, that appear when decomposing $\vec{21} \to \vec{7} \oplus \vec{14}$ have rank three.
Since the $7 \times 7$ block is proportional to the projector $(P_{(0,2)\subset\vec{7}})_{ab}{}^{cd}$, the off-diagonal blocks intertwine $(0,2)\subset\vec{7}$ and $(0,2)\subset\vec{14}$ and the $14\times 14$ block is proportional to $(P_{(0,2)\subset\vec{14}})_{ab}{}^{cd}$.
Working out the normalisation, we find
\begin{equation}
    (P_{(0,2)\subset\vec{14}})_{ab}{}^{cd}
    = \frac{75}{2} (P_{14} \varsigma P_{14})_{ab}{}^{cd}.
\end{equation}

Another useful building block is (c.f.\ \cref{eq:projector:one-form:H_second})
\begin{equation}
    (\sigma - 5 \varsigma)_{ab}{}^{cd}
    = (\delta_{[a|}^{[c|} - 5 s_{[a|}{}^i s^{[c|}{}_i)  s_{|b]}{}^j s^{|d]}_j,
    \quad\text{where}\quad \sigma_{ab}{}^{cd} = \delta_{[a|}^{[c|} s_{|b]}{}^i s^{|d]}_i.
\end{equation}
The off-diagonal $g_2$-blocks of $(\sigma - 5 \varsigma)_{ab}{}^{cd}$ vanish while the $7 \times 7$ and $14 \times 14$ blocks have ranks four and eight which are the dimensions of $(1,1)\subset\vec{7}$ and $(1,3)\subset\vec{14}$, respectively.
Hence, by demanding that the projectors square to themselves to fix the normalisation, we find
\begin{equation}
    (P_{(1,1)\subset\vec{7}})_{ab}{}^{cd} = 10 P_7 (\sigma - 5 \varsigma)_{ab}{}^{cd} P_7,\quad
    (P_{(1,3)\subset\vec{14}})_{ab}{}^{cd} = 10 P_{14} (\sigma - 5 \varsigma)_{ab}{}^{cd} P_{14}.
\end{equation}
The last projector is simply
\begin{equation}
    (P_{(2,0)\subset\vec{14}})_{ab}{}^{cd} = (P_{14} - P_{(0,2)\subset\vec{14}} - P_{(1,3)\subset\vec{14}})_{ab}{}^{cd}.
\end{equation}

These projection operators can be used to get an expression for the Weyl tensor.
Specifically, by using \cref{masterweyl} on two-forms we find that
\begin{equation}
    W_{ab}{}^{cd}
    = \frac{1}{5} (4 P_{14} + 8 P_{(0,2)\subset\vec{14}} - 10 P_{(1,3)\subset\vec{14}})_{ab}{}^{cd} \\
    = \frac{1}{5} P_{14} (4 - 100 \sigma + 800 \varsigma)_{ab}{}^{cd} P_{14}.
\end{equation}
In computations, we sometimes find it useful to rewrite this as
\begin{equation}
    W_{ab}{}^{cd}
    = \frac{4}{5} \delta_{ab}^{cd}
    - 20 \sigma_{ab}{}^{cd} + 160 \varsigma_{ab}{}^{cd}
    + 12 (P_7 \sigma)_{ab}{}^{cd} - 120 (P_7 \varsigma)_{ab}{}^{cd}
    - 60 (\varsigma P_7)_{ab}{}^{cd}.
\end{equation}

\subsubsection{Three-forms}
The three-form irrep $\vec{35}$ of $so(7)$ decomposes into $\vec{1} \oplus \vec{7} \oplus \vec{27}$ when restricted to $g_2$.
Clearly, $a_{abc}$ and $c_{a}{}^{bcd}$ can be viewed as intertwiners from three-forms to scalars and one-forms, respectively, which allows us to easily construct the projection operators onto $\vec{1}$ and $\vec{7}$.
The projector onto the $\vec{27}$ is then obtained by subtracting the previous two from the identity.
Fixing the normalisations through $P^2 = P$, this gives
\begin{align}
    & (P_1)_{abc}{}^{def} = \frac{1}{42} a_{abc}a^{def}, \\[2pt]
    & (P_{7})_{abc}{}^{def} = -\frac{1}{24} c_{abc}{}^{g} c_g{}^{def}
    = \frac{1}{24} (6\delta_{abc}^{def} + 9\delta_{[a}^{[d}c_{bc]}{}^{ef]} - a_{abc}a^{def}), \\
    & (P_{27})_{abc}{}^{def} = (1-P_1-P_7)_{abc}{}^{def} = \frac{1}{56}(42\delta_{abc}^{def} - 21\delta_{[a}^{[d}c_{bc]}{}^{ef]} + a_{abc}a^{def}).
\end{align}

\subsubsection{Spin 1/2}
The spinor $\vec{8}$ of $so(7)$ splits into a singlet $\vec{1}$, the Killing spinor, and a $\vec{7}$ when restricted to $g_2$.
The projection operators are
\begin{equation}
    P_1 = \eta \bar{\eta},\qquad
    P_7 = \Gamma_a \eta \bar{\eta} \Gamma^a,
\end{equation}
where $\eta$ is the normalised Killing spinor ($\bar{\eta} \eta = 1$) discussed in detail in \cref{app:octonions}.

The projectors can be written in terms of gamma-matrices as
\begin{equation}
    P_1 = \frac{1}{8} + \frac{i}{48} a_{abc} \Gamma^{abc},\qquad
    P_7 = \frac{7}{8} - \frac{i}{48} a_{abc} \Gamma^{abc},
\end{equation}
where we have used that $a_{abc} = i \bar{\eta} \Gamma_{abc} \eta$.
This is seen by computing the traces $\operatorname{Tr}(P \Gamma^{(n)})$ where $n=0,1,2,3$ indicates the number of indices on $\Gamma$.
The above proves the Fierz identity $\Gamma_a \eta \bar{\eta} \Gamma^a = 1 - \eta \bar{\eta}$ which is nothing but $P_1 + P_7 = 1$.

Lastly, we note that the projectors onto the $H$-irreps appearing in the decomposition of the spinor under $so(7) \rightarrow h$ are immediately obtained from the above as
\begin{equation}
    P_{(0,2)} = \Gamma^a \eta (5 s_a{}^i s^b{}_i) \bar{\eta} \Gamma_b,\qquad
    P_{(1,1)} = \Gamma^a \eta (\delta_a^b - 5 s_a{}^i s^b{}_i) \bar{\eta} \Gamma_b.
\end{equation}

\subsubsection{Spin 3/2}
Gamma-traceless vector-spinors, or spin-3/2 fields, transform in the $\vec{48}$ of $so(7)$, which decomposes into $\vec{7} \oplus \vec{14} \oplus \vec{27}$ under $so(7) \rightarrow g_2$.
From \crefrange{eq:eigs:spin-3/2:split}{eq:eigs:spin-3/2:irr_comps} one can read off the projection operators (with supressed spinor indices)
\begin{align}
    &(P_7)_a{}^b = \frac{1}{6} (6\delta_a^b \eta + i a_a{}^{bc} \Gamma_c \eta) \bar{\eta},\\
    &(P_{14})_a{}^b = -(P_{14})_{ac}{}^{db} \Gamma^c \eta \bar{\eta} \Gamma_d,\\
    &(P_{27})_a{}^b = \delta_{(a}^b \Gamma^c \eta \bar{\eta} \Gamma_{c)},
\end{align}
where $(P_{14})_{ac}{}^{db}$ on the second line is the two-form projection operator.

\subsection{Casimir conventions}
\label{app:gtwoprojectors:Casimirs}
In this subsection, we give our conventions for the Casimirs of the groups appearing in the analysis.
First, the isometry group ($G = Sp_2 \times Sp_1^C$) Casimir in the master formula for $\Delta$, \cref{masterdelta}, is
\begin{equation}
    C_g(p,q;r) = - T_A T^A = 2 C_{sp_2}(p,q) + 3 C_{\smash[t]{sp_1^C}}(r),
\end{equation}
where $(p,q)$ and $(r)$ are the Dynkin labels of $sp_2$ and $sp_1^C$, respectively, $C_{sp_2}$ and $C_{sp_1}$ are given in \cref{tab:Casimirs} and, as in \cref{sec:review_and_method}, we raise and lower $G$-indices using $-\kappa_{AB}/6$ and its inverse, where $\kappa_{AB}$ is the negative definite Cartan--Killing form.
The Casimir of the isotropy group, $H = Sp_1^A \times Sp_1^{B+C}$, is
\begin{equation}
    C_h = -T_i T^i = 2 C_{sp_1^A} + \frac{6}{5} C_{\smash[t]{sp_1^{B+C}}},
\end{equation}
where $H$-indices are raised and lowered using the restriction of $-\kappa_{AB}/6$ to $h$ and its inverse.

The Casimirs of $Spin(7)$ and $G_2$ also appear in \cref{masterdelta}.
These are defined by
\begin{equation}
    C_{so(7)} = - \Sigma_{ab} \Sigma^{ab},\qquad
    C_{g_2} = - (P_{14})_{abcd} \Sigma^{ab} \Sigma^{cd}.
\end{equation}
Formulas for their eigenvalues on irreps in terms of Dynkin labels are provided in \cref{tab:Casimirs}.

\begin{table}[H]
    \centering
    \begin{tabular}{ll}
        \toprule
        Casimir & Adjoint \\ \midrule \addlinespace[\belowrulesep]
        $C_{sp_1}(r) = \frac{1}{4}\+ r(r+2)$ &
        $C_{sp_1}(2) = 2$ \\ \addlinespace
        $C_{sp_2}(p,q) = \frac{1}{4}\+ p(p+2q+4) + \frac{1}{2}\+ q(q+3)$ &
        $C_{sp_2}(2,0) = 3$ \\ \addlinespace
        $C_{g_2}(p,q) = \frac{1}{3}\+ p(p+3q+5) + q(q+3)$ &
        $C_{g_2}(0,1) = 4$ \\ \addlinespace
        $C_{so(7)}(p,q,r) = \frac{1}{2}\+ p(p+2q+r+5) + q(q+r+4) + \frac{3}{8}\+ r(r+6)$ &
        $C_{so(7)}(0,1,0) = 5$ \\ \addlinespace[\aboverulesep]
        \bottomrule
    \end{tabular}
    \caption{Casimir eigenvalues on irreps specified by Dynkin labels with the adjoint representation given explicitly.}
    \label{tab:Casimirs}
\end{table}

\section{Non-transverse modes and two-form mode scalar products}
\label{app:non_transv_and_2-form_scalar_prods}

In this appendix we give the non-transverse eigenmodes of $\Delta_1$ and $\Delta_2$ that, just like the transverse ones, are obtained by diagonalising \cref{eq:modes:1-form:Delta_result,eq:modes:2-form:Delta_matrix}.
We also provide the two-form scalar products that are needed to derive the cross diagrams in \cref{fig:modes:2-form:cross_diagrams} and below.
In \cite{Nilsson:2018lof}, the isometry irrep spectra are first derived for the transverse and non-transverse modes combined by decomposing the $Spin(7)$-irreps into $H$-irreps and, then, the non-transverse pieces are subtracted to arrive at the transverse irrep spectra.
Deriving the cross diagrams for the non-transverse modes and combining these with the transverse ones of \cref{sec:eigenmodes} thus provides a consistency check of this last step in \cite{Nilsson:2018lof} upon comparison.

\begin{figure}[H]
    \centering
    \begingroup
    \renewcommand{\scale}{0.50cm}  %
    \footnotesize
    \newcommand{\fig}{%
        \begin{tikzpicture}[x=\scale, y=\scale]
            \crossdiagramlayout{p}
            \foreach \p in {1,...,\range} {
                \foreach \q in {0,...,\range} {
                    \node at (\p,\q) {$\times$};
                }
            }
            \foreach \q in {1,...,\range} {
                \node at (0,\q) {$\times$};
            }
        \end{tikzpicture}
    }
    \newcommand*{\eig}[1]{{\normalsize$\Delta_1^{#1}$}}%
    \begin{tabular}{c}
        \eig{(2)}   \\
        \fig
    \end{tabular}
    \endgroup
    \caption{Non-transverse one-form cross diagrams. Each cross corresponds to an isometry irrep $(p,q;r)$ of non-transverse eigenmodes of $\Delta_1$ generated by the differential operator in \cref{eq:non-transv_modes:1-form:mode} and with the eigenvalue in \cref{eq:non-transv_modes:1-form:eigenvalue}.}
    \label{fig:non-transv_modes:1-form:cross_diagrams}
\end{figure}
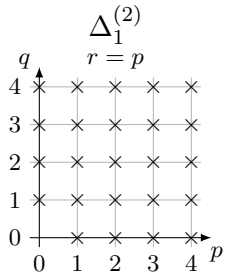

From \cref{eq:modes:1-form:Delta_result}, we immediately see that the non-transverse one-form eigenmode is generated by
\begin{equation}
\label{eq:non-transv_modes:1-form:mode}
    \tilde{\mode}^{(2)}_a = \mode^{(1)}_a = \check{D}_a
\end{equation}
with eigenvalue
\begin{equation}
\label{eq:non-transv_modes:1-form:eigenvalue}
    \Delta_1^{(2)} = C_g
\end{equation}
and norm
\begin{equation}
    \| \tilde{\mode}^{(2)} \phi \|^2
    = C_g^\phi \| \phi \|^2.
\end{equation}
Hence, this non-transverse eigenmode exists for $(p,q;r)=(p,q;p)$ with $(p,q) \neq (0,0)$ corresponding to the cross diagram in \cref{fig:non-transv_modes:1-form:cross_diagrams}.
Note that the cross-diagrams in \cref{fig:modes:1-form:cross_diagrams,fig:non-transv_modes:1-form:cross_diagrams} combined exactly match the isometry irrep spectrum of (transverse and non-transverse) one-froms obtained from the $(0,2)$ and $(1,1)$ $H$-diagrams in \cite{Nilsson:2018lof}.%
\footnote{Recall that $\vec{7} \to (0,2) \oplus (1,1)$ when $Spin(7) \to H$.}
This provides the first consistency check of the type described above.

Turning to two-forms, the non-transverse $\Delta_2$-eigenmodes obtained by diagonalising \cref{eq:modes:2-form:Delta_matrix} are generated by the following differential operators with associated eigenvalues:
\begin{align}
\label{eq:non-transv_modes:2-form:eigenmodes}
    &\begin{aligned}[b]
        \tilde{\mode}^{(4)i}_{ab}
        & =
        \frac{1}{20} \Bigl(15 \Delta C - 13 \pm \sqrtsmash[C_G]{20 C_g + 49}\Bigr) \mode^{(2)i}_{ab}
        + \frac{3}{2\sqrt{5}} \Bigl(13 \mp \sqrtsmash[C_G]{20 C_g + 49}\Bigr) \mode^{(3)i}_{ab}
        \\ &\phantom{{}={}}
        + \frac{1}{2\sqrt{5}} \Bigl(12 \mp \sqrtsmash[C_G]{20 C_g + 49}\Bigr) \mode^{(4)i}_{ab}
        - \frac{\sqrt{5}}{2} \mode^{(5)i}_{ab}
        + \mode^{(6)i}_{ab},
    \end{aligned}
    \\
\label{eq:non-transv_modes:2-form:eigenvalues}
    & \Delta_2^{(4)_\pm} = \frac{m^2}{9} \Bigl(20 C_g + 14 \pm 2 \sqrtsmash[C_G]{20 C_g + 49} \Bigr).
\end{align}

To compute the norms of these non-transverse eigenmodes, and also for the norms of the transverse eigenmodes in \crefrange{eq:modes:2-form:eigenmodes:first}{eq:modes:2-form:eigenmodes:last} leading to \cref{fig:modes:2-form:cross_diagrams}, we need the scalar products of the basic building blocks $\mode^{(\text{1--7})}_{ab}$ from \crefrange{eq:modes:2-form:building-modes:first}{eq:modes:2-form:building-modes:last}.
Hence, we compute
\begingroup
\allowdisplaybreaks
\begin{align}
\label{eq:2-form:scalar_products:first}
    & \langle \mode^{(1)} \phi, \mode^{(1)} \phi \rangle = 6 C_g^\phi \|\phi\|^2,\\
    & \langle \mode^{(1)} \phi, \mode^{(2)i} s_i \phi \rangle = 6 C_{\smash[t]{sp_1^C}}^\phi \|\phi\|^2,\\
    & \langle \mode^{(1)} \phi, \mode^{(3)i} s_i \phi \rangle = \frac{2}{\sqrt{5}} C_{\smash[t]{sp_1^C}}^\phi \|\phi\|^2,\\
    & \langle \mode^{(1)} \phi, \mode^{(4)i} s_i \phi \rangle = -\frac{3}{\sqrt{5}} C_{\smash[t]{sp_1^C}}^\phi \|\phi\|^2,\\
    & \langle \mode^{(1)} \phi, \mode^{(5)i} s_i \phi \rangle = 4 \langle \mode^{(1)} \phi, \mode^{(4)i} s_i \phi \rangle,\\
    & \langle \mode^{(1)} \phi, \mode^{(6)i} s_i \phi \rangle = \frac{27}{10} C_{\smash[t]{sp_1^C}}^\phi \|\phi\|^2,\\
    & \langle \mode^{(1)} \phi, \mode^{(7)ij} s_{ij} \phi \rangle = -\frac{1}{6} C_{\smash[t]{sp_1^C}}^\phi \bigl(4 C_{\smash[t]{sp_1^C}}^\phi - 3\bigr) \|\phi\|^2,\\
    & \langle \mode^{(2)i} \phi, \mode^{(2)j} \varphi \rangle = \frac{6}{5} \langle\phi, \delta^{ij} \varphi\rangle,\\
    & \langle \mode^{(2)i} \phi, \mode^{(3)j} \varphi \rangle = \frac{2}{5\sqrt{5}} \langle\phi, \delta^{ij} \varphi\rangle,\\
    & \langle \mode^{(2)i} \phi, \mode^{(4)j} \varphi \rangle = \frac{1}{\sqrt{5}} \langle\phi, \epsilon^{ijk} s_k \varphi\rangle,\\
    & \langle \mode^{(2)i} \phi, \mode^{(5)j} \varphi \rangle = 4 \langle \mode^{(2)i} \phi, \mode^{(4)j} \varphi \rangle,\\
    & \langle \mode^{(2)i} \phi, \mode^{(6)j} \varphi \rangle = \frac{1}{5} C_g^\phi \langle\phi, \delta^{ij} \varphi\rangle + \langle\phi, s^{ij} \varphi\rangle,\\
    & \langle \mode^{(2)i} \phi, \mode^{(7)jk} s_k \varphi \rangle = -\frac{1}{10} C_{\smash[t]{sp_1^C}}^\phi \langle\phi, \delta^{ij} \varphi\rangle - \frac{1}{12} \langle\phi, \epsilon^{ijk} s_k \varphi\rangle + \frac{1}{30} \langle\phi, s^{ij} \varphi\rangle,\\
    & \langle \mode^{(3)i} \phi, \mode^{(3)j} \varphi \rangle = \frac{2}{25} \langle\phi, \delta^{ij} \varphi\rangle,\\
    & \langle \mode^{(3)i} \phi, \mode^{(4)j} \varphi \rangle = \frac{1}{5} \langle\phi, \epsilon^{ijk} s_k \varphi\rangle,\\
    & \langle \mode^{(3)i} \phi, \mode^{(5)j} \varphi \rangle = \frac{2}{\sqrt{5}} \langle \mode^{(2)i} \phi, \mode^{(4)j} \varphi \rangle - 2 \langle \mode^{(3)i} \phi, \mode^{(4)j} \varphi \rangle,\\
    & \langle \mode^{(3)i} \phi, \mode^{(6)j} \varphi \rangle = \frac{1}{\sqrt{5}} C_{\smash[t]{sp_1^C}}^\phi \langle\phi, \delta^{ij} \varphi\rangle + \frac{1}{\sqrt{5}} \langle\phi, s^{ij} \varphi\rangle,\\
    & \langle \mode^{(3)i} \phi, \mode^{(7)jk} s_k \varphi \rangle = -\frac{1}{10\sqrt{5}} C_{\smash[t]{sp_1^C}}^\phi \langle\phi, \delta^{ij} \varphi\rangle - \frac{1}{12\sqrt{5}} \langle\phi, \epsilon^{ijk} s_k \varphi\rangle + \frac{1}{30\sqrt{5}} \langle\phi, s^{ij} \varphi\rangle,\\
    & \langle \mode^{(4)i} \phi, \mode^{(4)j} \varphi \rangle = \frac{1}{10} C_g^\phi \langle\phi, \delta^{ij} \varphi\rangle - \frac{1}{4} \langle\phi, \epsilon^{ijk} s_k \varphi\rangle + \frac{1}{2} \langle\phi, s^{ij} \varphi\rangle,\\
    & \langle \mode^{(4)i} \phi, \mode^{(5)j} \varphi \rangle = - \frac{2}{5} \langle\phi, \epsilon^{ijk} s_k \varphi\rangle,\\
    & \begin{aligned}[b]
        \langle \mode^{(4)i} \phi, \mode^{(6)j} \varphi \rangle
        & = \frac{1}{20\sqrt{5}} \bigl(C_g^\phi - 30 C_{\smash[t]{sp_1^C}}^\phi \bigr) \langle\phi, \delta^{ij} \varphi\rangle + \frac{1}{8\sqrt{5}} \bigl(4C_g^\phi - 9\bigr) \langle\phi, \epsilon^{ijk} s_k \varphi\rangle
        \\ &\phantom{{}={}}
        - \frac{11}{4\sqrt{5}} \langle\phi, s^{ij} \varphi\rangle,
    \end{aligned} \\
    & \begin{aligned}[b]
        \langle \mode^{(4)i} \phi, \mode^{(7)jk} s_k \varphi \rangle
        & = -\frac{3}{20\sqrt{5}} C_{\smash[t]{sp_1^C}}^\phi \langle\phi, \delta^{ij} \varphi\rangle - \frac{1}{12\sqrt{5}} \bigl(3 C_{\smash[t]{sp_1^C}}^\phi - 2\bigr) \langle\phi, \epsilon^{ijk} s_k \varphi\rangle
        \\ &\phantom{{}={}}
        - \frac{11}{30\sqrt{5}} \langle\phi, s^{ij} \varphi\rangle,
    \end{aligned} \\
    & \langle \mode^{(5)i} \phi, \mode^{(5)j} \varphi \rangle = 8 \langle \mode^{(4)i} \phi, \mode^{(4)j} \varphi \rangle + 2 \langle \mode^{(4)i} \phi, \mode^{(5)j} \varphi \rangle,\\
    & \langle \mode^{(5)i} \phi, \mode^{(6)j} \varphi \rangle = 2 \langle \mode^{(4)i} \phi, (3 P_7  - 1) \mode^{(6)j} \varphi \rangle,\\
    & \langle \mode^{(5)i} \phi, \mode^{(7)jk} s_k \varphi \rangle = 2 \langle \mode^{(4)i} \phi, (3 P_7  - 1) \mode^{(7)jk} s_k \varphi \rangle,\\
    & \begin{aligned}[b]
        \langle \mode^{(6)i} \phi, \mode^{(6)j} \varphi \rangle
        & = \frac{1}{200} C_g^\phi \bigl(20 C_g^\phi - 39\bigr) \langle\phi, \delta^{ij} \varphi\rangle - \frac{1}{80} \bigl(36 C_g^\phi - 171\bigr) \langle\phi, \epsilon^{ijk} s_k \varphi\rangle
        \\ &\phantom{{}={}}
        + \frac{1}{40} \bigl(20 C_g^\phi - 3\bigr) \langle\phi, s^{ij} \varphi\rangle,
    \end{aligned} \\
    & \begin{aligned}[b]
        \langle \mode^{(6)i} \phi, \mode^{(7)jk} s_k \varphi \rangle
        & = -\frac{1}{200} C_{\smash[t]{sp_1^C}}^\phi \bigl(10 C_g^\phi - 47\bigr) \langle\phi, \delta^{ij} \varphi\rangle
        \\ &\phantom{{}={}}
        - \frac{1}{120} \bigl(5 C_g^\phi - 19 C_{\smash[t]{sp_1^C}}^\phi + 7\bigr) \langle\phi, \epsilon^{ijk} s_k \varphi\rangle
        \\ &\phantom{{}={}}
        + \frac{1}{300} \bigl(5 C_g^\phi - 100 C_{\smash[t]{sp_1^C}}^\phi + 214 \bigr) \langle\phi, s^{ij} \varphi\rangle,
    \end{aligned} \\
    & \begin{aligned}[b]
        \langle \mode^{(7)ij} \phi, \mode^{(7)}_{kl} \varphi \rangle
        & = \frac{1}{50} \bigl(C_g^\phi - 5 C_{\smash[t]{sp_1^C}}^\phi\bigr) \langle \phi, \delta^{\{i}_{\{k} \delta^{j\}}_{l\}} \varphi \rangle + \frac{9}{100} \langle \phi, \delta^{\{i}_{\{k} \epsilon_{l\}}{}^{j\}m} s_m \varphi \rangle
        \\ &\phantom{{}={}}
        - \frac{1}{10} \langle \phi, \delta^{\{i}_{\{k} s_{l\}}{}^{j\}} \varphi \rangle,
    \end{aligned}
\label{eq:2-form:scalar_products:last}
\end{align}
\endgroup
where
\begin{align}
    &P_7 \mode^{(6)i}_{ab} = \frac{1}{6} \mode^{(1)}_{ab} s^i - \frac{1}{6} \mode^{(2)i}_{ab} \check{\Box} + \frac{1}{2} \epsilon^i{}_{jk} \mode^{(2)j}_{ab} s^k + \frac{1}{6\sqrt{5}} \mode^{(4)i}_{ab} + \frac{1}{12 \sqrt{5}} \mode^{(5)i}_{ab},\\
    &P_7 \mode^{(7)ij}_{ab} = \frac{1}{6} \mode^{(2)\{i}_{ab} s^{j\}}.
\end{align}

To compute scalar products and norms like $\langle P_r \mode^{m} \phi, P_r \mode^{(n)} \phi \rangle$, we need to use \cref{eq:modes:1-form:projector} in cases where the differential operators are not $sp_1^C$-singlets.
For the $\vec{5}_{sp_1^C}$ case, we need some $sp_1^C$ identities, such as
\begin{align}
\nonumber
    \epsilon^{(i}{}_{kl} \delta^{j)}_{(m} \delta_{n)}^k \epsilon^{(m|p}{}_r \mode_{pq} s^{|n)qr} s^l
    & =
    \frac{1}{3} Y^{ij} s^2
    + \frac{1}{4} \delta^{ij} \mode^{kl} s_{kl}
    - \frac{1}{4} \epsilon^{(i}{}_{kl} \mode^{j)k} s^l
    \\ &\phantom{{}={}}
    - \frac{1}{6} \mode^{k(i} s_{k}{}^{j)} (7 + 3 s^2)
    + \frac{3}{2} \epsilon^{(i|k}{}_{m} \mode_{kl} s^{|j)lm}
    + \frac{1}{2} \mode_{kl} s^{ijkl},
\end{align}
since \cref{eq:modes:1-form:projector} requires four successive applications of the Casimir
\begin{align}
\nonumber
    C_{\smash[t]{sp_1^C}} \mode^{ij} \phi
    & = -(T_k T^k \mode^{ij}) \phi - \mode^{ij} (T_k T^k \phi) - 2 (T_k \mode^{ij}) (T^k \phi) \\
    & = \bigl(C_{\smash[t]{sp_1^C}}^\phi + 6\bigr) \mode^{ij} \phi - 4 \epsilon^{(i}{}_{kl} \mode^{j)k} s^l,
\end{align}
where the $6$ comes from $C_{\smash[t]{sp_1^C}}(\vec{5}) = 6$.
This reduces the computation of $\langle P_r \mode^{m} \phi, P_r \mode^{(n)} \phi \rangle$ to \crefrange{eq:2-form:scalar_products:first}{eq:2-form:scalar_products:last} with $\varphi$ substituted by one of
\begin{equation}
    \delta_{ij} \phi, \qquad
    \epsilon_{ijk} s^k \phi, \qquad
    s_{ij},
\end{equation}
in the $\vec{3}_{sp_1^C}$ cases (of the form $\langle \mode^i \phi, \mode'^j \varphi \rangle$) and one of
\begin{equation}
    \delta_{(i}^{(k} \delta_{j)}^{l)} \phi, \quad
    \epsilon_{(i}{}^{(k|m|} \delta_{j)}^{l)} s_m \phi, \quad
    \delta_{(i}^{(k} s_{j)}{}^{l)} \phi, \quad
    \epsilon_{(i}{}^{(k|m|} s_{j)m}{}^{l)} \phi, \quad
    s_{ij}{}^{kl} \phi,
\end{equation}
in the $\vec{5}_{sp_1^C}$ cases (of the form $\langle \mode^{ij} \phi, \mode'_{kl} \varphi \rangle$).
Lastly, to put the above together, we need identities like \cref{eq:modes:1-form:0-form_scalar_product} in the $\vec{3}_{sp_1^C}$ case and, for instance,
\begin{equation}
    \langle \phi, s_{l}{}^{j} s_{ij}{}^{il} \phi\rangle
    = -  C_{\smash[t]{sp_1^C}}^\phi \biggl(\frac{5}{36} - \frac{23}{54}  C_{\smash[t]{sp_1^C}}^\phi +  \frac{4}{9} \bigl( C_{\smash[t]{sp_1^C}}^\phi\bigr)^2\biggr) \|\phi\|^2,
\end{equation}
in the $\vec{5}_{sp_1^C}$ case.

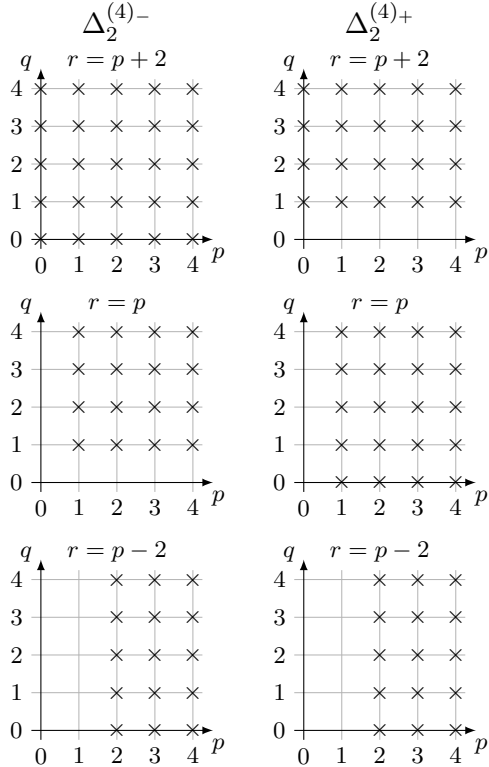
\begin{figure}[H]
    \centering
    \begingroup
    \renewcommand{\scale}{0.50cm}  %
    \footnotesize
    \newcommand{\figIppII}{%
        \begin{tikzpicture}[x=\scale, y=\scale]
            \crossdiagramlayout{p+2}
            \foreach \p in {0,...,\range} {
                \foreach \q in {1,...,\range} {
                    \node at (\p,\q) {$\times$};
                }
            }
        \end{tikzpicture}
    }
    \newcommand{\figIp}{%
        \begin{tikzpicture}[x=\scale, y=\scale]
            \crossdiagramlayout{p}
            \foreach \p in {1,...,\range} {
                \foreach \q in {0,...,\range} {
                    \node at (\p,\q) {$\times$};
                }
            }
        \end{tikzpicture}
    }
    \newcommand{\figIpmII}{%
        \begin{tikzpicture}[x=\scale, y=\scale]
            \crossdiagramlayout{p-2}
            \foreach \p in {2,...,\range} {
                \foreach \q in {0,...,\range} {
                    \node at (\p,\q) {$\times$};
                }
            }
        \end{tikzpicture}
    }
    \newcommand{\figIIppII}{%
        \begin{tikzpicture}[x=\scale, y=\scale]
            \crossdiagramlayout{p+2}
            \foreach \p in {0,...,\range} {
                \foreach \q in {0,...,\range} {
                    \node at (\p,\q) {$\times$};
                }
            }
        \end{tikzpicture}
    }
    \newcommand{\figIIp}{%
        \begin{tikzpicture}[x=\scale, y=\scale]
            \crossdiagramlayout{p}
            \foreach \p in {1,...,\range} {
                \foreach \q in {1,...,\range} {
                    \node at (\p,\q) {$\times$};
                }
            }
        \end{tikzpicture}
    }
    \newcommand{\figIIpmII}{%
        \begin{tikzpicture}[x=\scale, y=\scale]
            \crossdiagramlayout{p-2}
            \foreach \p in {2,...,\range} {
                \foreach \q in {0,...,\range} {
                    \node at (\p,\q) {$\times$};
                }
            }
        \end{tikzpicture}
    }
    \newcommand*{\eig}[1]{{\normalsize$\Delta_2^{#1}$}}%
    \begin{tabular}{cc}
        \eig{(4)_-} & \eig{(4)_+}   \\
        \figIIppII  & \figIppII     \\
        \figIIp     & \figIp        \\
        \figIIpmII  & \figIpmII
    \end{tabular}
    \endgroup
    \caption{Non-transverse two-form cross diagrams. Each cross corresponds to an isometry irrep $(p,q;r)$ of non-transverse eigenmodes of $\Delta_2$, listed in \cref{eq:non-transv_modes:2-form:eigenmodes}, with associated eigenvalues given in \cref{eq:non-transv_modes:2-form:eigenvalues}.}
    \label{fig:non-transv_modes:2-form:cross_diagrams}
  \end{figure}

Calculating the norms of the non-transverse $\Delta_2$-eigenmodes using the above scalar products, we arrive at the cross-diagrams in \cref{fig:non-transv_modes:2-form:cross_diagrams}.
Note that the cross diagrams in \cref{fig:non-transv_modes:2-form:cross_diagrams,fig:modes:2-form:cross_diagrams} combined exactly match the complete two-form isometry irrep spectrum obtained by combining the relevant $H$-diagrams in \cite{Nilsson:2018lof}, providing a non-trivial consistency check.%
\footnote{Recall that $\vec{21} \to (1,1) \oplus 2 \times (0,2) \oplus (2,0) \oplus (1,3)$ when $Spin(7) \to H$.}
\section{Spin-0 fields with  \texorpdfstring{$\frac{5}{2} < E_0 \leq 3$}{5/2 < E0 < 3} and spin-1/2  fields with \texorpdfstring{$2 < E_0 \leq 3 $}{2 < E0 < 3}}
\label{app:fields_in_ezero_range}

We start by listing the bosonic and fermionic fields in the $AdS_4$ spectrum of the round $S^7$  with $E_0\le 3$ \cite{Duff:1986hr}. These are, apart from the massless  graviton  (with $E_0=3$), spin-1 gauge fields (with $E_0=2$)
and massive $1^-$ vector fields from the levels $n=1$ and $n=2$  (with $E_0=\frac{5}{2}$ and  $E_0=3$), the scalar, pseudo-scalar  and spin-1/2 fermion fields listed in \cref{roundezerolessthanthree:spinzero,roundezerolessthanthree:spinhalf}.

\begin{table}[H]
    \centering
    \begin{tabular}{l llll llll}
        \toprule
        & \multicolumn{4}{c}{$0^+_{\scriptscriptstyle (-)}(\Delta_0)$}
        & \multicolumn{4}{c}{$0^-_{\scriptscriptstyle (-)}(Q)$}
        \\ \cmidrule(rl){2-5} \cmidrule(rl){6-9}
        $n$
        & $Spin(8)$ irrep & $M^2$ & $E_0$ & B.c.
        & $Spin(8)$ irrep & $M^2$ & $E_0$ & B.c.
        \\ \midrule
        $-1$
        & $(1,0,0,0)=\vec{8}_v$ & $3m^2$ & $\frac{1}{2}$ & sing.
        \\
        $0$
        & $(2,0,0,0)=\vec{35}_v$ & $0$ & $1$ & Neu.
        & $(0,0,2,0)=\vec{35}_c$ & $0$ & $2$ & Dir.
        \\
        $1$
        & $(3,0,0,0)=\vec{112}_v$ & $-m^2$ & $\frac{3}{2}$ & deg.
        & $(1,0,2,0)=\vec{112}_{cv}$ & $3m^2$ & $\frac{5}{2}$ & Dir
        \\
        $2$
        & $(4,0,0,0)$ & $0$ & $2$ & Dir.
        & $(2,0,2,0)$ & $8m^2$ & $3$ & Dir.
        \\
        $3$
        & $(5,0,0,0)$ & $3m^2$ & $\frac{5}{2}$ & Dir.
        \\
        $4$
        & $(6,0,0,0)$ & $8m^2$ & $3$ & Dir.
        \\ \bottomrule
    \end{tabular}
    \caption{Scalar and pseudo-scalar fields with $E_0\leq 3$ in the $AdS_4$ spectrum based on the round $S^7$ where $n$ is the level number. All massless (pseudo-)scalar fields in the spectrum
    are present in this list. These fields are dual to relevant operators in the boundary $CFT_3$. We also specify whether the fields have singleton ($E_0 = \frac{1}{2}$), Neumann ($\frac{1}{2} < E_0 < \frac{3}{2}$), degenerate ($E_0 = \frac{3}{2}$) or Dirichlet ($E_0 > \frac{3}{2}$) boundary conditions.}
    \label{roundezerolessthanthree:spinzero}
\end{table}

\begin{table}[H]
    \centering
    \begin{tabular}{l llll llll}
        \toprule
        & \multicolumn{4}{c}{$\frac{1}{2}_{\scriptscriptstyle (-)}(i\slashed{D}_{1/2})$}
        & \multicolumn{4}{c}{$\frac{1}{2}_{\scriptscriptstyle (-)}(i\slashed{D}_{3/2})$}
        \\ \cmidrule(rl){2-5} \cmidrule(rl){6-9}
        $n$
        & $Spin(8)$ irrep & $M$ & $E_0$ & B.c.
        & $Spin(8)$ irrep & $M$ & $E_0$ & B.c.
        \\ \midrule
        $-1$
        & $(0,0,1,0)=\vec{8}_c$ & $-m$ & $1$ & sing.
        \\
        $0$
        & $(1,0,1,0)=\vec{56}_s$ & $0$ & $\frac{3}{2}$ & deg.
        \\
        $1$
        & $(2,0,1,0)$ & $m$ & $2$ & Dir.
        & $(0,1,1,0)$ & $-3m$ & $3$ & Dir.
        \\
        $2$
        & $(3,0,1,0)$ & $2m$ & $\frac{5}{2}$ & Dir.
        \\
        $3$
        & $(4,0,1,0)$ & $3m$ & $3$ & Dir.
        \\ \bottomrule
    \end{tabular}
    \caption{Spin-1/2 fermion fields with $E_0\leq 3$ in the $AdS_4$ spectrum based on the round $S^7$ where $n$ is the level number. All massless spin-1/2 fields in the spectrum
    are present in this list. These fields are dual to relevant operators in the boundary $CFT_3$. We also specify whether the fields have singleton ($E_0 = 1$), Neumann ($1 < E_0 < \frac{3}{2}$), degenerate ($E_0 = \frac{3}{2}$) or Dirichlet ($E_0 > \frac{3}{2}$) boundary conditions.}
    \label{roundezerolessthanthree:spinhalf}
\end{table}

To facilitate the comparison between the squashed and the round cases, we present the  decomposition
of some of the $Spin(8)$ irreps under $Spin(8) \rightarrow Sp_2 \times Sp_1^C$, for both left- and right-squashing when this is relevant (indicated by $L$ and $R$ on the arrow), in \cref{decomp}. Recall from, e.g., \cite{Duff:1986hr} that the left- and right-squashings can be defined by how the three eight-dimensional
irreps of $Spin(8)$ behave. While ${\bf 8}_v=(1,0,0,0) \rightarrow (1,0;1)=({\bf 4}, {\bf 2})$ is true  in both cases the irreps ${\bf 8}_c=(0,0,1,0)$ and ${\bf 8}_s=(0,0,0,1)$
interchange their behaviour when skew-whiffed, see \cref{decomp}.
\begin{table}[H]
    \centering
    \begin{tabular}{lll}
        \toprule
        $Spin(8)$ & $\longrightarrow$ & $Sp_2 \times Sp_1^C$
        \\ \midrule
        ${\bf 8}_v=(1,0,0,0)$  &  $\overset{\phantom{L}}{\longrightarrow}$ &  $(1,0;1)=({\bf 4}, {\bf 2})$ \\
        ${\bf 35}_v=(2,0,0,0)$ &  $\overset{\phantom{L}}{\longrightarrow}$ &  $(2,0;2)\oplus (0,1;0)=({\bf 10}, {\bf 3})\oplus ({\bf 5}, {\bf 1})$ \\
        ${\bf 112}_v=(3,0,0,0)$  &  $\overset{\phantom{L}}{\longrightarrow}$ & $(3,0;3)\oplus (1,1;1)=({\bf 20}, {\bf 4})\oplus ({\bf 16}, {\bf 2})$ \\
        ${\bf 294}_v=(4,0,0,0)$  &  $\overset{\phantom{L}}{\longrightarrow}$ & $(4,0;4)\oplus (2,1;2) \oplus (0,2;0) = ({\bf 35}', {\bf 5}) \oplus ({\bf 35}, {\bf 3}) \oplus ({\bf 14}, {\bf 1})$ \\
        ${\bf 8}_c=(0,0,1,0)$  &  $\overset{L}{\longrightarrow} $ &  $  (0,1;0) \oplus (0,0;2) =({\bf 5}, {\bf 1})\oplus ({\bf 1}, {\bf 3}) $ \\
        ${\bf 8}_s=(0,0,0,1) $ & $\overset{L}{\longrightarrow} $&$    (1,0;1)=({\bf 4}, {\bf 2})$ \\
        ${\bf 56}_s=(1,0,1,0)$  &  $\overset{L}{\longrightarrow} $  &  $  (1,1;1) \oplus (1,0;3)  \oplus (1,0;1)=({\bf 16}, {\bf 2})\oplus ({\bf 4}, {\bf 4})\oplus ({\bf 4}, {\bf 2})$ \\
        ${\bf 8}_c=(0,0,1,0)$  &  $\overset{R}{\longrightarrow}  $&$  (1,0;1)=({\bf 4}, {\bf 2}) $ \\
        ${\bf 8}_s=(0,0,0,1)$  &  $\overset{R}{\longrightarrow}$ &  $  (0,1;0) \oplus (0,0;2) =({\bf 5}, {\bf 1})\oplus ({\bf 1}, {\bf 3}) $ \\
        ${\bf 56}_s=(1,0,1,0)$  &  $\overset{R}{\longrightarrow}$&  $(2,0;2) \oplus (2,0;0)\oplus (0,1;2) \oplus (0,0;0)=$ \\
        &&$({\bf 10}, {\bf 3})\oplus ({\bf 10}, {\bf 1})\oplus ({\bf 5}, {\bf 3}) \oplus ({\bf 1}, {\bf 1})$
        \\ \bottomrule
    \end{tabular}
    \caption{Decompositions of some $Spin(8)$-irreps into $Sp_2\times Sp_1^C$-irreps. The behaviour of the irreps ${\bf 8}_c=(0,0,1,0)$ and ${\bf 8}_s=(0,0,0,1)$
    define how the decompositions work in the left- and right-squashed seven-sphere vacua.}
    \label{decomp}
\end{table}

In the case of the $AdS_4$ spectra related to the squashed sphere, we here list the  scalar and pseudo-scalar fields in the range $\frac{5}{2} < E_0 \leq 3 $,
see \cref{tab:E0>5/2_scalars},
which complements \cref{tab:D_or_N:scalars,tab:D_or_N:pseudo-scalars} to provide a full account of all such fields with $E_0 \leq 3$.
These fields have  masses between $M^2=3m^2$ and $M^2=8m^2$ and can thus only have Dirichlet boundary conditions.
The analogous fermions have $2< E_0 \leq 3 $ and masses in the range $m < |M| \leq 3m$. These fields can be found  in \cref{tab:E0>2_spinors:LS,tab:E0>2_spinors:RS}, complementing \cref{{tab:D_or_N:spinors}}.

\begin{table}[H]
    \centering
    \setlength{\defaultaddspace}{3pt}
    \setlength{\tabcolsep}{10pt}
    \begin{tabular}{lllll}
        \toprule
        Field & $(p,q;r)$ & $C_g$ & $M^2/m^2$ & $E_0$
        \\ \midrule
        $0^+_{\scriptscriptstyle (-)}(\Delta_0^{(1)})$
        & $(1,3;1)$ & $103/4$ & $\frac{911}{9}-8 \sqrt{149}$ & $\frac{3}{2} + \frac{1}{3} \bigl(\sqrt{149}-9\bigr)$ \\ \addlinespace
        & $(2,2;2)$ & $26$ & $\frac{916}{9}-4 \sqrt{601}$ & $\frac{3}{2} + \frac{1}{6} \bigl(\sqrt{601}-18\bigr)$ \\ \addlinespace
        & $(0,4;0)$ & $28$ & $\frac{956}{9}-4 \sqrt{641}$ & $\frac{3}{2} + \frac{1}{6} \bigl(\sqrt{641}-18\bigr)$ \\ \addlinespace
        & $(3,1;3)$ & $115/4$ & $\frac{971}{9}-16 \sqrt{41}$ & $\frac{3}{2} + \frac{1}{3} \bigl(2 \sqrt{41}-9\bigr)$
        \\ \addlinespace[\aboverulesep] \midrule
        Field ($LS^7$) & $(p,q;r)$ & $C_g$ & $M_L^2/m^2$ & $E_0$
        \\ \midrule
        $0^-(Q^{(1)_-})$
        & $(0,0;0)$ & $0$ & $\frac{40}{9}$ & $\frac{3}{2} + \frac{7}{6}$ \\ \addlinespace
        & $(0,2;0)$ & $10$ & $\frac{16}{9} \bigl(16-\sqrt{201}\+\bigr)$ & $\frac{3}{2} + \frac{1}{6} \bigl(\sqrt{201}-8\bigr)$ \\ \addlinespace
        & $(2,0;2)$ & $12$ & $\frac{8}{9} \bigl(37 - 2 \sqrt{241}\+\bigr)$ & $\frac{3}{2} + \frac{1}{6} \bigl(\sqrt{241}-8\bigr)$ \\ \addlinespace
        $0^-(Q^{(2)_-})$
        & $(2,1;0)$ & $12$ & $\frac{40}{9}$ & $\frac{3}{2} + \frac{7}{6}$ \\ \addlinespace
        & $(3,0;1)$ & $51/4$ & $\frac{5}{9} \bigl(79-16 \sqrt{19}\+\bigr)$ & $\frac{3}{2} + \frac{1}{3} \bigl(2 \sqrt{19}-5\bigr)$ \\ \addlinespace
        & $(1,0;3)$ & $55/4$ & $\frac{55}{9}$ & $\frac{3}{2} + \frac{4}{3}$ \\ \addlinespace
        $0^-(Q^{(3)_-})$
        & $(1,2;1)$ & $67/4$ & $\frac{1}{9} \bigl(551-96 \sqrt{26}\+\bigr)$ & $\frac{3}{2} + \frac{2}{3} \bigl(\sqrt{26}-3\bigr)$ \\ \addlinespace
        & $(0,3;0)$ & $18$ & $8$ & $\frac{3}{2} + \frac{3}{2}$ \\ \addlinespace
        & $(2,1;2)$ & $18$ & $8$ & $\frac{3}{2} + \frac{3}{2}$
        \\ \addlinespace[\aboverulesep] \midrule
        Field ($RS^7$) & $(p,q;r)$ & $C_g$ & $M^2_R/m^2$ & $E_0$
        \\ \midrule
        $0^-(Q^{(1)_+})$
        & $(4,0;0)$ & $16$ & $\frac{4}{9} \bigl(103 - 5 \sqrt{321}\+\bigr)$ & $\frac{3}{2} + \frac{1}{6} \bigl(\sqrt{321}-10\bigr)$ \\ \addlinespace
        & $(1,2;1)$ & $67/4$ & $\frac{1}{9} \bigl(427-80 \sqrt{21}\+\bigr)$ & $\frac{3}{2} + \frac{1}{3} \bigl(2 \sqrt{21}-5\bigr)$ \\ \addlinespace
        & $(0,3;0)$ & $18$ & $8$ & $\frac{3}{2} + \frac{3}{2}$ \\ \addlinespace
        & $(2,1;2)$ & $18$ & $8$ & $\frac{3}{2} + \frac{3}{2}$ \\ \addlinespace
        $0^-(Q^{(2)_+})$
        & $(1,1;1)$ & $39/4$ & $\frac{1}{9} \bigl(299-32 \sqrt{61}\+\bigr)$ & $\frac{3}{2} + \frac{1}{3} \bigl(\sqrt{61}-4\bigr)$ \\ \addlinespace
        & $(0,1;2)$ & $10$ & $\frac{16}{9} \bigl(19-\sqrt{249}\+\bigr)$ & $\frac{3}{2} + \frac{1}{6} \bigl(\sqrt{249}-8\bigr)$ \\ \addlinespace
        & $(2,0;2)$ & $12$ & $8$ & $\frac{3}{2} + \frac{3}{2}$ \\ \addlinespace
        & $(2,1;0)$ & $12$ & $8$ & $\frac{3}{2} + \frac{3}{2}$ \\ \addlinespace
        $0^-(Q^{(3)_+})$
        & $(0,1;0)$ & $4$ & $\frac{4}{9} \bigl(47-3 \sqrt{161}\+\bigr)$ & $\frac{3}{2} + \frac{1}{6} \bigl(\sqrt{161}-6\bigr)$ \\ \addlinespace
        & $(1,0;1)$ & $19/4$ & $\frac{1}{9} \bigl(203-48 \sqrt{11}\+\bigr)$ & $\frac{3}{2} + \frac{1}{3} \bigl(2 \sqrt{11}-3\bigr)$
        \\ \addlinespace[\aboverulesep] \bottomrule
    \end{tabular}
    \caption{Scalars and pseudo-scalars with $\frac{5}{2} < E_0 \leq 3$ in both the left- and right-squashed vacuum. Note that the properties of the scalar fields are the same in the two vacua.}
    \label{tab:E0>5/2_scalars}
\end{table}

\begin{table}[H]
    \centering
    \setlength{\defaultaddspace}{3pt}
    \setlength{\tabcolsep}{10pt}
    \begin{tabular}{lllll}
        \toprule
        Field ($LS^7$) & $(p,q;r)$ & $C_g$ & $M_L/m$ & $E_0$
        \\ \midrule
        $\frac{1}{2}(i\slashed{D}_{1/2}^{(1)_-})$
        & $(1,2;1)$ & $67/4$ & $\frac{1}{3}\bigl(4 \sqrt{26}-15\bigr)$ & $\frac{3}{2} + \frac{1}{6} \bigl(4 \sqrt{26}-15\bigr)$ \\ \addlinespace
        & $(0,3;0)$ & $18$ & $2$ & $\frac{3}{2} + 1$ \\ \addlinespace
        & $(2,1;2)$ & $18$ & $2$ & $\frac{3}{2} + 1$ \\ \addlinespace
        & $(3,0;3)$ & $87/4$ & $\frac{1}{3}\bigl(2\sqrt{129}-15\bigr)$ & $\frac{3}{2} + \frac{1}{6} \bigl(2 \sqrt{129}-15\bigr)$ \\ \addlinespace
        $\frac{1}{2}(i\slashed{D}_{1/2}^{(2)_-})$
        & $(2,1;0)$ & $12$ & $\frac{4}{3}$ & $\frac{3}{2} + \frac{2}{3}$ \\ \addlinespace
        & $(3,0;1)$ & $51/4$ & $\frac{1}{3} \bigl(4 \sqrt{19}-13\bigr)$ & $\frac{3}{2} + \frac{1}{6} \bigl(4 \sqrt{19}-13\bigr)$ \\ \addlinespace
        & $(1,0;3)$ & $55/4$ & $\frac{5}{3}$ & $\frac{3}{2} + \frac{5}{6}$ \\ \addlinespace
        & $(0,2;2)$ & $16$ & $\frac{1}{3}\bigl(3\sqrt{41}-13\bigr)$ & $\frac{3}{2} + \frac{1}{6} \bigl(3 \sqrt{41}-13\bigr)$ \\ \addlinespace
        & $(1,2;1)$ & $67/4$ & $\frac{1}{3} \bigl(8 \sqrt{6}-13\bigr)$ & $\frac{3}{2} + \frac{1}{6} \bigl(8 \sqrt{6}-13\bigr)$ \\ \addlinespace
        & $(2,1;2)$ & $18$ & $\frac{1}{3} \bigl(\sqrt{409}-13\bigr)$ & $\frac{3}{2} + \frac{1}{6} \bigl(\sqrt{409}-13\bigr)$ \\ \addlinespace
        & $(1,1;3)$ & $75/4$ & $\frac{1}{3} \bigl(2 \sqrt{106}-13\bigr)$ & $\frac{3}{2} + \frac{1}{6} \bigl(2 \sqrt{106}-13\bigr)$ \\ \addlinespace
        & $(3,1;1)$ & $79/4$ & $\frac{1}{3}\bigl(2\sqrt{111}-13\bigr)$ & $\frac{3}{2} + \frac{1}{6} \bigl(2 \sqrt{111}-13\bigr)$ \\ \addlinespace
        & $(2,2;0)$ & $20$ & $\frac{1}{3} \bigl(\sqrt{449}-13\bigr)$ & $\frac{3}{2} + \frac{1}{6} \bigl(\sqrt{449}-13\bigr)$ \\ \addlinespace
        $\frac{1}{2}(i\slashed{D}_{3/2}^{(1)_-})$
        & $(0,0;0)$ & $0$ & $\frac{4}{3}$ & $\frac{3}{2} + \frac{2}{3}$ \\ \addlinespace
        $\frac{1}{2}(i\slashed{D}_{3/2}^{(3)_-})$
        & $(1,0;1)$ & $19/4$ & $-\frac{7}{3}$ & $\frac{3}{2} + \frac{7}{6}$ \\ \addlinespace
        & $(2,0;0)$ & $6$ & $-\frac{8}{3}$ & $\frac{3}{2} + \frac{4}{3}$ \\ \addlinespace
        $\frac{1}{2}(i\slashed{D}_{3/2}^{(4)_-})$
        & $(1,1;1)$ & $39/4$ & $-\frac{1}{3} \bigl(2 \sqrt{61} - 7\bigr)$ & $\frac{3}{2} + \frac{1}{6} \bigl(2 \sqrt{61}-7\bigr)$ \\ \addlinespace
        & $(0,1;2)$ & $10$ & $-\frac{1}{3}\bigl(\sqrt{249}-7\bigr)$ & $\frac{3}{2} + \frac{1}{6} \bigl(\sqrt{249}-7\bigr)$
        \\ \addlinespace[\aboverulesep] \bottomrule
    \end{tabular}
    \caption{Spinor fields with $2 < E_0 \leq 3$ in the left-squashed vacuum.}
    \label{tab:E0>2_spinors:LS}
\end{table}

\begin{table}[H]
    \centering
    \setlength{\defaultaddspace}{3pt}
    \setlength{\tabcolsep}{10pt}
    \begin{tabular}{lllll}
        \toprule
        Field ($RS^7$) & $(p,q;r)$ & $C_g$ & $M_R/m$ & $E_0$
        \\ \midrule
        $\frac{1}{2}(i\slashed{D}_{1/2}^{(1)_+})$
        & $(1,1;1)$ & $39/4$ & $\frac{2}{3} \bigl(\sqrt{69}-6\bigr)$ & $\frac{3}{2} + \frac{1}{3} \bigl(\sqrt{69}-6\bigr)$ \\ \addlinespace
        & $(0,2;0)$ & $10$ & $\frac{1}{3} \bigl(\sqrt{281} - 12\bigr)$ & $\frac{3}{2} + \frac{1}{6} \bigl(\sqrt{281}-12\bigr)$ \\ \addlinespace
        & $(2,0;2)$ & $12$ & $\frac{1}{3}\bigl(\sqrt{321}-12\bigr)$ & $\frac{3}{2} + \frac{1}{6} \bigl(\sqrt{321}-12\bigr)$ \\ \addlinespace
        & $(1,2;1)$ & $67/4$ & $\frac{4}{3} \bigl(\sqrt{26}-3\bigr)$ & $\frac{3}{2} + \frac{2}{3} \bigl(\sqrt{26}-3\bigr)$ \\ \addlinespace
        & $(0,3;0)$ & $18$ & $3$ & $\frac{3}{2} + \frac{3}{2}$ \\ \addlinespace
        & $(2,1;2)$ & $18$ & $3$ & $\frac{3}{2} + \frac{3}{2}$ \\ \addlinespace
        $\frac{1}{2}(i\slashed{D}_{1/2}^{(2)_+})$
        & $(0,2;2)$ & $16$ & $\frac{1}{3}\bigl(3\sqrt{41}-14\bigr)$ & $\frac{3}{2} + \frac{1}{6} \bigl(3 \sqrt{41}-14\bigr)$ \\ \addlinespace
        & $(3,0;1)$ & $51/4$ & $\frac{2}{3} \bigl(2 \sqrt{19}-7\bigr)$ & $\frac{3}{2} + \frac{1}{3} \bigl(2 \sqrt{19}-7\bigr)$ \\ \addlinespace
        & $(1,2;1)$ & $67/4$ & $\frac{2}{3} \bigl(4 \sqrt{6}-7\bigr)$ & $\frac{3}{2} + \frac{1}{3} \bigl(4 \sqrt{6}-7\bigr)$ \\ \addlinespace
        & $(2,1;2)$ & $18$ & $\frac{1}{3} \bigl(\sqrt{409}-14\bigr)$ & $\frac{3}{2} + \frac{1}{6} \bigl(\sqrt{409}-14\bigr)$ \\ \addlinespace
        & $(1,1;3)$ & $75/4$ & $\frac{2}{3} \bigl(\sqrt{106}-7\bigr)$ & $\frac{3}{2} + \frac{1}{3} \bigl(\sqrt{106}-7\bigr)$ \\ \addlinespace
        & $(3,1;1)$ & $79/4$ & $\frac{2}{3} \bigl(\sqrt{111}-7\bigr)$ & $\frac{3}{2} + \frac{1}{3} \bigl(\sqrt{111}-7\bigr)$ \\ \addlinespace
        & $(2,2;0)$ & $20$ & $\frac{1}{3} \bigl(\sqrt{449}-14\bigr)$ & $\frac{3}{2} + \frac{1}{6} \bigl(\sqrt{449}-14\bigr)$ \\ \addlinespace
        & $(3,0;3)$ & $87/4$ & $\frac{8}{3}$ & $\frac{3}{2} + \frac{4}{3}$ \\ \addlinespace
        & $(4,0;2)$ & $22$ & $\frac{1}{3} \bigl(\sqrt{489}-14\bigr)$ & $\frac{3}{2} + \frac{1}{6} \bigl(\sqrt{489}-14\bigr)$ \\ \addlinespace
        & $(0,3;2)$ & $24$ & $3$ & $\frac{3}{2} + \frac{3}{2}$ \\ \addlinespace
        $\frac{1}{2}(i\slashed{D}_{3/2}^{(1)_-})$
        & $(0,0;0)$ & $0$ & $\frac{5}{3}$ & $\frac{3}{2} + \frac{5}{6}$ \\ \addlinespace
        $\frac{1}{2}(i\slashed{D}_{3/2}^{(1)_+})$
        & $(0,1;0)$ & $4$ & $-\frac{5}{3}$ & $\frac{3}{2} + \frac{5}{6}$ \\ \addlinespace
        $\frac{1}{2}(i\slashed{D}_{3/2}^{(2)_+})$
        & $(1,1;1)$ & $39/4$ & $-\frac{2}{3}\bigl(\sqrt{51}-3\bigr)$ & $\frac{3}{2} + \frac{1}{3} \bigl(\sqrt{51}-3\bigr)$ \\ \addlinespace
        $\frac{1}{2}(i\slashed{D}_{3/2}^{(3)_+})$
        & $(0,0;2)$ & $6$ & $-3$ & $\frac{3}{2} + \frac{3}{2}$ \\ \addlinespace
        & $(2,0;0)$ & $6$ & $-3$ & $\frac{3}{2} + \frac{3}{2}$
        \\ \addlinespace[\aboverulesep] \bottomrule
    \end{tabular}
    \caption{Spinor fields with $2 < E_0 \leq 3$ in the right-squashed vacuum.}
    \label{tab:E0>2_spinors:RS}
\end{table}

\section{Derivation of the supermultiplets}
\label{sec:supermultiplets:details}
\settocdepth{section}
\setsecnumdepth{section}

In this appendix, we explain, in detail, how one can associate $E_0$ values  and cross diagrams  to each of the supermultiplets as summarized in
\cref{spinonetotwomultiplets,wesszuminopositivecgmultiplets,wesszuminocgequaltozero}.

In the process of explaining this we also obtain information of how to assign unique operator eigenvalues to the various cross diagrams, or single crosses in some special cases.
Based on the results of \cite{Nilsson:2018lof,Karlsson:2021oxd} we here  use supersymmetry to carry this analysis as far as possible but a
second goal is to  correlate these results with the ones obtained in \cref{sec:eigenmodes} where we   provide  a direct  connection between
cross diagrams and operator eigenvalues. This is particularly important for the novel results concerning
$\Delta_2$ which will give  new important insights into the structure of supermultiplets.
Thus, we will  use the notation for cross diagrams and their eigenvalues introduced in \cref{sec:eigenmodes} and in some cases compare them to the cross diagrams listed in the appendix of \cite{Nilsson:2018lof}.
Alternatively, one can start from the results of \cref{sec:eigenvalues,sec:eigenmodes} and arrive at the same results without using \cite{Nilsson:2018lof} except for the cross diagrams of the supermultiplets only associated to mode types that were not analysed in \cref{sec:eigenmodes}, specifically the results in \cref{fig:coloured_cross_diagram} below.

In what follows, we do not look for numerical coincidences that could give rise to exceptional supermultiplets for small $p$ or $q$.
Although we do not present the details, we have verified that no such exceptions are possible when accounting for the full spectrum of supermultiplets.

\subsection{Spin-2 supermultiplets}
We start from the top of the list of supermultiplets, \crefrange{spintwo}{wz} above, and follow it downwards. The first one is thus the spin $2^+$ supermultiplets
\begin{equation}
    D(E_0,2^+:\Delta_0)\oplus D(E_0-\tfrac{1}{2}, \tfrac{3}{2}:i\slashed{D}_{1/2})\oplus D(E_0+\tfrac{1}{2}, \tfrac{3}{2}:i\slashed{D}_{1/2})\oplus D(E_0, 1^+:\Delta_2),
\end{equation}
where we have included the operators responsible for the different spectra. The values of $E_0$ are computed from the operator eigenvalues
as follows: Using  \cref{table:massop} and  \cref{eq:eigs:summary:Delta_0:1} we get $M^2(2^+)=\Delta_0$ with eigenvalues $\Delta_0^{(1)}=\frac{20m^2}{9}C_g$ which gives (see \cref{ezerolist})
\begin{equation}
    E_0(2^+)=\frac{3}{2}+\frac{1}{2}\sqrt{\left(\frac{M}{m}\right)^2+9}=\frac{3}{2}+\frac{1}{2}\sqrt{\frac{20}{9}C_g+9}=\frac{3}{2}+\frac{1}{6}\sqrtsmash[C_G]{20C_g+81},
\end{equation}
applicable to all isometry irreps in the cross diagram of $\Delta_0$, that is,
all irreps $(p,q;r)$ with $p=r\ge 0$, $q\ge 0$ (see \cite{Nilsson:2018lof} or \cref{fig:modes:0-form:cross_diagrams}).
There is, however, a special case namely the short massless
supergravity multiplet in the irrep $(0,0;0)$. Looking more carefully at the cross diagrams for the different fields in the spin $2^+$ supermultiplets one finds that the cross
diagrams of \cite{Nilsson:2018lof} are related as follows\footnote{The signs appearing as indices on the 3/2 components refer to the sign in their energies $E_0\pm\frac{1}{2}$.}
\begin{gather}
\nonumber
    2^+\colon [\Delta_0](1),\
    (3/2)_{\scriptscriptstyle (-)}\colon [i\slashed{D}_{1/2}](1)=[\Delta_0](1),\
    (3/2)_{\scriptscriptstyle (+)}\colon [i\slashed{D}_{1/2}](2)=[\Delta_0](1)-(0,0;0), \\
    1^+\colon [\Delta_2](2+3-1)=[\Delta_0](1)-(0,0;0),
\end{gather}
where the bracket $[...]$ enclosing the operator indicates that it refers to a cross diagram and not an eigenvalue.
Here $[i\slashed{D}_{1/2}](1)$ refers to the first cross diagram in the list of $i\slashed{D}_{1/2}$ in the appendix of \cite{Nilsson:2018lof} while $[\Delta_2](2+3-1)$ means that one should add
the $\Delta_2$ cross diagrams number 2 and 3 and subtract diagram number 1. Thus we see that the last two fields in the supermultiplets have cross diagrams that do not contain the irrep
$(0,0;0)$ which therefore gives rise to a short supermultiplet, the supergravity multiplet. In particular this means that the Dirac eigenvalues are associated with cross diagrams as follows
\begin{align}
    & i\slashed{D}_{1/2}(1)=\frac{m}{2}+\frac{m}{3}\sqrtsmash[C_G]{20C_g+81} \quad\implies\quad
    E_0=\frac{3}{2}-\frac{1}{2}+\frac{1}{6}\sqrtsmash[C_G]{20C_g+81}, \\
    & i\slashed{D}_{1/2}(2)=\frac{m}{2}-\frac{m}{3}\sqrtsmash[C_G]{20C_g+81} \quad\implies\quad
    E_0=\frac{3}{2}+\frac{1}{2}+\frac{1}{6}\sqrtsmash[C_G]{20C_g+81},
\end{align}
which follows from the old results of \cite{Nilsson:1983ru} or the results in this paper.\footnote{Note  that the  sign in front of
the first term  cannot be deduced from the eigenvalue calculations  in \cref{sec:eigenvalues} of this paper, which are based on computing $(i{\slashed D}_{1/2})^2$, but follows here instead from
implementing supersymmetry. The mode calculation in \cref{sec:eigenmodes}, on the other hand, does provide the sign.}

Similarly, we conclude that, in analogy with $\Delta_0(1)=\frac{m^2}{9}\,20C_g$, we have
\begin{equation}
    \Delta_2(2+3-1)=\frac{m^2}{9}(20C_g+72) \quad\implies\quad
    E_0(1^+)=\frac{3}{2}+\frac{1}{6}\sqrtsmash[C_G]{20C_g+81},
\end{equation}
with the same $E_0$ as for the spin $2^+$ component, as required by supersymmetry.

Using the notation of \cref{sec:eigenmodes} we can summarize the result as follows: The four fields in the $2^+$ supersmultiplet make use of the following cross diagrams and eigenvalues
\begin{equation}
    2^+(\Delta_0^{(1)}),\quad
    \frac{3}{2}(i\slashed{D}_{1/2}^{(1)_+}),\quad
    \frac{3}{2}(i\slashed{D}_{1/2}^{(1)_-}),\quad
    1^+(\Delta_2^{(1)}),
\end{equation}
given in \cref{fig:modes:0-form:cross_diagrams,fig:modes:spinor:cross_diagrams,fig:modes:2-form:cross_diagrams}.
It is interesting to note that the
combination of diagrams from the appendix of \cite{Nilsson:2018lof} appearing in $[\Delta_2](2+3-1)=[\Delta_0](1)-(0,0;0)$ comes out automatically in the mode calculation
presented in \cref{sec:eigenmodes} as $\Delta_2^{(1)}$. This will happen many times in the following when comparing these two approaches but from now on we do not give any details on this issue.
We should mention that apart from combining entire cross diagrams it is also necessary in some cases to move columns or rows, or just single crosses, between diagrams
in  \cite{Nilsson:2018lof} in order to generate the ones that come out of the mode function analysis in \cref{sec:eigenmodes}.
This is to be expected since the analysis of \cite{Nilsson:2018lof} only provides the mutliplicity of the isometry irreps for the various operators; the split into separate cross diagrams in \cite{Nilsson:2018lof} is to some extent arbitrary.
\subsection{Spin-3/2 supermultiplets}
Turning to the next set of supermultiplets, the six spin-3/2  ones, we have
\begin{equation}
D(E_0,\tfrac{3}{2}:i\slashed{D}_{1/2})\oplus D(E_0-\tfrac{1}{2}, 1^{\pm}:\Delta_{2\,\text{or}\,1})\oplus D(E_0+\tfrac{1}{2}, 1^{\mp}:\Delta_{1\,\text{or}\,2})\oplus D(E_0, \tfrac{1}{2}:i\slashed{D}_{3/2}).
\end{equation}
Here we must relate each of the two spin 1 parity assignments  to three cross diagrams, a fact that  follows by checking how supersymmetry works starting from the
six cross diagrams for the spin-3/2 component.
We conclude immediately that these are determined by the six Dirac cross
diagrams not already used above,  all associated to the eigenvalues with a $\sqrtsmash[C_G]{20C_g+49}$. The reason for this is that we must  use up all the
Dirac modes associated to spin-3/2 fields, which is the first component of the spin-3/2 supermultiplet.

All the forty $i\slashed{D}_{3/2}$ cross diagrams in the appendix of \cite{Nilsson:2018lof} must appear a single time each when looking at the entire spectrum. The fact that the spin-1/2 fields
in the spin-3/2  supermultiplets are related to the $i\slashed{D}_{3/2}$ and not $i\slashed{D}_{1/2}$ is  a consequence of supersymmetry which uniquely gives this answer.

From the results derived in \cite{Nilsson:1983ru} (or in \cref{sec:eigenmodes} of this paper) we have
\begin{equation}
    i\slashed{D}_{1/2}(3, 5, 7)=-\frac{m}{6}-\frac{m}{3}\sqrtsmash[C_G]{20C_g+49}\ \implies\ E_0(3/2)=\frac{3}{2}+\frac{5}{6}+\frac{1}{6}\sqrtsmash[C_G]{20C_g+49},
\end{equation}
where the first spin-1 field in the supermultiplet  is $1^-$ (see below), and
\begin{equation}
    i\slashed{D}_{1/2}(4, 6, 8)=-\frac{m}{6}+\frac{m}{3}\sqrtsmash[C_G]{20C_g+49}\ \implies\ E_0(3/2)=\frac{3}{2}-\frac{5}{6}+\frac{1}{6}\sqrtsmash[C_G]{20C_g+49}
\end{equation}
where the first spin-1 field is $1^+$ (see below). In the notation of \cref{sec:eigenmodes} these are, respectively, denoted $i\slashed{D}_{1/2}^{(2)_-}$ and $i\slashed{D}_{1/2}^{(2)_+}$,
see \cref{fig:modes:spinor:cross_diagrams}.

To establish the  results for the spin-1 components  is here a bit more intricate than for the $2^+$ multiplets above.
The spin 1 content of these six branches (cross diagrams) is three each of $1^+$ and $1^-$. However, the six $\Delta_1$ cross diagrams  are used twice since the mass square operator has
two square root branches, see \cref{table:massop}. The second occurrence of these six sets of
cross diagrams will appear in the six sets of $1^-$ supermultiplets discussed below.  The two sets of $\pm$ signs gives four different sets of
$E_0$ values and all four are needed for everything to work out. To see this we recall the derivation of the $E_0(1^-)$ values here.

To obtain the $E_0(1^-)$ values we  start from  \cref{ezerolist},
\begin{equation}
    E_0(1^-)=\frac{3}{2}+\frac{1}{2}\sqrt{M^2/m^2+1},
\end{equation}
where  $M^2$ has two branches%
(see \cref{table:massop})%
\footnote{Recalling from \cite{Duff:1986hr} that $\Delta_1\ge 7m^2$ we see that $\sqrt{\Delta_1+4m^2}\ge 3m$.}
\begin{equation}
    M^2(1^-_{\scriptscriptstyle (\pm)})=\Delta_1+12m^2\pm6m\sqrt{\Delta_1+4m^2}=(\sqrt{\Delta_1+4m^2}\pm 3m)^2-m^2,
\end{equation}
where also $\Delta_1$ has two branches, denoted $\Delta_1^{(1)_{\pm}}$ in \cref{eq:eigs:summary:Delta_1:1},
\begin{equation}
    \Delta_1=\frac{m^2}{9} \bigl(20C_g+14 \pm 2 \sqrtsmash[C_G]{20C_g+49}\bigr)=\frac{m^2}{9}\bigl(\sqrtsmash[C_G]{20C_g+49}\pm 1\bigr)^2-4m^2.
\end{equation}
Inserting the last expression into the one immediately above it gives
\begin{equation}
    M^2(1^-)=\frac{m^2}{9}\bigl(\sqrtsmash[C_G]{20C_g+49}\pm 1\pm 9\bigr)^2 - m^2.
\label{massspinoneminus_doubleplusminus}
\end{equation}
The sign choice associated with $\Delta_1$ is the one that must be correlated to the different cross diagrams as will be done shortly.

The four choices of signs in \cref{massspinoneminus_doubleplusminus} give $-10, -8, 8, 10$ to be added to the square root in $M^2$. Using $10$ and $-10$ we
get\footnote{Note that $\sqrtsmash[C_G]{20C_g+49}\ge 10$ for all irreps in the relevant Dirac cross diagrams.}
\begin{equation}
\label{spinonezerofivethirds}
    E_0(1^-)=\frac{3}{2}\pm\frac{5}{3}+\frac{1}{6}\sqrtsmash[C_G]{20C_g+49},
\end{equation}
while $8$ and  $-8$  give
\begin{equation}
\label{spinonezerofourthirds}
    E_0(1^-)=\frac{3}{2}\pm\frac{4}{3}+\frac{1}{6}\sqrtsmash[C_G]{20C_g+49}.
\end{equation}

The first set of $E_0(1^-)$ values will be needed in the spin $1^-$ supermultiplets as demonstrated below while the second set fit into the spin $3/2$ ones analysed here.
This last statement follows since the values to be added to $\frac{3}{2}$ are $\pm\frac{5}{6}$ which arise as $\frac{4}{3}-\frac{1}{2}=\frac{5}{6}$ and $-\frac{4}{3}+\frac{1}{2}=-\frac{5}{6}$ when relating the different values of $E_0$ in the multiplet.
We therefore conclude that the content of the six 3/2 supermultiplets is as given above in terms of the numbered   Dirac cross diagrams.

Note that the six cross diagrams of $\Delta_1$ are identical to the $\sqrtsmash[C_G]{20C_g+49}$ Dirac ones. This is clear from
\cref{fig:modes:1-form:cross_diagrams,fig:modes:spinor:cross_diagrams}
which also correlates the signs in their operator
eigenvalues.\footnote{The same result is obtained  by supersymmetry alone, namely that the    $\Delta_1$ cross diagrams
numbered 1, 3 and 5 in \cite{Nilsson:2018lof} have eigenvalues from the minus branch
and  diagrams 2, 4 and 6 from the  plus branch.}
This assignment of signs must be a consistent choice also in the $1^-$ supermultiplets
which it is as we will see below. %

The $1^+$ field in this supermultiplet has to come from modes with eigenvalues given by $\Delta_2^{(2)_\pm}$ due to the $\sqrtsmash[C_G]{20 C_g + 49}$.
For these, we have
\begin{equation}
    \Delta_2^{(2)_\pm} = \frac{m^2}{9} \Bigl(20 C_g + 44 \pm 4 \sqrtsmash[C_G]{20 C_g + 49}\Bigr) \ \implies\
    E_0(1^+) = \frac{3}{2} \pm \frac{2}{6} + \frac{1}{6} \sqrtsmash[C_G]{20 C_g + 49}.
\end{equation}
Hence, the fields associated to $\Delta_2^{(2)_\pm}$ fit into the supermultiplet containing the spin-3/2 field associated to $i\slashed{D}_{1/2}^{(2)_\mp}$.
Note that the corresponding cross diagrams in \cref{fig:modes:spinor:cross_diagrams,fig:modes:2-form:cross_diagrams} agree.

The $i\slashed{D}_{3/2}$-eigenvalues that should be used for the spin $1/2$ components to fit into these six spin $3/2$ supermultiplets are (see \cref{eq:eigs:summary:D_3/2:3})
\begin{equation}
    i\slashed{D}_{3/2}^{(3)\pm}= \frac{m}{3}\biggl(\frac{1}{2}\pm \sqrtsmash[C_G]{20C_g+49}\biggr)\ \implies\
    E_0(1/2)=\frac{3}{2}\pm\frac{5}{6}+\frac{1}{6}\sqrtsmash[C_G]{20C_g+49},
\end{equation}
where the eigenvalues with a plus and a minus are associated with, respectively, the same cross diagrams as those with eigenvalues $ i\slashed{D}_{1/2}^{(2)-}$ and  $ i\slashed{D}_{1/2}^{(2)+}$
(see \cref{eq:eigs:summary:D_1/2:2} and \cref{fig:modes:spinor:cross_diagrams}).

We can summarise the situation for the six spin $3/2$ supermultiplets, using the notation of \cref{sec:eigenmodes}, as follows (ordered as $E_0, E_0-\frac{1}{2}, E_0+\frac{1}{2}, E_0$)
\begin{align}
    & \frac{3}{2}_1\colon \quad
    \frac{3}{2}(i\slashed{D}_{1/2}^{(2)_-}), \
    1^+(\Delta_2^{(2)_+}),\
    1^-_{\scriptscriptstyle (+)}(\Delta_1^{(1)_-}),\
    \frac{1}{2}(i\slashed{D}_{3/2}^{(3)_+}),\\[6pt]
    & \frac{3}{2}_2\colon \quad
    \frac{3}{2}(i\slashed{D}_{1/2}^{(2)_+}),\
    1^-_{\scriptscriptstyle (-)}(\Delta_1^{(1)_+}),\
    1^+(\Delta_2^{(2)_-}),\
    \frac{1}{2}(i\slashed{D}_{3/2}^{(3)_-}).
\end{align}
We emphasise  that since for the spin $3/2$ operator $i\slashed{D}_{3/2}$ we have no independent way to relate eigenvalues to cross
diagrams (as was done in \cref{sec:eigenmodes} for $i\slashed{D}_{1/2}$, $\Delta_1$ and $\Delta_2$) we have here relied entirely on supersymmetry and used the cross diagrams of \cite{Nilsson:2018lof}.

\subsection{The \texorpdfstring{$1^-$}{1-} supermultiplets}
The six sets of $1^-$ supermultiplets  are of the form
\begin{equation}
    D(E_0,1^-:\Delta_1)\oplus D(E_0-\tfrac{1}{2}, \tfrac{1}{2}:i\slashed{D}_{1/2\,\text{or}\,3/2})\oplus D(E_0+\tfrac{1}{2}, \tfrac{1}{2}:i\slashed{D}_{3/2\,\text{or}\,1/2})\oplus D(E_0, 0^-:Q),
\end{equation}
where  $i\slashed{D}_{1/2}$ and $i\slashed{D}_{3/2}$ occur in a  similar way to how in the previous case there were three supermultiplets with spin-1 fields coming from $\Delta_1$ and $\Delta_2$.
That this happens here is a direct consequence of supersymmetry and is easily established using  the known eigenvalues of the various operators.

The eigenvalues of the  $3+3$ cross diagrams of $\Delta_1$ were in this case identified above as the ones coming from 10 and -10 in the expression for $E_0(1^-)$ in  \cref{spinonezerofivethirds}. Thus
\begin{align}
    \Delta_1^{(1)_+}\colon\ E_0(1^-_{\scriptscriptstyle (+)})=\frac{3}{2}+\frac{5}{3}+\frac{1}{6}\sqrtsmash[C_G]{20C_g+49}, \\[6pt]
    \Delta_1^{(1)_-}\colon\ E_0(1^-_{\scriptscriptstyle (-)})=\frac{3}{2}-\frac{5}{3}+\frac{1}{6}\sqrtsmash[C_G]{20C_g+49},
\end{align}
where the cross diagrams associated to the two sets of $\Delta_1^{(1)\pm}$-eigenvalues are specified in  \cref{fig:modes:1-form:cross_diagrams}.

The last component of these supermultiplets is the pseudo-scalar $0^-$. Using the $\sqrtsmash[C_G]{20C_g+49}$ eigenvalues of $Q$ in \cref{eq:eigs:summary:Q:2}
\begin{equation}
    Q^{(2)\pm}=\frac{m}{3} \bigl(1\pm\sqrtsmash[C_G]{20C_g+49}\bigr),
\end{equation}
and the $0^-$ mass operator from \cref{table:massop}
\begin{equation}
    M^2(0^-)=(Q+3m)^2-m^2,
\end{equation}
we find as expected that  all six  cross diagrams satisfy
\begin{equation}
    E_0(0^-)=E_0(1^-).
\end{equation}

Finally, we must check that also the spin-1/2 fields fit into these $1^-$ supermultiplets.
It is interesting to note that we need to use both $i\slashed{D}_{1/2}$ and $i\slashed{D}_{3/2}$
eigenvalues for  this to work.
Let us start with the spin-1/2 energies $E_0=\frac{3}{2}\pm\frac{1}{2}|M/m|$ where the minus sign cannot occur in these supermultiplets due to the unitarity bound $E_0(1) \ge 2$.
The mass related to the $i\slashed{D}_{1/2}$ and $i\slashed{D}_{3/2}$ operators are (see \cref{table:massop}, \cref{eq:eigs:summary:D_1/2:2} and \cref{eq:eigs:summary:D_3/2:4})
\begin{align}
    & M(1/2, i\slashed{D}_{1/2}^{(2)_{\pm}})
    =- i\slashed{D}_{1/2}-\frac{9m}{2}
    =-\frac{13m}{3}\mp\frac{m}{3}\sqrtsmash[C_G]{20C_g+49}, \\[2pt]
    & M(1/2, i\slashed{D}_{3/2}^{(4)_{\pm}})
    = i\slashed{D}_{3/2}+\frac{3m}{2}
    =\frac{7m}{3}\pm\frac{m}{3}\sqrtsmash[C_G]{20C_g+49},
\end{align}
where we should note that it is the second set of $\sqrtsmash[C_G]{20C_g+49}$ eigenvalues of $i\slashed{D}_{3/2}$ that must be used here. This produces the following two sets of energy
values (for Dirichlet boundary conditions)
\begin{align}
    & E_0(1/2, i\slashed{D}_{1/2}^{(2)_{\pm}})=\frac{3}{2}\mp\frac{13}{6}+\frac{1}{6}\sqrtsmash[C_G]{20C_g+49},\\
    & E_0(1/2, i\slashed{D}_{3/2}^{(4)_{\pm}})=\frac{3}{2}\pm\frac{7}{6}+\frac{1}{6}\sqrtsmash[C_G]{20C_g+49}.
\end{align}
Hence the supermultiplets are (ordered as $(E_0, E_0-1/2, E_0+1/2, E_0)$),
in terms of the notation of \cref{sec:eigenmodes},
\begin{align}
    1^-_1\colon\quad
    1_{\scriptscriptstyle (+)}^-(\Delta_1^{(1)_+}), \
    \frac{1}{2}(i\slashed{D}_{3/2}^{(4)_+}), \
    \frac{1}{2}(i\slashed{D}_{1/2}^{(2)_+}), \
    0^-(Q^{(2)_+}), \\[6pt]
    1^-_2\colon\quad
    1_{\scriptscriptstyle (-)}^-(\Delta_1^{(1)_-}), \
    \frac{1}{2}(i\slashed{D}_{1/2}^{(2)_-}), \
    \frac{1}{2}(i\slashed{D}_{3/2}^{(4)_-}), \
    0^-(Q^{(2)_-}),
\end{align}
where as before the lower sign in the bracket $1^-_{\scriptscriptstyle (\pm)}$ indicates which branch of $M^2$ is used.
These results for the  cross diagrams imply that there are two short gauge multiplets when $C_g=6$, and hence $E_0=2$, connected to the
two adjoint irreps $(2,0;0)$  and $(0,0;2)$ of the isometry group $Sp_2\times Sp^C_1$.

\subsection{The \texorpdfstring{$1^+$}{1+} supermultiplets}
So far, $\mN=1$ supersymmetry, together with the operator eigenvalues derived  in previous works and the result for the spin-3/2 operator obtained in \cite{Karlsson:2021oxd},
have  been sufficient to uniquely fix all relations between operator eigenvalues and isometry irreps (that is, cross diagrams). This nice state of affairs will no longer
persist when we now turn to the eight spin $1^+$ supermultiplets, each with the content
\begin{equation}
    D(E_0,1^+:\Delta_2)\oplus D(E_0-\tfrac{1}{2} , \tfrac{1}{2}: i\slashed{D}_{3/2})\oplus D(E_0+\tfrac{1}{2}, \tfrac{1}{2}:i\slashed{D}_{3/2})\oplus D(E_0, 0^+:\Delta_L),
\end{equation}
where the potential  ambiguity for the last entry, the $0^+$ scalars, which could be given by either  $\Delta_0$ or $\Delta_L$ is settled by supersymmetry.

It is clear that the remaining eight cross diagrams for $\Delta_2$ in   \cref{fig:modes:2-form:cross_diagrams}
account for the ${\bf 3}$ and ${\bf 5}$ spin $1^+$ components of these supermultiplets.
These must therefore all be associated with the eigenvalues \cref{eq:eigs:summary:Delta_2:3} and energies
\begin{equation}
    \Delta_2=\frac{m^2}{9}\,20 C_g \quad\implies\quad
    E_0(1^+,\Delta_2)=\frac{3}{2}+\frac{1}{6}\sqrtsmash[C_G]{20C_g+9}.
\end{equation}
Now we understand that the spin-1/2 fields
must be associated with $i\slashed{D}_{3/2}$ since the eigenvalues of $i\slashed{D}_{1/2}$ do not contain $\sqrtsmash[C_G]{20C_g+9}$.

Similarly, the scalar fields $0^+$ are related to $\Delta_L$-eigenvalues \cref{eq:eigs:summary:Delta_L:1} and the energies
\begin{equation}
    \Delta_L=\frac{m^2}{9}\,(20 C_g + 36)\quad\implies\quad
    E_0(0^+,\Delta_L)=\frac{3}{2}+\frac{1}{6}\sqrtsmash[C_G]{20C_g+9}.
\end{equation}

The unexpected feature that appears here is that neither of the two sets   of  eigenvalues, related to  the ${\bf 3}$ and ${\bf 5}$ cross diagrams,  have a negative square root branch (cf.\ $\Delta_2^{(2)_\pm}$). This means that
there is a degeneracy in the spectrum since different modes in the same isometry irrep must pairwise be associated with the same operator
eigenvalue. This must happen for all the fields in these $1^+$ supermultiplets whose irreps are of the kind $(p,q;r)$ with either $r=p$ or $r=p\pm 2$.
As we will discover below when discussing the Wess--Zumino supermultiplets,
the ones containing scalar fields related to $\Delta_L$, and only these ones, will display the same phenomenon.

This being concluded here, we should now recall the results of \cref{sec:eigenmodes} on the $\Delta_2$ eigenmodes and their eigenvalues which support this conclusion.
In fact, in \cref{sec:eigenmodes}  we identified  explicitly the different mode functions  that have the same $\Delta_2$-eigenvalues, $ \Delta_2=\frac{m^2}{9}\,20 C_g$, and the modes are precisely the ones
in the ${\vec 3}$ and ${\vec 5}$ cross diagrams given above. The notation used in \cref{sec:eigenmodes} to make this clear was to denote eigenvalues associated with the mode functions (and thus the cross diagrams)
related to ${\vec 3}$ by
$\Delta_2^{(3)}$ while those related to ${\vec 5}$ was denoted $\Delta_2^{(3)'}$ despite the fact that their eigenvalues (and masses and $E_0$) are the same.
Unfortunately, why this happens is not  clear from the results we have obtained so far. We mention here the fact (which is further discussed in the Conclusions, \cref{sec:conclusions})
that this degeneracy occurs only in supermultiplets related to modes of  $\Delta_L$.%
\footnote{However, not in all supermultiplets containing $\Delta_L$ since the ones in ${\bf 5}$ with $r=p\pm 4$ are not degenerate.}

To conclude the discussion of the $1^+$ multiplets we need also the  spin-3/2 eigenvalues
\begin{equation}
    i\slashed{D}_{3/2}^{(2)_{\pm}}=-\frac{m}{2}\pm\frac{m}{3}\sqrtsmash[C_G]{20C_g+9}\ \implies\
    E_0(1/2,i\slashed{D}_{3/2}^{(2)_{\pm}})=\frac{3}{2}\pm\frac{1}{2}+\frac{1}{6}\sqrtsmash[C_G]{20C_g+9},
\end{equation}
which completes the two $1^+$ lines in \cref{spinonetotwomultiplets}. The relevant cross diagrams are the same for all four entries in each line $1^+$  in \cref{spinonetotwomultiplets},
and coincide with those specified in \cref{fig:modes:2-form:cross_diagrams}.
It is now clear that there are, as expected,  no short multiplets in this case.

The situation for spin $1^+$ can be summarised as follows:
\begin{align}
    & 1^+_1\colon\quad
    1^+(\Delta_2^{(3)}),\
    \frac{1}{2}(i\slashed{D}_{3/2}^{(2)_-}),\
    \frac{1}{2}(i\slashed{D}_{3/2}^{(2)_+}),\
    0^+(\Delta_L^{(1)}), \\[6pt]
    & 1^+_2\colon\quad
    1^+(\Delta_2^{(3)'}),\
    \frac{1}{2}(i\slashed{D}_{3/2}^{(2)'_-}),\
    \frac{1}{2}(i\slashed{D}_{3/2}^{(2)'_+}),\
    0^+(\Delta_L^{(1)'}),
\end{align}
where we emphasise again that we in \cref{sec:eigenmodes} have derived the eigenvalues for the cross diagrams of ${\vec 3}$ and ${\vec 5}$ and found them to coincide, hence the notation
$\Delta_2^{(3)}$ and $\Delta_2^{(3)'}$.
For the rest of the spins, on the other hand, the corresponding mode calculations have not been done but it follows from supersymmetry that their cross diagrams are identical to the $\vec{3} \oplus \vec{5}$ ones of $\Delta_2^{(3),(3)'}$.

\subsection{The Wess--Zumino supermultiplets}
There are 14 sets of Wess--Zumino supermultiplets (see \cref{wz}) of which two sets must be related to the remaining modes of the Dirac operator with eigenvalues $\sqrtsmash[C_G]{20C_g+81}$. This fits nicely
with the fact that also the two $M_{\scriptscriptstyle (\pm)}^2(0^+)$ branches of scalars connected to $\Delta_0$ are left to account for. This explains the first two lines, denoted WZ1 and WZ2, of
\cref{wesszuminopositivecgmultiplets}.

Thus the WZ1 and WZ2 multiplets have $E_0$ and cross diagrams governed  by the spin-1/2 fields.  These have
$M(1/2)=-i\slashed{D}_{1/2}^{(1)_{\pm}}-\frac{9m}{2}$ with $i\slashed{D}_{1/2}^{(1)_{\pm}}=\frac{m}{2}\pm\frac{m}{3}\sqrtsmash[C_G]{20C_g+81}$ (see  \cref{table:massop} and  \cref{eq:eigs:summary:D_1/2:1})
where the modes for the upper sign fills a whole cross diagram and similarly for the modes with the lower sign except that the singlet is missing, see  \cref{fig:modes:spinor:cross_diagrams}.

The $E_0$ spectrum, valid for the upper sign on $i\slashed{D}_{1/2}^{(1)_{\pm}}$ and for all $C_g\ge 0$, is
\begin{equation}
    E_0(1/2, i\slashed{D}_{1/2}^{(1)_{+}})=\frac{3}{2}+\frac{1}{2} |M/m|=\frac{3}{2}+\frac{5}{2}+\frac{1}{6}\sqrtsmash[C_G]{20C_g+81},
\end{equation}

For the lower sign  the situation is slightly different due to the absolut value on the mass in $E_0$.  Now $C_g>0$ but $|M|=\frac{m}{3}|-15\mp\sqrtsmash[C_G]{20C_g+81}|$  depends on the relative size of the two terms under the absolut sign. We get the following two cases
\begin{alignat}{2}
    & E_0(1/2, i\slashed{D}_{1/2}^{(1)_{-}})=\frac{3}{2}+\frac{1}{2} |M/m|=\frac{3}{2}-\frac{5}{2}+\frac{1}{6}\sqrtsmash[C_G]{20C_g+81},\quad &&C_g\ge \frac{36}{5},\\[4pt]
    & E_0(1/2, i\slashed{D}_{1/2}^{(1)_{-}})=\frac{3}{2}+\frac{1}{2} |M/m|=\frac{3}{2}+\frac{5}{2}-\frac{1}{6}\sqrtsmash[C_G]{20C_g+81},\quad &&C_g < \frac{36}{5},
\end{alignat}
where the limit in the last case is only satisfied by the  two values $C_g=4, \frac{19}{4}$ (recall that the spectrum here involves only irreps $(p,q;p)$).
Note that we have here used only the upper (Dirichlet) sign in $ E_0(1/2)=\frac{3}{2}\pm\frac{1}{2} |M/m|$. The other case, related to Neumann boundary conditions,  is studied separately in \cref{spectrum_summary} for the five possible values of $C_g$ allowed by unitarity
(which occur only for the lower sign in $i\slashed{D}_{1/2}^{(1)_{\pm}}$  and not necessarily for all the members of the supermultiplets). Further details on the cases allowing for Neumann boundary conditions are given in \cref{sec:supermultiplets:skew-whiffing}.

It is now interesting to compare the situation above for the spin-1/2 fields with the one for the scalars.  From \cref{ezerolist} and \cref{table:massop}  we see  that $\Delta_0^{(1)}=\frac{m^2}{9}20C_g$ gives
a dependence of $M^2(0^+)$ on the absolut value of $\sqrtsmash[C_G]{20C_g+81}\pm 18$. This gives, for the plus branch (and for Dirichlet boundary conditions),
\begin{equation}
    E_0(0^+_{\scriptscriptstyle (+)})=\frac{3}{2}+ 3 +\frac{1}{6}\sqrtsmash[C_G]{20C_g+81},
\end{equation}
while for the minus branch we again find two cases
\begin{align}
    & E_0(0^+_{\scriptscriptstyle (-)})=\frac{3}{2}-3 +\frac{1}{6}\sqrtsmash[C_G]{20C_g+81},\quad C_g \ge \frac{243}{20},\\[4pt]
    & E_0(0^+_{\scriptscriptstyle (-)})=\frac{3}{2}+ 3 -\frac{1}{6}\sqrtsmash[C_G]{20C_g+81},\quad C_g < \frac{243}{20},
\end{align}
which contains the five values listed in \cref{wzcgvalueswithdandnbc}, that is, $C_g=4,\frac{19}{4}, \frac{39}{4}, 10, 12$.
Here the signs in the bracket on $0^+_{\scriptscriptstyle (\pm)}$ refer to the two branches of $M^2(0^+)$ in \cref{table:massop}.

 This also fits with the $\sqrtsmash[C_G]{20C_g+81}$ branch of eigenvalues for $Q$ (see \cref{eq:eigs:summary:Q:3} and \cref{table:massop})
 \begin{equation}
    Q^{(3)_{\pm}}=m\pm\frac{m}{3}\sqrtsmash[C_G]{20C_g+81} \ \implies\
    M^2(Q^{(3)_{\pm}})=\Bigl(4m\pm\frac{m}{3}\sqrtsmash[C_G]{20C_g+81}\Bigr)^2-m^2,
 \end{equation}
and hence $E_0$  depends on the absolut value of $\sqrtsmash[C_G]{20C_g+81}\pm 12$. This gives
$E_0$ values (for Dirichlet boundary conditions), noting that $\sqrtsmash[C_G]{20C_g+81} > 12$ for all irreps in WZ2,
\begin{equation}
    E_0(0^-,  Q^{(3)_{\pm}})=\frac{3}{2}+\frac{1}{2} \sqrt{(M/m)^2+1}=\frac{3}{2}\pm 2 +\frac{1}{6}\sqrtsmash[C_G]{20C_g+81}.
\end{equation}
Again the lower sign in $E_0$ is not discussed here and the reader is referred to \cref{spectrum_summary} for more details.

These results lead to the  two Wess--Zumino supermultiplets WZ1 and WZ2 in the Wess--Zumino table for $C_g>0$ in \cref{spectrum_summary} which turn out to have a slightly different
structure (note the order of the fields $(E_0, E_0-\frac{1}{2}, E_0+\frac{1}{2}))$:
\begin{alignat}{3}
    & \text{WZ1}_{C_g\ge 0} \colon\quad
    &&\frac{1}{2}(i\slashed{D}^{(1)_+}_{1/2}), \
    0^-(Q^{(3)_+}),\
    0_{\scriptscriptstyle (+)}^+(\Delta_0^{(1)}),\quad
    &&E_0=\frac{3}{2}+\frac{5}{2}+\frac{1}{6}\sqrtsmash[C_G]{20C_g+81},
    \\
    & \text{WZ2}_{C_g\ge \frac{243}{20}}\colon\quad
    &&\frac{1}{2}(i\slashed{D}^{(1)_-}_{1/2}), \
    0^+_{\scriptscriptstyle (-)}(\Delta_0^{(1)}),\
    0^-(Q^{(3)_-}),\quad
    &&E_0=\frac{3}{2}-\frac{5}{2}+\frac{1}{6}\sqrtsmash[C_G]{20C_g+81},
\end{alignat}
plus the special cases listed above to complete WZ2 for $0<C_g<\frac{243}{20}$. We note here that without involving Neumann boundary conditions there are no
Wess--Zumino supermultiplets with  $C_g=4,\frac{19}{4}, \frac{39}{4}, 10, 12$. These five cases are studied in detail in \cref{spectrum_summary}.

The $Q$ cross diagrams relevant for  fields in   WZ1 and WZ2  can be found in the appendix of \cite{Nilsson:2018lof}.%
\footnote{The cross diagrams in the appendix of \cite{Nilsson:2018lof} are related by $[i\slashed{D}_{1/2}](1)=[\Delta_0](1)=[Q](1)$, $[i\slashed{D}_{1/2}](2)=[\Delta_0](1)-(0,0;0)=[Q](6+7-5)$.}
By supersymmetry, they have to be identical to the $i\slashed{D}^{(1)_\pm}_{1/2}$ cross diagrams in \cref{fig:modes:spinor:cross_diagrams}.
Note that the singlet irrep $(0,0;0)$ should be removed from the irrep spectrum of $0^+_{\scriptscriptstyle (-)}(\Delta_0^{(1)})$ even though it appears in the cross diagram of $\Delta_0^{(1)}$, as explained in \cite{Duff:1986hr}, which is consistent with the cross diagrams of $i\slashed{D}^{(1)_\pm}_{1/2}$.
One should note that there is only one mode with $C_g=0$ here, and that this mode belongs to the multiplet WZ1. The multiplet WZ2 lacks
the $C_g=0$ mode which it must to be consistent with unitarity. The $C_g=0$ mode is discussed further in \cref{spectrum_summary}.

The final twelve sets of Wess--Zumino multiplets are in some sense less intriguing since  the only special
case is $C_g=0$ (see \cite{Nilsson:2023ctq}).%
\footnote{Note that in all three expressions for the mass discussed here the square root term is larger then the term without a square root when $C_g>0$.}
Leaving this case for a separate discussion we now explain the rows WZ3--6 in \cref{wesszuminopositivecgmultiplets}. Checking which operator eigenvalues that have not been used so far we
find that they are all of the $\sqrtsmash[C_G]{20C_g+1}$ form and occur only for $\Delta_L$, $Q$ and $i\slashed{D}_{3/2}$. The remaining cross diagrams in \cite{Nilsson:2018lof} are also
precisely the ones that are needed to form the Wess--Zumino supermultiplets denoted WZ3--6 in \cref{wesszuminopositivecgmultiplets}.
From the results of \cite{Nilsson:2018lof} we conclude that these should be grouped into two $Sp_1^C$ sets of ${\bf 1}+{\bf 5}$. To verify  this we need the  eigenvalues for these three operators
$\Delta_L$, $Q$ and $i\slashed{D}_{3/2}$. We start with the spin-1/2  ones (see \cref{eq:eigs:summary:D_3/2:1} and \cref{table:massop})
\begin{equation}
    i\slashed{D}_{3/2}^{(1)_{\pm}}=\frac{m}{6}\pm\frac{m}{3}\sqrtsmash[C_G]{20C_g+1} \ \implies\
    M(1/2)=i\slashed{D}_{3/2}+\frac{3m}{2}=4m\pm\frac{m}{3}\sqrtsmash[C_G]{20C_g+1}
\end{equation}
which give the spin-1/2 $E_0$ values in  all the WZ3--6 multiplets as
\begin{equation}
    E_0(1/2)=\frac{3}{2}+\frac{1}{2} |M/m|=\frac{3}{2}\pm \frac{5}{6}+\frac{1}{6}\sqrtsmash[C_G]{20C_g+1},
\end{equation}
where only the sign for Dirichlet boundary conditions is used.
One can check that Neumann boundary conditions  are impossible due to
unitarity.\footnote{Recall that Neumann boundary conditions are related to irreps with $E_0\le \frac{3}{2}$.}

The eigenvalues of the modes associated with the scalar  fields  are determined by \cref{eq:eigs:summary:Delta_L:2}
\begin{equation}
    \Delta_{L}=\frac{m^2}{9}\,\bigl(20C_g+32\pm 4\sqrtsmash[C_G]{20C_g+1}\bigl)=\frac{m^2}{9}\,\bigl(\sqrtsmash[C_G]{20C_g+1}\pm2\bigl)^2+3m^2,
\end{equation}
and with $M^2=\Delta_L-4m^2$ \cref{table:massop} we find
\begin{equation}
    E_0(0^+,\Delta_L^{(2)_{\pm}})=\frac{3}{2}\pm\frac{1}{3}+\frac{1}{6} \sqrtsmash[C_G]{20C_g+1},
\end{equation}
and the ones of the pseudo-scalars (see \cref{eq:eigs:summary:Q:1} and \cref{table:massop})
\begin{equation}
    Q^{(1)_{\pm}}=-\frac{m}{3}\pm \frac{m}{3} \sqrtsmash[C_G]{20C_g+1}\ \implies\
    M^2(0^-)=\biggl(\frac{8m}{3}\pm \frac{m}{3}\sqrtsmash[C_G]{20C_g+1}\biggl)^2-m^2,
\end{equation}
which leads to
\begin{equation}
    E_0(0^-)=\frac{3}{2}\pm\frac{4}{3}+\frac{1}{6} \sqrtsmash[C_G]{20C_g+1}.
\end{equation}
For the scalars $0^{\pm}$ Neumann boundary conditions are possible for the minus branch in their expressions for $E_0$ above but since Neumann boundary conditions
are  impossible for the spin-1/2 fields, as we saw above, supersymmetry implies that this is the case also for the scalars.
A full analysis of possible Neumann boundary conditions is carried out in \cref{sec:supermultiplets:skew-whiffing}, then for both the  left- and right-squashed vacua.

The above facts clearly  give rise to $C_g > 0$ Wess--Zumino supermultiplets of the  following  two kinds (ordered as $(E_0, E_0-\frac{1}{2}, E_0+\frac{1}{2})$)
\begin{align}
    &\text{WZ3--4}\colon\quad
    \frac{1}{2}(i\slashed{D}_{3/2}^{(1)_+}),\
    0^+(\Delta_L^{(2)_{+}}),\
    0^-(Q^{(1)_{+}}),\quad
    E_0=\frac{3}{2}+\frac{5}{6}+\frac{1}{6}\sqrtsmash[C_G]{20C_g+1},\\[6pt]
    &\text{WZ5--6}:\quad
    \frac{1}{2}(i\slashed{D}_{3/2}^{(1)_-}),\
    0^-(Q^{(1)_{-}}),\
    0^+(\Delta_L^{(2)_{-}}),\quad
    E_0=\frac{3}{2}-\frac{5}{6}+\frac{1}{6}\sqrtsmash[C_G]{20C_g+1}.
\end{align}
The remaining cross diagrams cannot, with the available information from \cite{Nilsson:1983ru,Nilsson:2018lof,Ekhammar:2021gsg,Karlsson:2021oxd,Nilsson:2023ctq} and this paper,  be associated with WZ3--4 or WZ5--6 in a unique way
so we have put all their irreps into one coloured cross diagram in \cref{fig:coloured_cross_diagram}.
However, from the lists of cross diagrams in the
appendix of \cite{Nilsson:2018lof} it seems clear that
each pair, i.e., WZ3--4 and WZ5--6, must be connected to ${\vec 1}+{\vec 5}$ worth of cross diagrams.
This completes the explanation of all the entries
in the Wess--Zumino table for $C_g>0$.

The analysis of the Wess--Zumino multiplets with $C_g=0$ is carried out in detail in \cref{spectrum_summary}, following closely \cite{Nilsson:2023ctq}, and is hence not discussed here.

As should now be clear, using the results obtained in this section all supermultiplets appearing in the left-squashed $S^7$ spectrum (listed in the beginning of \cref{sec:review_and_method:review} and \cref{spectrum_summary}),
except the twelve Wess--Zumino multiplets that involve the Lichnerowicz operator, can be uniquely associated with operator eigenvalues and groups of cross diagrams
from the appendix of \cite{Nilsson:2018lof}. Thus it is possible to decide exactly which cross diagrams these twelve  Wess--Zumino multiplets
must be connected to in order that all the cross diagrams are used in a consistent way. However, to be able to divide these cross diagrams in two groups and associate them with
the two branches of $E_0$ values the information in this paper is not enough. Fortunately, using input from \cite{Duboeuf:2022mam}  this can be done, i.e., the results of \cite{Duboeuf:2022mam} allows us to split \cref{fig:coloured_cross_diagram} (which gives the isometry irrep content of WZ3--4 and WZ5--6 combined) into the two rows in \cref{fig:colour-split}.
\begin{figure}[H]
  \centering
  \begingroup %
  \renewcommand{\scale}{0.39cm} %
  \definecolor{color1}{rgb}{1,0.60,0.75}
  \definecolor{color2}{rgb}{0.65,0.93,1}
  \definecolor{color3}{rgb}{1,0.75,0.50}
  \definecolor{color4}{rgb}{0.72,1,0.56}
  \footnotesize %
  \newcommand{\figMIV}{%
      \begin{tikzpicture}[x=\scale, y=\scale]
          \fill[color2] (3.5,-0.5) rectangle (\range+0.5,\range+0.5);
          \colouredcrossdiagramlayout{p-4}
      \end{tikzpicture}%
  }%
  \newcommand{\figMII}{%
      \begin{tikzpicture}[x=\scale, y=\scale]
          \fill[color1] (2.5,-0.5) rectangle (\range+0.5,0.5);
          \fill[color2] (2.5,0.5) rectangle (\range+0.5,\range+0.5);
          \colouredcrossdiagramlayout{p-2}
      \end{tikzpicture}%
  }%
  \newcommand{\fig}{%
      \begin{tikzpicture}[x=\scale, y=\scale]
          \fill[color1] (-0.5,-0.5) rectangle (0.5,0.5);
          \fill[color1] (-0.5,0.5) rectangle (1.5,1.5);
          \fill[color2] (-0.5,1.5) rectangle (1.5,\range+0.5);
          \fill[color2] (1.5,-0.5) rectangle (\range+0.5,0.5);
          \fill[color3] (1.5,0.5) rectangle (\range+0.5,1.5);
          \fill[color4] (1.5,1.5) rectangle (\range+0.5,\range+0.5);
          \colouredcrossdiagramlayout{p}
      \end{tikzpicture}%
  }%
  \newcommand{\figPII}{%
      \begin{tikzpicture}[x=\scale, y=\scale]
          \fill[color2] (0.5,0.5) rectangle (\range+0.5,\range+0.5);
          \colouredcrossdiagramlayout{p+2}
      \end{tikzpicture}%
  }%
  \newcommand{\figPIV}{%
      \begin{tikzpicture}[x=\scale, y=\scale]
          \fill[color1] (-0.5,-0.5) rectangle (\range+0.5,1.5);
          \fill[color2] (-0.5,1.5) rectangle (\range+0.5,\range+0.5);
          \colouredcrossdiagramlayout{p+4}
      \end{tikzpicture}%
  }%
  \newcommand*{\legendsize}{0.5}%
  \newcommand*{\legendspacing}{2}%
  \hspace*{0.75cm} %
  \begin{tikzpicture}[x=\scale, y=\scale]
    \foreach \i in {1,...,4} {
        \filldraw[draw=black,ultra thin,fill=color\i] ({(\i-1)*(\legendsize+\legendspacing)},0) rectangle ({(\i-1)*(\legendsize+\legendspacing)+\legendsize},\legendsize);
        \node[right] at ({(\i-1)*(\legendsize+\legendspacing)+\legendsize},{\legendsize/2}) {$\i$};
    }
  \end{tikzpicture}
  \begin{tabular}{ccccc}%
    \figMIV  & \figMII  & \fig  & \figPII  & \figPIV
  \end{tabular}%
  \endgroup
  \caption{All isometry irreps of the twelve Wess--Zumino supermultiplet towers WZ3--6 (including the Page supermultiplet), corresponding to two sets of ${\vec 1} \oplus {\vec 5}$. The colours at the intersections of the grid lines indicate the multiplicities of the isometry irreps.}
  \label{fig:coloured_cross_diagram}
\end{figure}
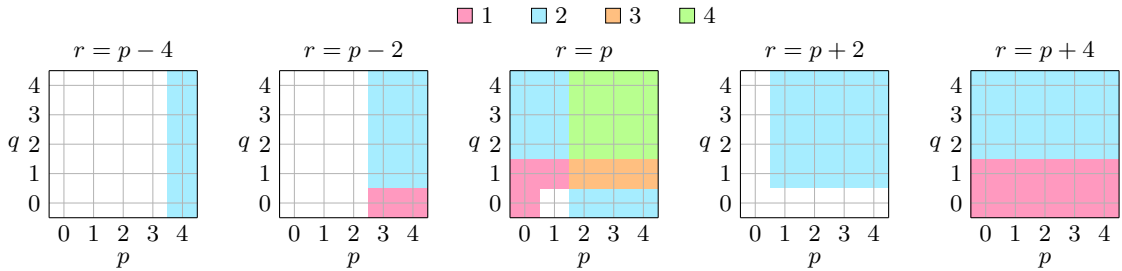

\resettocdepth
\resetsecnumdepth

\phantomsection
\addcontentsline{toc}{section}{References}
\bibliographystyle{myJHEP}
\bibliography{refs}

\end{document}